\documentclass[fleqn,usenatbib]{mnras}

\usepackage{newtxtext,newtxmath}

\usepackage[T1]{fontenc}

\DeclareRobustCommand{\VAN}[3]{#2}
\let\VANthebibliography\thebibliography
\def\thebibliography{\DeclareRobustCommand{\VAN}[3]{##3}\VANthebibliography}

\usepackage{graphicx}	
\usepackage{amsmath}	
\usepackage{bm}
\usepackage{booktabs}
\usepackage{threeparttable}
\usepackage{longtable}
\usepackage{supertabular}
\usepackage{xcolor}
\usepackage{gensymb}
\usepackage[normalem]{ulem}
\usepackage[export]{adjustbox}
\usepackage{hyperref}
\usepackage{csquotes}
\usepackage{subcaption}
\usepackage{multirow}
\renewcommand{\mkbegdispquote}[2]{\itshape}

\usepackage{graphicx}	
\usepackage{amsmath}	

\defcitealias{Mandel22}{M20}
\defcitealias{Thorp21}{T21}
\defcitealias{Avelino19}{Avelino}

\title[$R_V$ Distributions from SN Ia Peak Colours]{
\textsc{Bird-Snack}: Bayesian Inference of dust law $R_V$ Distributions using SN Ia Apparent Colours at peaK
}

\author[S. M. Ward et al.
]{
\parbox{\textwidth}{
\Large
Sam~M.~Ward,$^{1}$\thanks{E-mail:smw92@cam.ac.uk}
Suhail~Dhawan,$^{1}$
Kaisey~S.~Mandel,$^{1,2,3}$
Matthew~Grayling,$^{1}$
and Stephen~Thorp$^{1,4}$
}
\\
\\
\noindent
$^{1}$ Institute of Astronomy and Kavli Insititute for Cosmology, Madingley Road, Cambridge, CB3 0HA, UK\\
$^{2}$Statistical Laboratory, DPMMS, University of Cambridge, Wilberforce Road, Cambridge, CB3 0WB, UK\\
$^{3}$The Alan Turing Institute, Euston Road, London, NW1 2DB, UK\\
$^{4}$The Oskar Klein Centre, Department of Physics, Stockholm University, AlbaNova University Centre, SE 106 91 Stockholm, Sweden
}

\date{Accepted XXX. Received YYY; in original form ZZZ}
\pubyear{2023}
\begin{document}
\label{firstpage}
\pagerange{\pageref{firstpage}--\pageref{lastpage}}
\maketitle

\begin{abstract}
To reduce systematic uncertainties in Type Ia supernova (SN Ia) cosmology, the host galaxy dust law shape parameter, $R_V$, must be accurately constrained. 
We thus develop a computationally-inexpensive pipeline, \href{https://github.com/sam-m-ward/birdsnack/tree/main}{\textsc{Bird-Snack}}, to rapidly infer dust population distributions from optical-near infrared SN colours at peak brightness, and determine which analysis choices significantly impact the population mean $R_V$ inference, $\mu_{R_V}$. 
Our pipeline uses a 2D Gaussian process to measure peak $BVriJH$ apparent magnitudes from SN light curves, 
and a hierarchical Bayesian model to simultaneously constrain population distributions of intrinsic and dust components. 
Fitting a low-to-moderate-reddening sample of 65 low-redshift SNe yields $\mu_{R_V}=2.61^{+0.38}_{-0.35}$, with $68\%(95\%)$ posterior upper bounds on the population dispersion, $\sigma_{R_V}<0.92(1.96)$. 
This result is robust to various analysis choices, including: 
the model for intrinsic colour variations, 
fitting the shape hyperparameter of a gamma dust extinction distribution, 
and cutting the sample based on the availability of data near peak. 
However, these choices may be important if statistical uncertainties are reduced. 
With larger near-future optical and near-infrared SN samples, 
\textsc{Bird-Snack} can be used to better constrain dust distributions, 
and investigate potential correlations with host galaxy properties. 
\textsc{Bird-Snack} is publicly available; 
the modular infrastructure facilitates rapid exploration of custom analysis choices, 
and quick fits to simulated datasets, 
for better interpretation of real-data inferences.
\end{abstract}



\section{Introduction}
\label{S:Intro}
Type Ia supernovae (SNe~Ia) are stellar explosions with standardisable peak luminosities.  They are used as precise distance indicators over a wide range of redshifts ($0.01\lesssim z\lesssim1$), and are thus central to studies of dark energy, and for constraining the Hubble constant~\citep{Riess98:lambda, Perlmutter99, Brout22, Jones22, Riess22}. Standardisation typically involves linear corrections for brightness correlations with SN light curve shape, and apparent colour~\citep{Phillips93, Tripp98}. More recently, a step-function correction for a correlation with host galaxy stellar mass has been applied~\citep{Kelly10, Sullivan10}. However, systematic uncertainties in this SN~Ia standardisation procedure will soon dominate inferences in SN cosmology~\citep{Betoule14,Scolnic18:ps1,Brout22}.

To reduce these uncertainties, the root cause of empirical correlations between SNe~Ia and their host galaxies must be understood. For example, SN Ia luminosities have been reported to correlate with host galaxy stellar mass, star formation rate, stellar age, and metallicity~\citep{Kelly10, Sullivan10,DAndrea11, Childress13,Rigault13, Pan14}. While these effects can be standardised with step-functions for nearby ($z\lesssim 0.1$) SNe Ia, the redshift evolution of galaxies means the SN population --  and hence the standardisation corrections -- may also evolve with redshift. Therefore, the astrophysics that drives SN-host correlations must be robustly modelled, to prevent biasing estimates of cosmological parameters~\citep{Foley12:sdss,Childress14, Scolnic19, Nicolas21}. 

Modelling of host galaxy dust may play a role in empirical SN-host correlations. \cite{Mandel17} show the linear \cite{Tripp98} standardisation formula can lead to biased distance estimates in the tails of the apparent colour distribution. This is because the Tripp formula implicitly models two physically-distinct effects, an intrinsic SN colour-luminosity correlation, and extrinsic host galaxy dust reddening and extinction, together as a single linear relation. This systematic is further complicated by population variations in the host galaxy dust law shape parameter, $R_V$. These variations are expected, given that \cite{Schlafly16} measure an $R_V$ dispersion of $\sigma_{R_V}\approx 0.2$ within the Milky Way. Moreover, a wide range $R_V\approx 1-5$ has been reported in SN~Ia hosts, from analyses of SN~Ia observations, e.g.~\citep{Nobili08,Amanullah15, Cikota16}, or host galaxy SED fitting, e.g.~\citep{Salim18, Meldorf23}. However, there is still limited consensus regarding the host dust distributions. In this paper, we present the \textsc{Bird-Snack} model, a new method for rapid inference of host galaxy dust distributions from optical-NIR SN Ia peak apparent colours, to better understand the systematic uncertainties affecting host dust inferences.

Recent investigations into SN-host correlations have studied the role of the dust law $R_V$ in the `mass step'. The mass step is the empirical correlation that SNe Ia in high stellar mass host galaxies ($\log_{10} M/M_{\odot}\gtrsim 10$) appear optically brighter post-standardisation (with mean luminosity differences between low- and high-mass populations typically $\approx 0.04-0.1$~mag; ~\citealp{Kelly10, Lampeitl10:host, Sullivan10}). On the one hand, extrinsic effects may be the cause, with lower $R_V$ values potentially found in higher stellar mass hosts; when accounted for, these differences may explain some or all of the mass step. Typical population mean $R_V$ values of $\mu_{R_V}\approx 1.5-2.1$ and $\mu_{R_V}\approx 2.8-3.0$ were found in high and low stellar mass host galaxies, respectively, with differences in means $\Delta \mu_{R_V}\approx 0.7-1.3$, by~\cite{Brout21, Popovic23, Meldorf23}. \citealt{Johansson21} also report the non-detection of a mass step at near-infrared (NIR) wavelengths, which is consistent with a dust-based explanation of the mass step (given that NIR photons are weakly sensitive to dust compared to the optical). On the other hand, the mass step could result from intrinsic differences in the SN populations in low and high stellar mass host galaxies, e.g. ~\cite{Briday22}. This hypothesis is supported by the consistency between host-dependent $R_V$ population distributions in \citealp{Thorp21,Thorp22}, with population mean $R_V$ values typically $\mu_{R_V}\approx 2.4-2.8$. Non-zero mass step measurements at NIR wavelengths have also been reported in~\citealp{Uddin20, Ponder21, Jones22}, which further supports this scenario. Or, there may be a mixture of these effects, with host mass potentially tied to both the dust distributions, and the underlying SN Ia population (for further discussion see reviews in e.g. \citealt{Thorp22, Meldorf23}).

It is uncertain then what role $R_V$ plays in producing the mass step. More generally, it is unclear to what extent the modelling of intrinsic and extrinsic effects, or lack thereof, is responsible for empirical SN-host correlations. Therefore, accurately constraining dust population distributions is of central importance in SN cosmology research. 

This motivates that we develop new data-driven methods to rapidly infer $R_V$ distributions in SN Ia hosts, while adopting as few modelling assumptions as possible. This allows us to vary the remaining analysis choices, and discern which assumptions have the largest impact on $R_V$ inferences (e.g. dust parameter priors, intrinsic SN model, preprocessing choices etc.).

We thus build the \href{https://github.com/sam-m-ward/birdsnack/tree/main}{\textsc{Bird-Snack}} model, to perform \textbf{B}ayesian \textbf{I}nference of $\bm{R}_V$ \textbf{D}istributions using \textbf{SN} Ia \textbf{A}pparent \textbf{C}olours at pea\textbf{K}. \textsc{Bird-Snack} fits SN Ia light curves with data near peak time, extracts measurements of peak apparent magnitudes, and then hierarchically infers host galaxy $R_V$ population distributions. This pipeline is largely independent of any SN light curve model. The idea of using multi-band SN~Ia data to constrain dust properties without using distance-luminosity information has been used in previous studies, e.g.~\cite{Nobili08,Burns14,Thorp22}. In particular, the wide wavelength range probed by optical-NIR colours provides more stringent constraints on dust~\citep{Krisciunas07}. Our fiducial result from fitting 65 SNe Ia is a population mean $R_V$, $\mu_{R_V}=2.61^{+0.38}_{-0.35}$, and a Gaussian $R_V$ population dispersion, $\sigma_{R_V}<0.92(1.96)$, with $68\%(95\%)$ posterior upper bounds, respectively. Leveraging our fast inference scheme, we test the sensitivity of this fiducial result to various analysis choices. \textsc{Bird-Snack} also enables us to generate and fit many simulated datasets, to better interpret the real-data inferences. Our analysis pipelines are publicly available at \href{https://github.com/sam-m-ward/birdsnack/tree/main}{\texttt{https://github.com/birdsnack}}. 

In \S\ref{S:DataMethods}, we describe our fiducial sample of SN~Ia light curves, and the preprocessing pipeline for measuring rest-frame apparent magnitudes at peak brightness. In \S\ref{S:BayesianModel}, we detail the hierarchical Bayesian model we use to infer the population distributions of intrinsic chromatic variations, host galaxy dust extinction, and dust law shape. We perform our analysis in \S\ref{S:Analysis}, and discuss and conclude in \S\ref{S:Conclusions}.

\section{Datasets \& Preprocessing}
\label{S:DataMethods}

\subsection{SN Ia Sample}
\label{S:SNSample}
We compile photometry from three literature surveys of SNe Ia with both optical and NIR data near peak brightness. Firstly, we include the set of well-calibrated $uBgVriYJH$ light curves from the first stage of the Carnegie Supernova Project (CSP-I;~\citealt{Krisciunas17}). This comprises high-cadence observations of 134 SNe Ia in the redshift range $0.0037\leq z\leq 0.0835$. Next, we include $uBgVriYJHK$ light curves of 94 SNe Ia from the CfA3, CfA4 and CfAIR2 surveys ($0.0027\leq z\leq 0.0745$;~\citealt{Wood-Vasey08,Hicken09:lc, Hicken12, Friedman15}). We also include the `RATIR' sample: $uBgVriYJHK$ light curves of 42 SNe~Ia ($0.0007\leq z\leq 0.1324$) from the intermediate Palomar Transient Factory survey (iPTF;~\citealt{Johansson21}).

To increase the sample further, we perform a literature search, and compile a `Miscellaneous' sample of SN Ia photometry from various sources:~\cite{Jha99:98bu,Krisciunas00,Krisciunas01,Krisciunas03,Valentini03,Krisciunas04:4sn,Krisciunas04:7sn,Elias-Rosa06,Krisciunas07,Stanishev07:03du,Pignata08:02dj,Matheson12,Cartier14,Marion15,Zhang16:11fe,Burns20}. We use photometry of all SNe Ia referenced therein, except for SNe 1991T, 1991bg, and 1999aa because they are spectroscopically peculiar~\citep{Krisciunas00, Krisciunas04:7sn}, and SNe 1999da, 1999dk and 2013aa because they lack NIR observations~\citep{Krisciunas01}. The `Miscellaneous' sample totals $uBgVriJHK$ observations of 25 spectroscopically normal SNe Ia in the redshift range $0.0007\leq z\leq0.0296$.

Our total literature sample comprises 269 unique SNe Ia ($0.0007<z<0.1324$), 88 of which have light curves from multiple literature sources (for these SNe we introduce a pecking order system to select a single dataset; see $\S$~\ref{S:DataPreProcessing}). 

We further compile metadata of the spectroscopic sub-classification, and host galaxy stellar mass, from the following literature sources: \cite{Neill09,Kelly10,Friedman15,Krisciunas17,Rose19, Uddin20,Ponder21,Johansson21}. Following \cite{Johansson21}, we set all SNe Ia from the iPTF survey to be spectroscopically normal, except for: SNe~iPTF13abc, iPTF13ebh, iPTF14ale, iPTF14apg, iPTF14atg, iPTF14bdn. Where there are multiple metadata entries for a given SN, we set the SN class to be normal only if all the entries are normal, and we take the sample average of host galaxy stellar masses.

\subsection{Data Preprocessing \& Cuts}
\label{S:DataPreProcessing}

\begin{figure}
    \centering
    \includegraphics[width=1\linewidth]{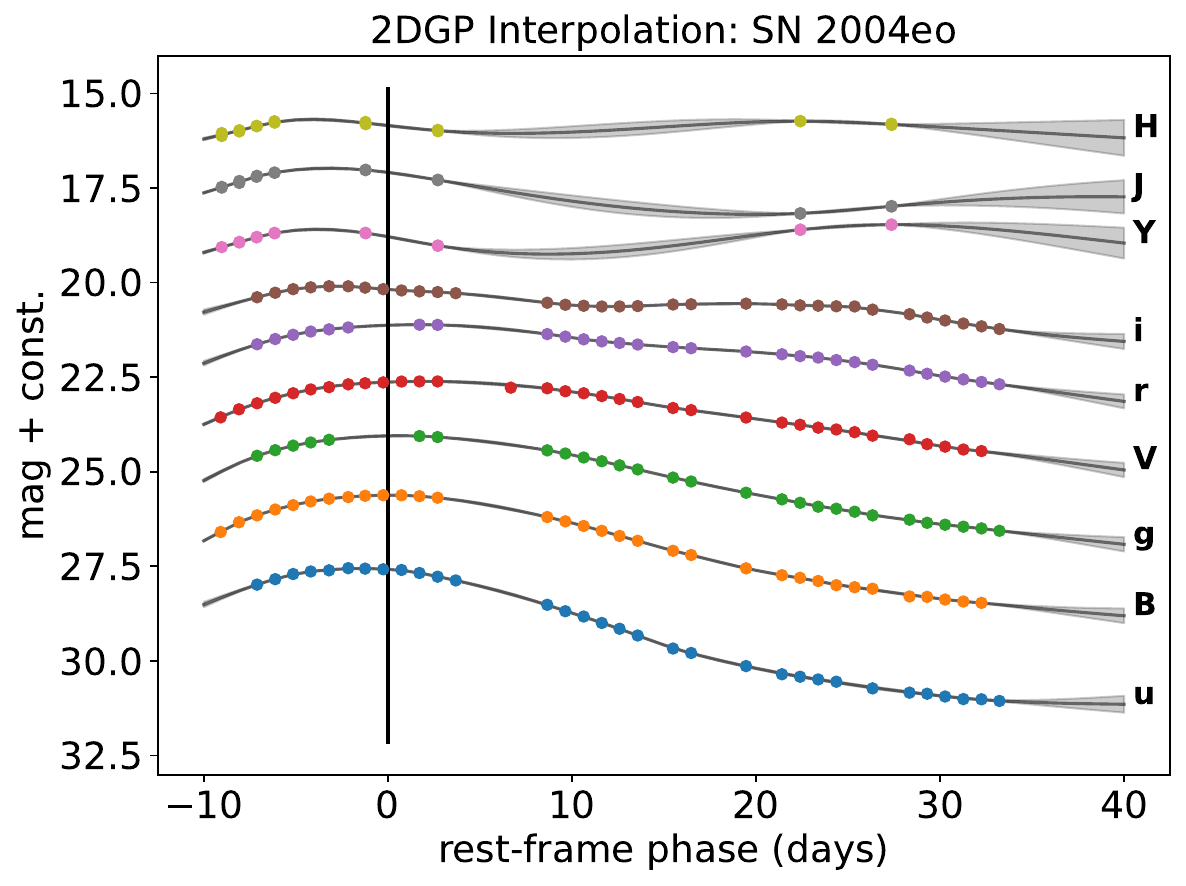}
    \caption{Example fit to $uBgVriYJH$ rest-frame apparent magnitudes data of SN~2004eo, using a 2D Gaussian Process in the phase range [-10, 40]~days. To compute these data, we keep observer-frame data with SNR$>$3, fit \textsc{SNooPy} to the interpolation filters ($uBgVriYJH$), then apply a Milky Way extinction correction, and mangled K-corrections. We then estimate $T_{B;\,\rm{max}}$ using a 1DGP fit to the $B$-band. Finally, we extract 2DGP $BVriJH$ measurements at the epoch of $B$-band maximum brightness, $T_{B;\,\rm{max}}$, depicted by the vertical solid line (more details in $\S$~\ref{S:DataPreProcessing}).}
    \label{fig:ExampleLC}
\end{figure}

We detail our data preprocessing pipeline, which we use to perform selection cuts, and measure rest-frame peak apparent magnitudes from the light curves. The number of SNe retained after each cut is recorded in Table~\ref{tab:NSNcuts}, and an example fit to rest-frame data is shown in Fig.~\ref{fig:ExampleLC}. The final fiducial sample comprises 69 SNe Ia, of which 62 have rest-frame peak apparent $|B-V|<0.3$~mag.

Firstly, we require each SN to have at least one observation in the peak-magnitude passbands: $BVriJH$. Next, we apply an SNR>3 cut, to remove noisy data points. We then fit the observer-frame data with \textsc{SNooPy}\footnote{\url{https://csp.obs.carnegiescience.edu/data/snpy}}~\citep{Burns11,Burns14}, to apply 
both Milky Way extinction corrections (with $R_{V;\,\rm{MW}}=3.1$) and
K-corrections to the observed data. These corrections are affected by the set of observer-frame filters that we fit (because the multi-band data are fitted simultaneously by \textsc{SNooPy}). By default, we fit all observer-frame filters that \textsc{SNooPy} naturally maps back to the rest-frame CSP filters. This choice is defined by the `interpolation filters', which we set to be: $uBgVriYJH$\footnote{To boost the sample size, we re-assign any observer-frame filters that map to the filter keys: \texttt{[Bs,Vs,Rs,Is]}, to map instead to their respective CSP filter keys: \texttt{[B,V,r,i]} (provided the latter is not already listed in an SN's set of rest-frame filters); we also map all \texttt{[J\_K,H\_K]} keys to \texttt{[J,H]}. Implementing these two rules boost the fiducial sample size from 28 to 69 SNe; passband transmission functions and filter keys can be found at \url{https://github.com/obscode/snpy/tree/master/snpy/filters}.}. The time of maximum, $T_{B;\,\rm{max}}$ (defined as the time of $B$-band maximum brightness in the rest-frame) is a free parameter during \textsc{SNooPy} fitting. We apply mangled K-corrections, which means \textsc{SNooPy} warps the \cite{Hsiao09} SED template to match the measured observer-frame apparent magnitudes. The SN is dropped from the sample if the \textsc{SNooPy} fit and/or the K-correction computation failed.

With rest-frame data computed, we proceed to re-estimate $T_{B;\,\max}$. We use the \texttt{george} package\footnote{\url{https://george.readthedocs.io/en/latest/}}, and fit a 1D Gaussian Process (1DGP) to the rest-frame $B$-band magnitude data. The time of maximum defines rest-frame phase, $t$, via:
\begin{equation}
\label{eq:phasecomp}
    t = \frac{T_{\rm{MJD}}-T_{B;\,\rm{max}}}{1+z_{\rm{Helio}}},
\end{equation}
where $z_{\rm{Helio}}$ is the observed (heliocentric) redshift. We make an initial estimate by fitting all the $B$-band data. We then cut data outside the phase range [-10, 40]~days, re-fit, draw 1000 GP samples (estimating $T_{B;\,\rm{max}}$ for each), and use the sample mean. 

Next, we trim the SN sample by imposing that there must be `enough data near peak' (where peak is defined as the epoch of $B$-band maximum brightness, $T_{B;\,\rm{max}}$, for all passbands, i.e. $t=0$). This ensures our peak magnitude estimates are data-driven. For the reference $B$-band, we make stringent cuts, imposing that there must be at least 2 data points before peak, and 2 data points after peak. For the remaining passbands in $BVriJH$, we require 1 data point before peak, and 1 data point after peak. 

At this stage, we remove light curves of SNe from multiple literature sources. For these SNe, we select the dataset that passes the above cuts, and is highest in a pecking order. From highest to lowest, this pecking order is: CSP, CfA, RATIR, Miscellaneous.

We then use a 2DGP to simultaneously fit the rest-frame magnitude data, and extract apparent magnitude measurements at peak time\footnote{For simplicity, we ignore covariances between the fitted magnitudes in different passbands that arise from the simultaneous 2DGP fit.}. For 2DGP interpolation, we use the methodologies in \cite{Boone19}. For some SNe in the RATIR sample, we scale the measurement errors by a pre-defined factor to prevent the time scale hyperparameter from becoming too small (i.e. $\ll$1~day variability); we trial factors equal to 1.5, then 2, and then increment by 1 until the fit looks reasonable upon visual inspection.
The new scaling factors of SNe in the final fiducial sample are 1.5 for iPTF13azs and iPTF16abc, and 5 for iPTF16auf. 

Finally, we cut the sample to retain only spectroscopically normal SNe Ia, with measurement errors $<0.3$~mag (corresponding to an SNR>3). Our fiducial sample thus comprises 69 objects (see Table~\ref{tab:NSNcuts}), of which there are 33 SNe from CSP, 16 from CfA, 5 from RATIR, and 15 from miscellaneous sources. Each SN in this sample has at least one data point within the phase window $-6<t<4$~days in each $BVriJH$ passband. We also perform additional cuts. High reddening SNe Ia are typically excluded from a cosmological sample, so we apply an apparent colour cut of $|B-V|<0.3$~mag, which reduces the fiducial sample to 62 SNe. Alternatively, the sample is split at a host galaxy stellar mass of $\log_{10} M/M_{\odot} = 10$; we are missing host mass metadata for 5 SNe (Table~\ref{tab:NSNcuts}), so the high/low sample split is 47/17, respectively. Applying both cuts yields a sample split of 42/15.

\begin{table}
\begin{threeparttable}
    \caption{Sample Cuts.}
    \label{tab:NSNcuts}
    \begin{tabular}{l c} 
    \toprule
        Cuts & No. of SNe After Cut\\
\midrule
Initial Sample & 269 \\ 
Has $BVriJH$ & 192 \\
Successful \textsc{SNooPy} Fit/K-corrections & 174 \\ 
$B$-band points near peak (2 before; 2 after) & 100 
\\ 
$Vri$ points near peak (1 before; 1 after) & 100 
\\
$JH$ points near peak (1 before; 1 after) & 79 
\\
Spectroscopically Normal & 70 
\\ 
Mag. Errors $<0.3$~mag \textbf{[Fiducial Sample]} & \textbf{69} \\
\midrule
\midrule
Additional Cuts \\
\midrule
$|B-V|<0.3$~mag & 62 \\
\midrule
High, Low, N/A Host Galaxy Stellar Mass\tnote{a} & 47, 17, 5 \\
\midrule
$|B-V| \&$ Mass Cuts & 42, 15, 5\\

\bottomrule
    \end{tabular}
    \begin{tablenotes}
        \item [a] Host galaxy stellar mass metadata is missing for the five following supernovae: SNe 2001bt, 2001cz, 2011by, 2011fe, 2017cbv.
    \end{tablenotes}
    \end{threeparttable}
\end{table}

\section{Modelling}
\label{S:BayesianModel}
We construct a hierarchical Bayesian model (HBM) to infer the $R_V$ population distribution from the peak apparent magnitude measurements. Using MCMC to fit the HBM to measurements of $S$ unique supernovae results in posterior inferences of the intrinsic and extrinsic population hyperparameters~\citep{Mandel09, Mandel11}.

\subsection{Intrinsic Deviations}
\label{S:IntrinsicDev}
A priori, the choice of intrinsic colours that are modelled may affect the dust hyperparameter inferences. For example, we can choose to model adjacent, $B-X$, or $X-H$ intrinsic colours using a multivariate Gaussian distribution\footnote{The transformation of colours \textit{data} is arbitrary. Different colour datasets, e.g. adjacent, $B-X$, $X-H$ colours etc., are linear transformations of one another; therefore, they contain the same information, so the inferences are the same for a fixed choice of model. It is the hyperpriors on intrinsic chromatic hyperparameters that can affect dust inferences. The latent intrinsic parameters can be transformed to fit an arbitrary set of colours data.}. However, there are many other colour combinations, so there is a degree of arbitrariness to this choice.

\begin{figure*}
    \centering
    \includegraphics[width=1\linewidth]{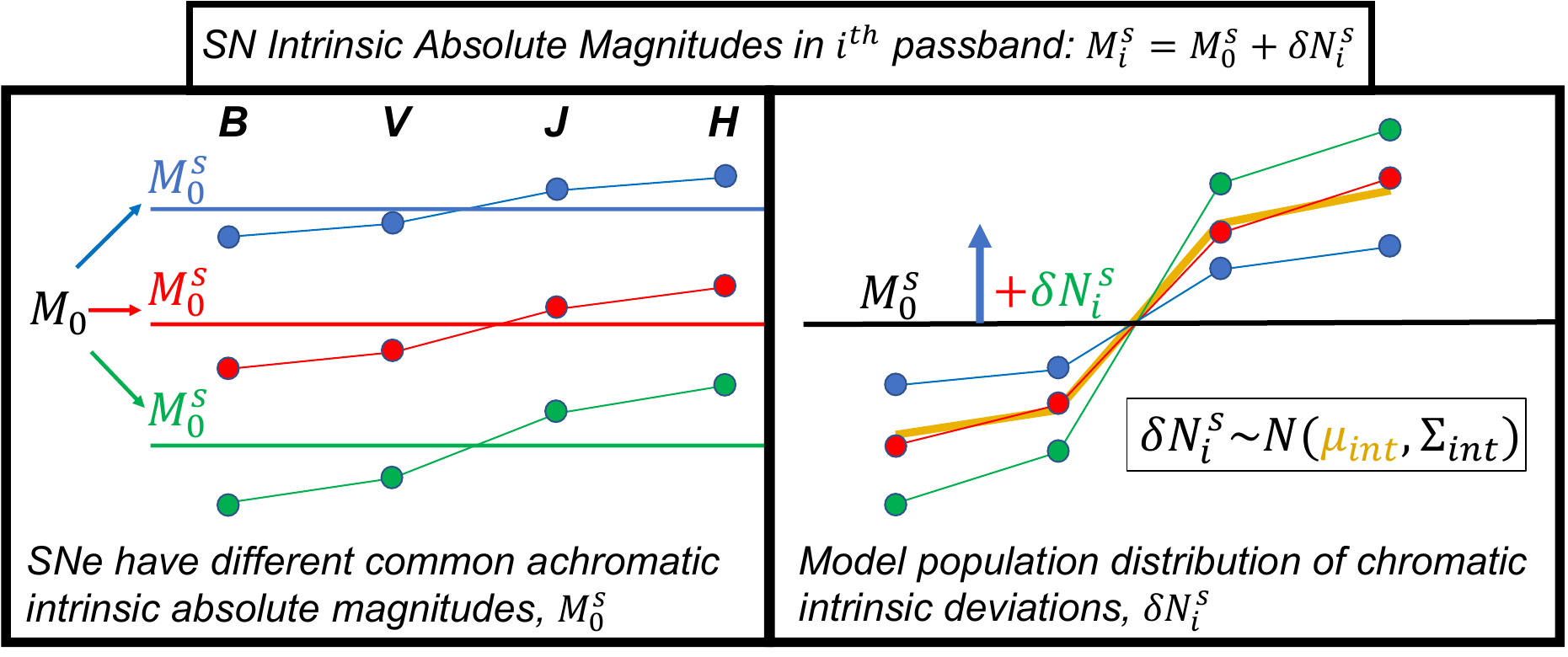}
    \caption{Schematic depicts the key features of our default intrinsic deviations hierarchical model. For $N$ passbands, there are $N$ modes of chromatic variation about the common achromatic component. (left panel) Intrinsic absolute magnitudes in $BVJH$ passbands are depicted for three different SNe. The intrinsic absolute magnitudes in each passband, $M_i^s$, are the sum of a common achromatic intrinsic absolute magnitude, $M_0^s$, and additional chromatic intrinsic deviations, $\delta N_i^s$. (right panel) The intrinsic deviations, $\delta N_i^s$, are modelled using a multivariate Gaussian distribution; the population mean intrinsic deviation vector, $\bm{\mu}_{\rm{int}}$, is highlighted in yellow.}
    \label{fig:MagDeviationsCartoon}
\end{figure*}

Our default choice is to work in magnitude space, and model population distributions of chromatic intrinsic deviations from each supernova's common achromatic magnitude component. This choice bypasses the need to pick an arbitrary set of colours. We model the intrinsic absolute magnitude in the $i$th passband, $M^s_{i}$, as the sum of an achromatic intrinsic absolute magnitude that is common to all passbands, $M_0^s$, and a chromatic intrinsic deviation, $\delta N^s_{i}$.
\begin{equation}
    M_i^s = M_0^s + \delta N_i^s
\end{equation}
Fig.~\ref{fig:MagDeviationsCartoon} visualises this intrinsic deviations component of our model. The $M_0^s$ component is degenerate with distance, and is strongly dependent on SN light curve shape, and any (partial) achromatic correlations with the host galaxy, such as the mass step. We group these effects together with the distance by modelling and marginalising over a nuisance parameter, $m_0^s$, the common apparent magnitude,
\begin{equation}
    m_0^s = \mu^s + M_0^s,
\end{equation}
where $\mu^s$ is the distance modulus. We do not impose an external constraint on the distance (e.g. a redshift-based distance estimate).

\begin{figure*}
    \centering
    \includegraphics[width=1\linewidth]{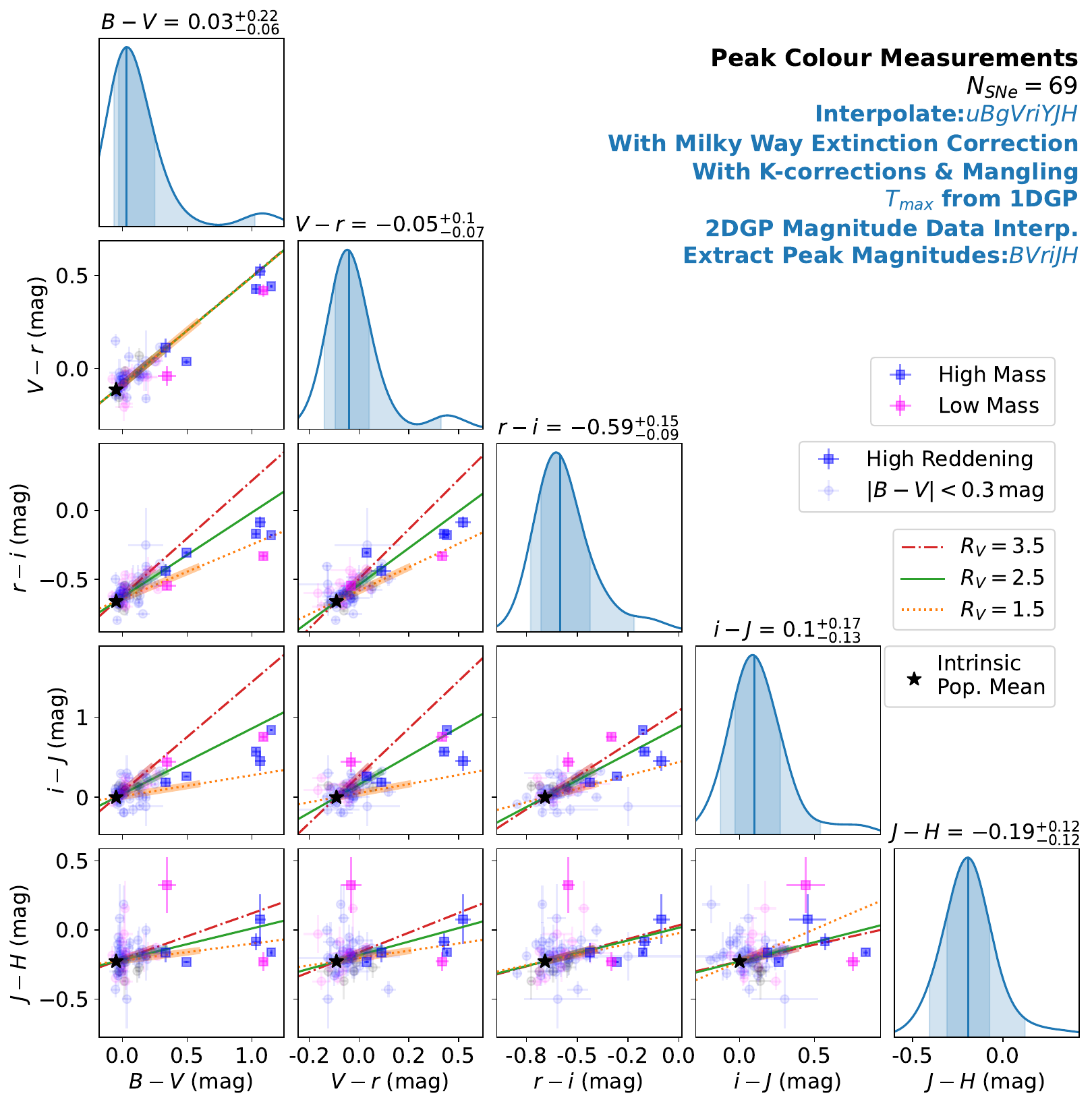}
    \caption{2DGP estimates of rest-frame peak apparent colours in our fiducial sample of 69 SNe Ia. 1D kernel density estimates (KDEs) show how the sample's colours are distributed, with the sample mean, 16\% and 84\% quantiles quoted. Scatter points are coloured according to their host galaxy stellar mass (cut at $\log_{10} M/M_{\odot}=10$), and marked according to their $B-V$ apparent colour. We also display \citealt{Fitzpatrick99} dust law projections for three choices of $R_V$, and highlight the regions from $A_V=0$ to $A_V=1$; the intersection point at $A_V=0$ is chosen to coincide with the population mean intrinsic colour inferred in the fiducial analysis (\S\ref{S:FiducialResults}). The $B-V$ vs. $V-r$ panel strongly indicates a population distribution of intrinsic colours, given that the dust contribution in this plane is weakly dependent on $R_V$, and that the scatter points are distributed in a locus rather than along a line; similarly, the small coefficients on $A_V$ in the $i-J$ vs. $J-H$ panel indicate dust cannot be responsible for the large sample dispersion. The high-reddening objects typically reside in the $R_V \approx 1.5-3.5$ estimated regions. The $B-V$ vs. $V-r$ panel shows there are 4 highly reddened objects with $B-V>1.0$~mag, which produce distinct bumps in the 1D KDEs.
    }
    \label{fig:ColourCorner}
\end{figure*}

We then model population distributions of the intrinsic and extrinsic chromatic deviations from the common apparent magnitudes,
\begin{equation}
\label{eq:magsintermsofdeviations}
    m_i^s = m_0^s + \delta N_i^s + A^s_V \xi_i(R_V^s),
\end{equation}
where $m_i^s$ is the apparent magnitude in the $i$th passband, and $A_V^s, R_V^s, \xi$ are the dust extinction and dust law shape parameters, and the \cite{Fitzpatrick99} dust law, respectively. In each passband, the dust law is evaluated at an effective wavelength (see \S\ref{S:EffectiveLambda}). The apparent colours are the differences between apparent deviations, i.e.
\begin{equation}
    c^s_{ij} = m_i^s - m_j^s = \delta N_i^s - \delta N_j^s + A_V^s\cdot[\xi_i(R_V^s) -\xi_j(R_V^s)].
\end{equation}
Fig.~\ref{fig:ColourCorner} visualises the estimates of adjacent apparent colours at peak for our fiducial sample of 69 SNe~Ia. This plot provides an intuition for how photometric information alone can be used to constrain dust population distributions (i.e. without assuming any cosmology).

\subsection{Generative Model}
\label{S:GenerativeModel}

\renewcommand{\arraystretch}{1}
\setlength{\tabcolsep}{3pt}
\begin{table}
\begin{threeparttable}
    \caption{Population hyperparameters and SN-level parameters for default intrinsic deviations model.}
    \label{tab:modelparams}
    \begin{tabular}{l c c l} 
    \toprule
    Type & Hyperparameters & Parameters & Pop. Dist. / Prior\\
    \midrule
    Extrinsic & $\tau_A$ & $A_V^s$ & Exponential\\
    - & $\mu_{R_V}, \sigma_{R_V}$ & $R_V^s$ & Truncated Gaussian\\
    Intrinsic & $\bm{\mu}_{\rm{int}}, \bm{\Sigma}_{\rm{int}}$ & $\delta N_i^s$ & Multivariate Gaussian \\
    Achromatic & - & $m_0^s$ & Uniform \\
\bottomrule
    \end{tabular}
    \end{threeparttable}
\end{table}

\begin{table}
\begin{threeparttable}
    \caption{Choice of intrinsic SN model. The deviations model gives equal weighting to all passbands, constraining $N$ intrinsic deviations per SN for $N$ passbands. The intrinsic colours model can be specified in an arbitrary reference frame, e.g. adjacent, $B-X$, $X-H$ colours etc., and is used to constrain $N-1$ colours per SN.}
    \label{tab:modelchoice}
    \begin{tabular}{l l l} 
    \toprule
    Model & Intrinsic Distribution & Data Fitted \\
    \midrule
    Deviations & $\bm{\delta N}^s
    \sim \mathcal{N}(\bm{\mu}_{\rm{int}},\bm{\Sigma}_{\rm{int}})$ & Apparent Magnitudes or Colours\\
    Colours & $\bm{c}_{\rm{int}}^s
    \sim \mathcal{N}(\bm{\mu}_{c,\rm{int}},\bm{\Sigma}_{c,\rm{int}})$ & Apparent Colours\\
\bottomrule
    \end{tabular}
    \end{threeparttable}
\end{table}

\subsubsection{Intrinsic Deviations Model (Magnitudes-Level Modelling)}
\label{S:ModelM2M}
We detail our default intrinsic deviations HBM, which we use to fit apparent magnitudes data. The intrinsic deviations are drawn from a multivariate Gaussian distribution, which is characterised by two sets of hyperparameters: the mean intrinsic deviation vector, $\bm{\mu}_{\rm{int}}$, and an intrinsic deviation covariance matrix, $\bm{\Sigma}_{\rm{int}}$. 
\begin{equation}
    \bm{\delta N}^s \sim \mathcal{N}(\bm{\mu}_{\rm{int}},\bm{\Sigma}_{\rm{int}})
\end{equation}
We model the dust extinction, $A_V^s$, as being drawn from an exponential distribution, which is characterised by the dust extinction hyperparameter, $\tau_A$.
\begin{equation}
\label{eq:drawAV}
     A^s_V \sim \rm{Exp}(\tau_A)
\end{equation}
The dust law shapes, $R_V^s$, are assumed to be drawn from a truncated Gaussian distribution, with a lower bound of $R_V^s=1$. This distribution is characterised by the mean, $\mu_{R_V}$, and dispersion, $\sigma_{R_V}$, hyperparameters.
\begin{equation}
\label{eq:drawRV}
    R^s_V \sim \text{Trunc-}\mathcal{N}(\mu_{R_V}, \sigma_{R_V},
    \rm{Min.}=1,\rm{Max.}=\infty)
\end{equation}
Finally, the common apparent magnitude parameters, $m_0^s$, have an uninformative prior,
\begin{equation}
    m_0^s \sim U(0,100).
\end{equation}
The latent parameters are then combined to yield the latent vector of extinguished apparent magnitudes, $\bm{m}^s = \{m_i^s\}_{i=1}^N$, as in Eq.~\ref{eq:magsintermsofdeviations}. Noisy measurements, $\{ \hat{m}^s_i, \hat{\sigma}^s_i \}_{i=1}^N $, are then made of each $m_i^s$, so the  measurement-likelihood function is:
\begin{equation}
\label{eq:measurement}
    \hat{m}^s_i \sim \mathcal{N} (m^s_i,\hat{\sigma}_i^s).
\end{equation}

To set the zero-point of the common apparent magnitudes, $m_0^s$, or equivalently, to break the achromatic degeneracy between the intrinsic deviations and $m_0^s$, we require that a component of the population mean intrinsic deviation vector, $\bm{\mu}_{\rm{int}}$, is fixed to a constant. A simple solution, and our default choice, is to fix the $B$-band component to zero, $\mu_{\rm{int};\,B}=0$. Comparing this to the right panel in Fig.~\ref{fig:MagDeviationsCartoon}, this is equivalent to shifting all the magnitude points up by  $|\mu_{\rm{int};\,B}|$. Our hyperprior on the complementary vector that excludes the reference band, $\bm{\mu}_{\rm{int}, \backslash\rm{ref}}$, is a wide Gaussian distribution,
\begin{equation}
    \bm{\mu}_{\rm{int}, \backslash\rm{ref}}\sim \mathcal{N}(\bm{0},10\cdot\mathbb{I}),
\end{equation}
where $\mathbb{I}$ is the identity matrix. We stress that the choice of reference band is arbitrary, and does not affect the dust hyperparameter inferences. This is because the choice of reference band is equivalent to a linear achromatic shift, $\mu_{\rm{int},\,\rm{ref}}$, that is added to all the common apparent magnitudes, thus setting the zero-point; meanwhile, the chromatic deviations from this zero-point are used to constrain the dust population distributions\footnote{Another choice, which leads to $\bm{\mu}_{\rm{int}}$ centred on zero like in Fig.~\ref{fig:MagDeviationsCartoon}, is to sample a unit $N$-simplex from a Dirchlet(1) distribution, $\bm{\eta}_{\rm{int}}$, and transform it using $\bm{\mu}_{\rm{int}} = 100(\bm{\eta}_{\rm{int}}-1/N)$. This yields identical results to the default hyperprior, and this $\bm{\mu}_{\rm{int}}$ has a passbands-average equal to zero, but we judge the Gaussian prior is more interpretable than the Dirichlet prior; \url{https://mc-stan.org/docs/functions-reference/dirichlet-distribution.html.}}.

The intrinsic deviations covariance matrix, $\bm{\Sigma}_{\rm{int}}$, is separated into a correlation matrix, $\bm{\Sigma}^{\rm{corr}}_{\rm{int}}$, and a vector of dispersions, $\bm{\sigma}_{\rm{int}}$, which are related via:
\begin{equation}
    \bm{\Sigma}_{\rm{int}} = \rm{diag}(\bm{\sigma}_{\rm{int}})\, \bm{\Sigma}^{\rm{corr}}_{\rm{int}} \, \rm{diag}(\bm{\sigma}_{\rm{int}}).
\end{equation}
The hyperpriors on these components are:
\begingroup
\allowdisplaybreaks
\begin{align}
\label{eq:LKJ}
    \bm{\Sigma}^{\rm{corr}}_{\rm{int}} &\sim LKJ(1),
    \\
    \bm{\sigma}_{\rm{int}} &\sim \text{Half-Cauchy}(0,1).
\end{align}
\endgroup
The $LKJ$ prior from \cite{LKJ09} places a uniform prior on positive semi-definite correlation matrices. The unit half-Cauchy prior reflects our expectations that each element in $\bm{\sigma}_{\rm{int}}$ is of order tenths of a magnitude, while also placing relatively little prior probability at values $\gtrsim 2$~mag. Following \cite{Thorp21}, the hyperpriors on the dust hyperparameters are:
\begingroup
\allowdisplaybreaks
\begin{align}
    \tau_A &\sim \text{Half-Cauchy}(0,1), 
    \\
    \mu_{R_V} &\sim U(1,5),
    \\
    \sigma_{R_V} &\sim \textrm{Half-}\mathcal{N}(0,2^2).
\end{align}
\endgroup

We build this model using the Stan probabilistic programming language, which uses Hamiltonian Monte Carlo (HMC) methods to perform posterior inferences~\citep{Hoffman14, Betancourt16, Carpenter17,Stan20}. Table~\ref{tab:modelparams} summarises the population hyperparameters, SN parameters, and their population distributions. The posterior probability distribution is decomposed as the product of the measurement-likelihood functions, and the (hyper)prior probability distributions:
\begin{equation}
\begin{split}
    &P(\{\bm{\phi}_s\}_{s=1}^S,\tau_A,\mu_{R_V},\sigma_{R_V},\bm{\mu}_{\rm{int}},\bm{\Sigma}_{\rm{int}} | \{\bm{\hat{m}}_s\}_{s=1}^S) \propto \\
    &\Big[\prod_{s=1}^{S} P(\bm{\hat{m}}_s| \bm{\delta N}_s, A_V^s, R_V^s,m_0^s) \times \\
    &P(\bm{\delta N}_s | \bm{\mu}_{\rm{int}}, \bm{\Sigma}_{\rm{int}})\times P(A_V^s | \tau_A) \times P(R_V^s | \mu_{R_V}, \sigma_{R_V}) \times P(m_0^s)\Big] \times  \\
    &P(\bm{\mu}_{\rm{int}}) \times P(\bm{\Sigma}_{\rm{int}}) \times P(\tau_A) \times P(\mu_{R_V}) \times P(\sigma_{R_V}),
\end{split}
\end{equation}
where $\bm{\phi}_s = \{\bm{\delta N}_s, A_V^s, R_V^s, m_0^s\}$ are SN-level latent parameters. For analysis in \S\ref{S:Analysis}, we use standard procedures and diagnostics to run and assess the quality of our MCMC chains. We run 4 independent chains, and randomly initialise the parameter locations. We use the Gelman-Rubin statistic to assess the mixing and convergence of chains~\citep{Gelman92, Vehtari21}, and confirm that there are no divergent transitions~\citep{Betancourt14,BetancourtGirolami15,Betancourt17}.

\subsubsection{Intrinsic Colours Model (Colours-Level Modelling)}
\label{S:ModelC2C}
A separate methodology is to directly model the population distribution of intrinsic colours~\citep{Mandel14}, rather than intrinsic deviations. This alternative model requires that a specific set of intrinsic colours is selected, on which the priors are placed. We find this \textit{modelling} choice affects the posterior inferences (regardless of the transformation of colours \textit{data}).

For this intrinsic colours model, we define a multivariate Gaussian population distribution for the intrinsic colours, $\bm{c}^s_{\rm{int}}$, just like the deviations:
\begin{equation}
    \bm{c}^s_{\rm{int}} \sim \mathcal{N} (\bm{\mu}_{c,\rm{int}},\bm{\Sigma}_{c,\rm{int}}).
\end{equation}
The colours are distance independent, so the achromatic zero point does not need to be defined like in the deviations model; therefore, we place a weakly informative Gaussian hyperprior on all elements of the population mean intrinsic colour vector,
\begin{equation}
\label{eq:mucintprior}
    \bm{\mu}_{c,\rm{int}} \sim \mathcal{N}(\bm{0},10\cdot \mathbb{I}).
\end{equation}
The hyperpriors on the intrinsic colour covariance matrix are the same as for the intrinsic deviations.

The latent apparent colours are
\begin{equation}
    \bm{c}^s = \bm{c}^s_{\rm{int}} + A_V^s \bm{\Delta \xi}(R_V^s),
\end{equation}
where $\Delta \xi_{ij} = \xi_i - \xi_j$. The measurement-likelihood function is
\begin{equation}
    \hat{\bm{c}}^s \sim \mathcal{N} (\bm{c}^s, \hat{\bm{\Sigma}}^s_{c}),
\end{equation}
where $\hat{\bm{\Sigma}}^s_{c}$ is the measurement error covariance matrix,
\begin{equation}
    \hat{\bm{\Sigma}}^s_{c} = \mathbf{Q}\cdot \text{diag}(\hat{\bm{\sigma}}^2_{s})\cdot \mathbf{Q} ^T,
\end{equation}
and $\mathbf{Q}$ is an $N-1 \times N$ matrix that transforms apparent magnitudes to a choice of colours; for example, to create adjacent apparent colours from $BVriJH$ apparent magnitudes ($N=6$), the transformation is:
\begin{equation}
\mathbf{Q} = 
\begin{bmatrix}
1 & -1 &  0 &  0 &  0 &  0 \\
0 &  1 & -1 &  0 &  0 &  0 \\
0 &  0 &  1 & -1 &  0 &  0 \\
0 &  0 &  0 &  1 & -1 &  0 \\
0 &  0 &  0 &  0 &  1 & -1 \\
\end{bmatrix}
. 
\end{equation}
Table~\ref{tab:modelchoice} contrasts the intrinsic deviations and intrinsic colour models.

\subsubsection{Light Curve Shape Modelling}
\label{S:ModelLCShape}
Correlations between intrinsic colours and light curve shape are reported in e.g.~\cite{Jha07, Nobili08, Burns14, Mandel22}. Therefore, we include an optional extension to the intrinsic deviation and intrinsic colour models that incorporates each SN's dependence on light curve shape. To do this, we estimate $\Delta m^s_{15}(B) = m^s_{15}(B)-m^s_0(B)$ for each SN using the GP fit with \textsc{Bird-Snack}, and record the point estimate and the measurement error. We place a flat prior on the $\Delta m^s_{15}(B)$ parameters, and use a Gaussian measurement-likelihood function. We then model a linear dependence on the latent $\Delta m^s_{15}(B)$ parameters using a slope vector $\bm{\alpha}_{\rm{int;\,shape}}$(which has the same hyperpriors as $\bm{\mu}_{\rm{int}}$; see \S\ref{S:ModelM2M},~\ref{S:ModelC2C}). This is added either to the latent apparent magnitudes, $\bm{m}^s$, via:
\begin{equation}
    \bm{m}^s \to \bm{m}^s + [\Delta m_{15}^s(B)-1.05] \cdot \bm{\alpha}_{\rm{int;\,shape}},
\end{equation}
or to the latent apparent colours, $\bm{c}^s$. The arbitrary \cite{Tripp98} 1.05 zero-point does not affect inferences. 

\subsection{Effective Wavelengths}
\label{S:EffectiveLambda}

\subsubsection{SED Approximation}
To ensure $R_V$ inferences are fast, we do not model an SED surface in the hierarchical inference, and instead evaluate the dust law at a set of effective wavelengths. The extinction in the generic $X$-band, $A^s_X$, transforms the intrinsic magnitudes, $m^{\rm{int}}_{X,s}$, into extinguished magnitudes, $m^{\rm{ext}}_{X,s}$:
\begin{equation}
    m^{\rm{ext}}_{X,s} = m^{\rm{int}}_{X,s} + A^s_X. 
\end{equation}
Therefore, $A_X^s$ depends not only on the properties of the dust, but also on the intrinsic SN Ia flux surface: ${\rm{SED}}_s^{\rm{int}}(\lambda)$. 
The correct transformation is:
\begin{equation}
    A^s_X = -2.5\log_{10}\Bigg( \frac{ \int {\rm{SED}
    }_s^{\rm{int}}(\lambda) \cdot 10^{-0.4 \, A^s_V \, \xi(\lambda, R^s_V)} \cdot T_X(\lambda)\lambda\, d\lambda }{ \int {\rm{SED}}_s^{\rm{int}}(\lambda) \cdot T_X(\lambda)\lambda \, d\lambda}\Bigg).
\end{equation}
However, we make the approximation that the dust extinction is constant in wavelength over the $X$-band transmission function, $T_X(\lambda)$. Adopting this SED approximation, we evaluate the dust law at an effective wavelength, $\lambda^X_{\rm{eff}}$: 
\begin{equation}
    A^s_X \approx A_V^s\, \xi(\lambda^X_{\rm{eff}}, R_V^s). 
\end{equation}

\subsubsection{Pre-computation of Effective Wavelengths using Simulations}
\label{S:lameffsims}
To pre-compute effective wavelengths, we simulate extinguished and intrinsic (SED-integrated) apparent magnitudes, compute the effective dust law value via:
\begin{equation}
\label{eq:xieff}
    \xi^X_{\rm{eff}} = \frac{m^{\rm{ext}}_{X,s} - m^{\rm{int}}_{X,s}}{A_V^s},
\end{equation}
then find the $\hat{\lambda}^X_{\rm{eff}}$ that minimises $|\xi(\hat{\lambda}^X_{\rm{eff}},R_V^s) - \xi^X_{\rm{eff}}|$, using a grid with $\Delta \hat{\lambda}_{\rm{eff}} = 0.1$\AA~resolution. 
Each simulated supernova has a unique set of effective wavelengths; therefore, the default set of effective wavelengths is obtained by averaging over many sets of simulated SNe. In turn then, these average effective wavelengths depend on the assumed population distributions of the intrinsic SEDs, host galaxy dust extinction, and dust law shape. 

To perform the simulations, we use the \textsc{BayeSN} SED integration scheme, and borrow various SED components, via the publicly available \citetalias{Mandel22} version of 
\textsc{BayeSN}\footnote{\url{https://github.com/bayesn/bayesn-public}}. In particular, we use the population mean intrinsic SED template, and implement light curve shape variations, and residual perturbations. Full details on the simulation distributions are provided in Appendix~\ref{S:appendixlameffcomp}. The resulting effective wavelengths are our default choice, and are recorded in Table~\ref{tab:leffs}. They have significant non-zero offsets with respect to the passband central wavelengths; nonetheless, in \S\ref{s:senstolameffpreproc}, 
we show the choice of either these default effective wavelengths, or the passband central wavelengths, has a negligible impact on dust hyperparameter inferences.

\renewcommand{\arraystretch}{1}
\setlength{\tabcolsep}{13pt}
\begin{table}
\begin{threeparttable}
    \centering
    \caption{The effective wavelengths, computed using simulations of SED-integrated extinguished and intrinsic apparent magnitudes, and Eq.~\ref{eq:xieff} (more in \S\ref{S:EffectiveLambda}; Appendix~\ref{S:appendixlameffcomp}).}
    \label{tab:leffs}
    \begin{tabular}{c c c c }
    \toprule
    Passband & $\lambda_{c}$ (\AA)\,\tnote{a}& $\lambda_{\rm{eff}}$ (\AA)\,\tnote{b}& $\Delta \lambda$ (\AA)\,\tnote{c}  \\
\midrule
$B$ &  $4402.1$ & $4366.5$& $-35.6\pm5.6$ \\ 
$V$ &  $5389.3$ & $5361.4$ & $-27.9\pm4.1$ \\ 
$r$ &  $6239.9$ & $6149.1$ & $\,\,\,-90.8\pm24.3$ \\ 
$i$ &  $7631.1$ & $7546.5$ & $\,\,\,-84.6\pm14.0$ \\ 
$J$ &  $12516.3$ & $12438.2$ & $\,\,\,-78.1\pm13.8$ \\ 
$H$ &  $16277.2$ & $16139.1$ & $-138.1\pm10.7$ \\
\bottomrule
    \end{tabular}
    \begin{tablenotes}
    \item[a] Passband central wavelengths.
    \item[b] Effective wavelengths (the default choice for population inferences).
    \item[c] The differences between the passband central wavelengths and effective wavelengths. Uncertainties denote the sample standard deviation of effective wavelengths over 1000 simulations.
    \end{tablenotes}
    \end{threeparttable}
\end{table}

\subsection{Simulation-Based Calibration}
\label{S:SBC}
We now perform simulation-based calibration~\citep{Talts18}. We simulate SED-integrated SN peak apparent magnitudes data using \textsc{BayeSN}~\citep{Mandel22}, apply our \textsc{Bird-Snack} HBM to fit the resulting data, and then assess recovery of the input dust hyperparameters: $(\tau_A, \mu_{R_V}, \sigma_{R_V})$. 
We assess recovery on a grid of hyperparameter knots, specifically: $\tau_A = (0.2, 0.5)$~mag, $\mu_{R_V} = (1.5, 2.5, 3.5)$ and $\sigma_{R_V} = (0.1, 0.5, 1.0)$. For each set of $(\tau_A, \mu_{R_V}, \sigma_{R_V})$, we perform 100 simulations, for a total of $100\times2\times3^2 = 1800$ simulations/fits. Each simulation comprises synthetic data of 100 SNe.

For each synthetic SN, we simulate extinguished rest-frame peak apparent magnitudes using \textsc{BayeSN}, using the same simulation distributions as in \S\ref{S:lameffsims}; Appendix~\ref{S:appendixlameffcomp}. We also include measurement errors (Eq.~\ref{eq:measurement}); after inspecting our real data, we choose to simulate error dispersions, $\hat{\sigma}^s_i$, from a truncated-normal distribution, with a mean and dispersion of 0.02~mag, and a lower bound at 0.005~mag.

To assess hyperparameter recovery, we record the posterior median hyperparameter in each simulation, then collate these medians across the 100 simulations, and quote the resulting median and 68\% credible interval. We also group the posterior samples from all 100 simulations together, and summarise this `Simulation-Averaged Posterior', using the median and 68\% interval of samples. Finally, we quote $(N_{68}, N_{95})$, which are the numbers of simulations where the true hyperparameter value falls within the 68\% and 95\% credible intervals, respectively; if these numbers are significantly less than 68\% or 95\%, respectively, it indicates a problem with the model.

\begin{figure}
    \centering
    \includegraphics[width=1\linewidth]{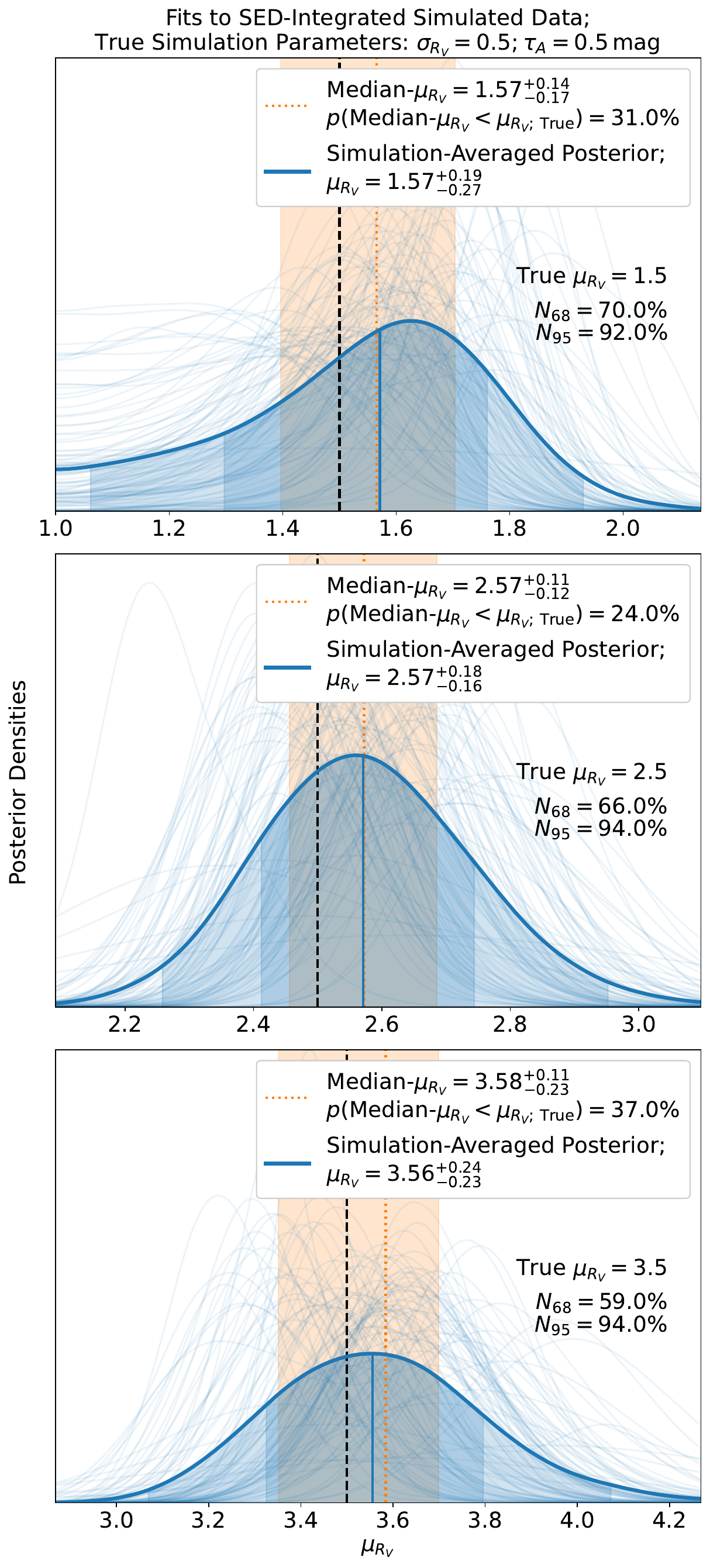}
    \caption{Recovery of simulation hyperparameters, $\mu_{R_V} = (1.5,2.5,3.5)$, from fitting synthetic SED-integrated data using our intrinsic deviations hierarchical Bayesian model (\S\ref{S:ModelM2M}). Each synthetic dataset comprises fake $BVriJH$ peak apparent magnitude measurements of 100 \textsc{BayeSN}-simulated SNe Ia (with variations in light-curve shape and residual perturbations included). For each input $\mu_{R_V}$, we perform 100 simulations. Each faint blue line is the posterior from one simulation; the thick blue line is the `Simulation-Averaged Posterior', which collates the posterior samples across all simulations; orange line (band) marks the median (16\% to 84\% quantiles) of posterior medians across simulations. The $(N_{68}, N_{95})$ statistics are the numbers of simulations where the true hyperparameter resides within the 68\% or 95\% credible intervals, respectively. Results show recovery of $\mu_{R_V}$ is successful, indicating the model is robust. The systematic offset $\Delta \mu_{R_V}\approx 0.07$ is insignificant compared to uncertainties.
    }
    \label{fig:sbcmuRV}
\end{figure}

For all 18 sets of simulations, recovery of dust hyperparameters is successful. In Fig.~\ref{fig:sbcmuRV}, we show the recovery of $\mu_{R_V}=(1.5,2.5,3.5)$ under $\sigma_{R_V} = 0.5$ and $\tau_A=0.5$~mag. There is a small systematic bias of $\Delta \mu_{R_V}\approx 0.07$, which is insignificant compared to uncertainties. Recovery of $(\tau_A, \sigma_{R_V})$ is also robust. We conclude the default intrinsic deviations model is validated for samples of $\approx$100 SNe. This justifies that our model can be applied to the real data to robustly infer $R_V$ population distributions.

\section{Analysis}
\label{S:Analysis}

\subsection{Choice of $A_V^s$ Population Distribution and Censored Data}
\label{S:AVspriorCens}

\renewcommand{\arraystretch}{1.5}
\setlength{\tabcolsep}{3pt}
\begin{table*}
\begin{threeparttable}
    \centering
    \caption{
    Dust hyperparameter inferences for different choices of $A_V^s$ population distribution, cuts on $B-V$ apparent colour, and censored data. Summaries are the posterior medians and 68\% credible intervals.
    }
    \label{tab:AVspriorCens}
    \begin{tabular}{l c c c c c c}
    \toprule
        Hyperparameters 
        & \multicolumn{2}{c}{$\nu_A\cdot\tau_A$ (mag)\,\tnote{a}}  
        & \multicolumn{2}{c}{$\mu_{R_V}$}
        & \multicolumn{2}{c}{$\sigma_{R_V}$\,\tnote{b}}\\
        \cmidrule(lr){2-3}\cmidrule(lr){4-5}\cmidrule(lr){6-7}
\midrule
\midrule
\bf{No Censored SNe} \\
\midrule
        Colour Cuts\,\tnote{c}
        &  No Cut & $|B-V|<0.3$~mag 
        &  No Cut & $|B-V|<0.3$~mag  
        &  No Cut & $|B-V|<0.3$~mag  \\
$N_{\rm{SNe}}$ & 69 & 62  & 69 & 62  & 69 & 62 \\
\midrule
$A_V^s$ Distribution \\
\midrule
$A^s_V \sim \rm{Exp}(\tau_A)$ & $0.44^{+0.06}_{-0.06}$ & $0.28^{+0.05}_{-0.05}$ & $2.42^{+0.37}_{-0.31}$ & $2.61^{+0.41}_{-0.35}$ 
& $< 0.79 (1.68)$
& $< 0.80 (1.97)$ \\ 
$A_V^s \sim $ Gamma$(\nu_A,\tau_A)$\,\tnote{d} & 
$0.38^{+0.08}_{-0.07}$ & $0.49^{+0.34}_{-0.20}$
&  $2.15^{+0.29}_{-0.18}$ & $3.43^{+0.85}_{-0.83}$ & $< 0.44 (1.11)$ & 
$< 1.10 (2.03)$ \\
\midrule
\midrule
\bf{With Censored SNe} \\
\midrule
        Censored $B-V$ Range (mag)\,\tnote{e}
        &  $[0.3,\infty]$ & $[0.3,1.0]$
        &  $[0.3,\infty]$ & $[0.3,1.0]$ 
        &  $[0.3,\infty]$ & $[0.3,1.0]$ \\
$N_{\rm{SNe}}\,(N_{\rm{Censored}})$\,\tnote{f} & 62 (7) & 62 (3)  & 62 (7) & 62 (3)  & 62 (7) & 62 (3) \\
\midrule
$A_V^s$ Distribution w/ Censored Data \\
\midrule
$A^s_V \sim \rm{Exp}(\tau_A)$ & $0.38^{+0.08}_{-0.06}$ & $0.33^{+0.06}_{-0.05}$ & $2.43^{+0.40}_{-0.54}$ & $2.56^{+0.41}_{-0.37}$& $< 1.25 (2.34)$ & $< 1.06 (2.17)$ \\
$A_V^s \sim $ Gamma$(\nu_A,\tau_A)$\,\tnote{g} & 
$0.40^{+0.13}_{-0.08}$ & $0.42^{+0.28}_{-0.13}$
& $2.54^{+0.59}_{-0.54}$ & $3.01^{+0.94}_{-0.69}$ & $< 1.43 (2.56)$ & $< 1.39 (2.46)$ \\
\bottomrule
    \end{tabular}
    \begin{tablenotes}
    \item[a] The combination of hyperparameters, $\nu_A\cdot\tau_A$, is the gamma $A_V^s$ distribution population mean dust extinction. For the exponential distribution, $\nu_A \cdot \tau_A = \tau_A$. 
    \item[b] The 68\% (95\%) quantiles are tabulated for posteriors that peak near the lower prior boundary.
    \item[c] We fit either the full sample of 69 SNe Ia, or the low-reddening $|B-V|<0.3$~mag sub-sample of 62 SNe Ia.
    \item[d] $\nu_A=0.55^{+0.21}_{-0.16}$, and 
    $\nu_A< 5.3 (14.9)$,
    from fits to the full and low-reddening sample, respectively. 
    \item[e] The censored SNe are those with rest-frame $B-V$ apparent colours in the range in the column heading. For $B-V \in [0.3,\infty]$~mag, there are 7 censored SNe.    
    For $B-V \in [0.3,1.0]$~mag, there are 3 censored SNe with $0.3<B-V<1.0$~mag: SNe~ 2002bo, 2008fp and iPTF13azs. The remaining 4 SNe, with $B-V>1.0$~mag, are excluded from the sample entirely: SNe~1999cl, 2003cg, 2006X and 2014J.
    \item[f] The low-reddening sample comprises 62 SNe Ia, and the number of censored SNe is either 7 or 3, depending on the censored $B-V$ range. 
    \item[g] 
    $\nu_A=1.03^{+1.08}_{-0.36}$, and
    $\nu_A< 3.07 (8.49)$, from fits with 7 or 3 censored SNe, respectively.
    \end{tablenotes}
    \end{threeparttable}
\end{table*}

\begin{figure}
    \centering
    \includegraphics[width=1\linewidth]{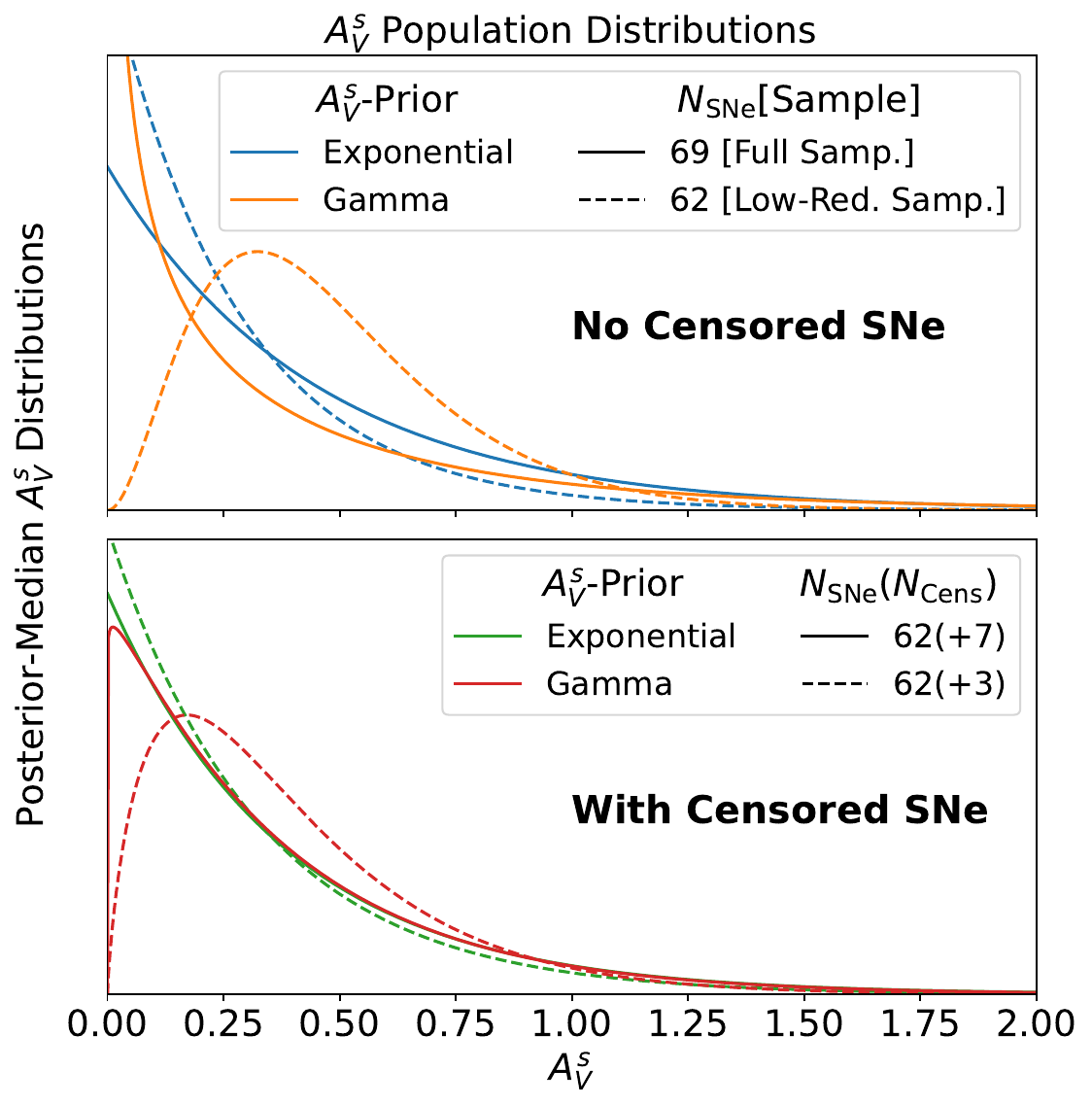}
    \includegraphics[width=1\linewidth]{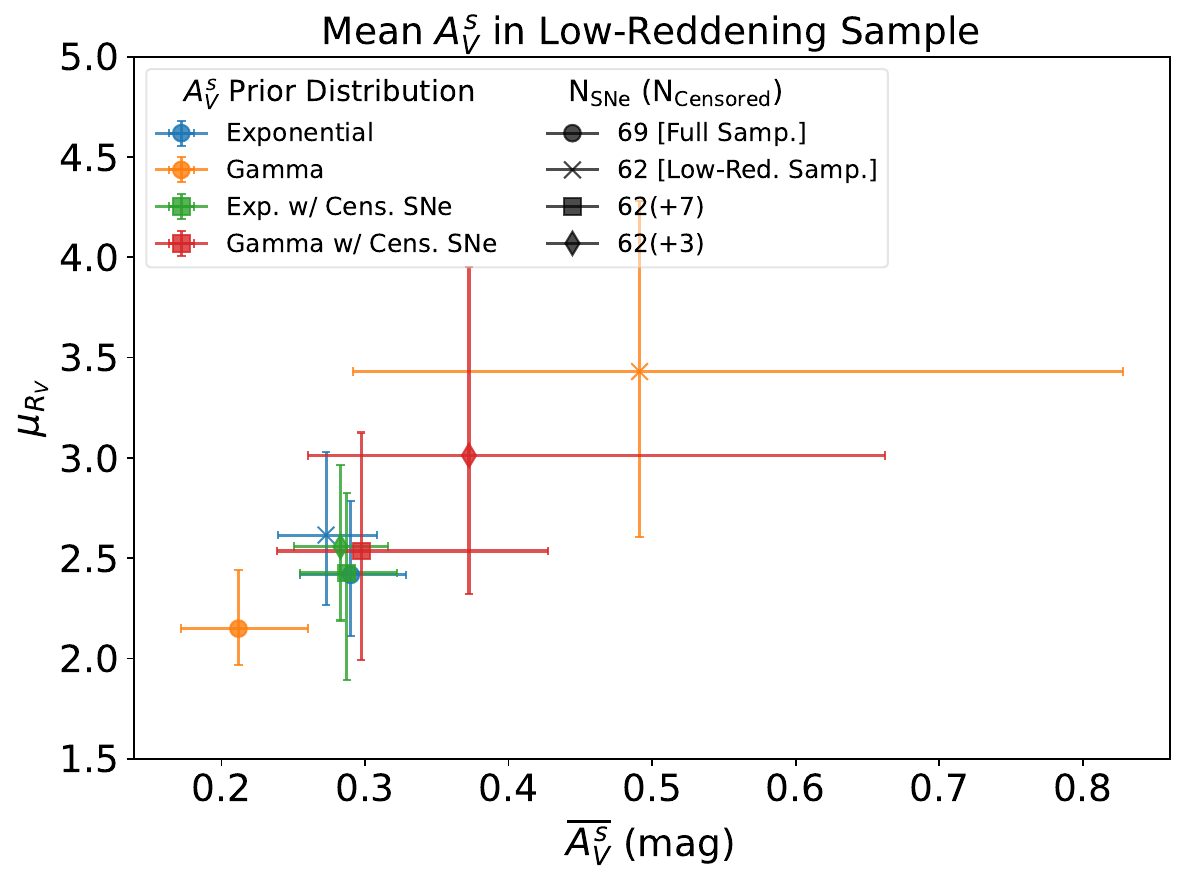}
    \caption{(top panels) The $A_V^s$ population distributions computed using the posterior median hyperparameters, for different choices of $A_V^s$ distribution, sample cuts, and censored data. When fitting the full sample, the flexible gamma distribution has a shape hyperparameter $\nu_A<1$; but when the high-reddening $B-V>0.3$~mag objects are excluded, the gamma distribution peaks at $A_V^s\approx 0.35$~mag. When modelling censored data to fit the low-reddening sample, the gamma distribution peaks at lower values, $A_V^s\approx 0.03, 0.15$~mag, under 7 or 3 censored SNe, respectively.
    (lower panel) The population mean $R_V$ hyperparameter, $\mu_{R_V}$, is plotted against the sample mean $A_V^s$ in the low-reddening sample, $\overline{A^s_V}$. The increase in $\overline{A^s_V}$ is correlated with an increase in $\mu_{R_V}$. These results indicate robust constraints on the dust extinction population distribution are important to estimate $\mu_{R_V}$. 
    }
    \label{fig:AVmeansNoCensCens}
\end{figure}

\subsubsection{Initial Results}
\label{S:initialresults}

We begin our analysis by testing the sensitivity of $R_V$ inferences to the choice of $A_V^s$ population distribution, and the inclusion of high-reddening $B-V>0.3$~mag objects (Fig.~\ref{fig:AVmeansNoCensCens}). In addition to the default exponential distribution, we test the more flexible gamma distribution, with shape hyperparameter $\nu_A$. The gamma distribution is equivalent to the exponential distribution when $\nu_A=1$, but peaks at non-zero $A_V^s$ values when $\nu_A>1$. This distribution was recently applied in \cite{Wojtak23}, who constrained the shape hyperparameter of the reddening distribution to be $\approx  3\pm 1$, indicating a non-zero-peaked $A_V^s$ distribution ($\nu_A>1$) is preferred. They note that the larger extinction values lead to bluer intrinsic colours, which may in turn affect our $\mu_{R_V}$ inferences. 

We thus individually test two $A_V^s$ population distributions:
\begingroup
\allowdisplaybreaks
\begin{align}
&A_V^s \sim \rm{Exp}(\tau_A), \\
&A_V^s \sim \rm{Gamma}(\nu_A,\tau_A), 
\end{align}
\endgroup
with hyperpriors on the hyperparameters:
\begin{equation}
    \tau_A, \nu_A \sim \textrm{Half-Cauchy}(0,1),
\end{equation}
noting that Gamma$(\nu_A=1,\tau_A)\equiv\,$Exp$(\tau_A)$.

The $\mu_{R_V}$ inferences in Table~\ref{tab:AVspriorCens} are strongly dependent on the choice of $A_V^s$ population distribution, and the sample being fitted (either the full sample of 69 SNe, or the low-reddening sample of 62 SNe with $B-V<0.3$~mag). In the full sample fits, applying the exponential $A_V^s$ distribution yields $\mu_{R_V}=2.42^{+0.37}_{-0.31}$, but the more flexible gamma distribution yields a \textit{lower} value, $\mu_{R_V} = 2.15^{+0.29}_{-0.18}$. Excluding the high-reddening objects, the exponential distribution fit yields $\mu_{R_V}=2.61^{+0.41}_{-0.35}$, while the gamma distribution returns a \textit{higher} $\mu_{R_V}$ inference: $\mu_{R_V}=3.43^{+0.85}_{-0.83}$.  

The gamma distribution fits thus return a  
$\Delta \mu_{R_V} \approx 1.35$ shift in the posterior medians depending on whether the high-reddening objects are included in the fit. Fig.~\ref{fig:AVmeansNoCensCens} shows that the inferred $\mu_{R_V}$ is correlated with the sample mean of the $A_V^s$ estimates in the low-reddening sample, which also shifts to higher values when the high-reddening objects are removed. These results show the treatment of high-reddening objects should be carefully considered when constraining dust distributions in low-reddening cosmological samples. We return to evaluate these real-data inferences at the end of \S\ref{S:FiducialResults}.

\subsubsection{Censored Data Modelling}
\label{S:CensoredDataInferences}

The sensitivity to modelling high-reddening objects motivates that we model the $B-V<0.3$~mag censoring process. This information is important, because it tells the model that the lack of high reddening objects could be a result of the $B-V$ cut, rather than, for example, the $A_V^s$ population distribution tapering off at high values\footnote{The importance of the data collection process in model inferences is described in Chapter 8.1 of \cite{Gelman13}.}.

We thus include the censoring process in the model\footnote{\url{https://mc-stan.org/docs/stan-users-guide/censored-data.html}}. This means we additionally model information about SNe that were removed as a result of the $B-V<0.3$~mag cut. The censored data comprises the number of SNe cut from the sample, and their $B-V$ measurement errors: $(N_{\rm{Cens}}, \bm{\hat{\sigma}}_{BV})$. In the model inference, we draw additional censored parameters, $(\bm{\delta N}^s, A_V^s, R_V^s)$, from the population distributions, one set for each censored SN. These transform to make a latent $B-V$ colour, which is then modelled using a latent $\hat{BV}$ parameter:
\begin{equation}
    \hat{BV}^s \sim \mathcal{N}(B-V^s, \hat{\sigma}_{BV}^s).
\end{equation}
We place a lower bound of $0.3$~mag on the $\hat{BV}^s$ parameters. This means the censored SNe's measured $B-V$ colours are each constrained to exceed $0.3$~mag. Consequently, the censored SNe are drawn from the same population distributions 
as the remainder of the sample, but are constrained to be cut as a result of the censoring process. The model thus takes into account that the lack of high-reddening objects could be a result of the data collection process, rather than the red tails of the population distributions tapering off. This improves inferences of the population hyperparameters. 

Simulations in Appendix~\ref{S:appcensoreddata} confirm censored-data modelling leads to robust dust hyperparameter recovery. We also show analysing low-reddening sub-samples in isolation can affect population inferences; for example, the mean dust extinction inference is biased low in our toy simulations, $\Delta \tau_A\approx -0.2$~mag, while $\mu_{R_V}$ inferences are less affected, $\Delta \mu_{R_V}\approx 0.3$. The size and significance of this effect depends on the simulation hyperparameters and sample size.

We test modelling either 7 or 3 censored SNe. The 7 censored SNe are all the moderate-to-high-reddening objects, whereas the 3 censored SNe excludes the 4 highly reddened objects with $B-V>1.0$~mag. In Fig.~\ref{fig:ColourCorner}, these 4 objects form a cluster in the $B-V$ vs. $V-r$ panel, which appears as bumps in the kernel density estimates of the $B-V$ and $V-r$ empirical distributions. We consider that these 4 SNe are unlikely to be drawn from the same population distributions as the remaining 65 SNe, given that they form a distinct high-reddening group in colour space. This contrasts with the 3 moderate-reddening SNe with $0.3\lesssim B-V\lesssim0.5$~mag, which reside much closer to the bulk of the low-reddening sample. Results in Table~\ref{tab:AVspriorCens} and Fig.~\ref{fig:AVmeansNoCensCens} show $\mu_{R_V}$ inferences are less sensitive to the choice of $A_V^s$ population distribution when censored data are modelled. 

\subsubsection{Posterior Predictive Checks}
\label{S:posteriorpredictive}

We perform posterior predictive checks to determine which $B-V>0.3$~mag objects should be included in the sample, for robust $R_V$ inferences in the low-reddening cosmological sample. In Appendix~\ref{S:appendixadditionalppc}, we use the posterior samples from fits to real data to predict the number of SNe in high $B-V$ bins. These simulations do not predict any SNe with $B-V>1.0$~mag, unlike the real data, which has 4 objects in this bin; however, the simulations successfully predict there should be 3 objects in the $0.3<B-V<1.0$~mag range, like the real data. Consequently, we exclude the 4 high-reddening objects with $B-V>1.0$~mag from the sample for the remainder of the analysis, while retaining the 3 moderate-reddening SNe.

\begin{table*}
\setlength{\tabcolsep}{9pt}
\begin{threeparttable}
    \centering
    \caption{Fiducial dust population inferences from fitting a sample of 65 SNe~Ia, which includes 3 moderate reddening objects with $0.3\lesssim B-V\lesssim0.5$~mag. We test fitting the gamma dust extinction distribution hyperparameter, $\nu_A$.
    }
    \label{tab:AVExpGammafiducial}
    \begin{tabular}{l c c c c}
    \toprule
        Hyperparameters 
        & $\nu_A\cdot\tau_A$ (mag)\,\tnote{a}
        & $\mu_{R_V}$
        & $\sigma_{R_V}$\,\tnote{b}&$\nu_A$\\
\midrule

$A^s_V \sim \rm{Exp}(\tau_A)$ & $0.32^{+0.06}_{-0.05}$ & $2.61^{+0.38}_{-0.35}$ & $< 0.92 (1.96)$ & 1$^*$ \\ 
$A_V^s \sim $ Gamma$(\nu_A,\tau_A)$ & 
$0.42^{+0.25}_{-0.11}$
& $3.00^{+0.78}_{-0.53}$ & $< 1.04 (1.90)$ & $< 2.85 (8.02)$ \\ 
\bottomrule
    \end{tabular}
    \begin{tablenotes}
    \item[a] Summaries are the posterior medians and 68\% credible intervals.
    \item[b] The 68\% (95\%) quantiles are recorded for posteriors that peak near the lower prior boundary.
    \end{tablenotes}
    \end{threeparttable}
\end{table*}

\subsection{Fiducial Results}
\label{S:FiducialResults}
\begin{figure}
    \centering
    \includegraphics[width=1\linewidth]{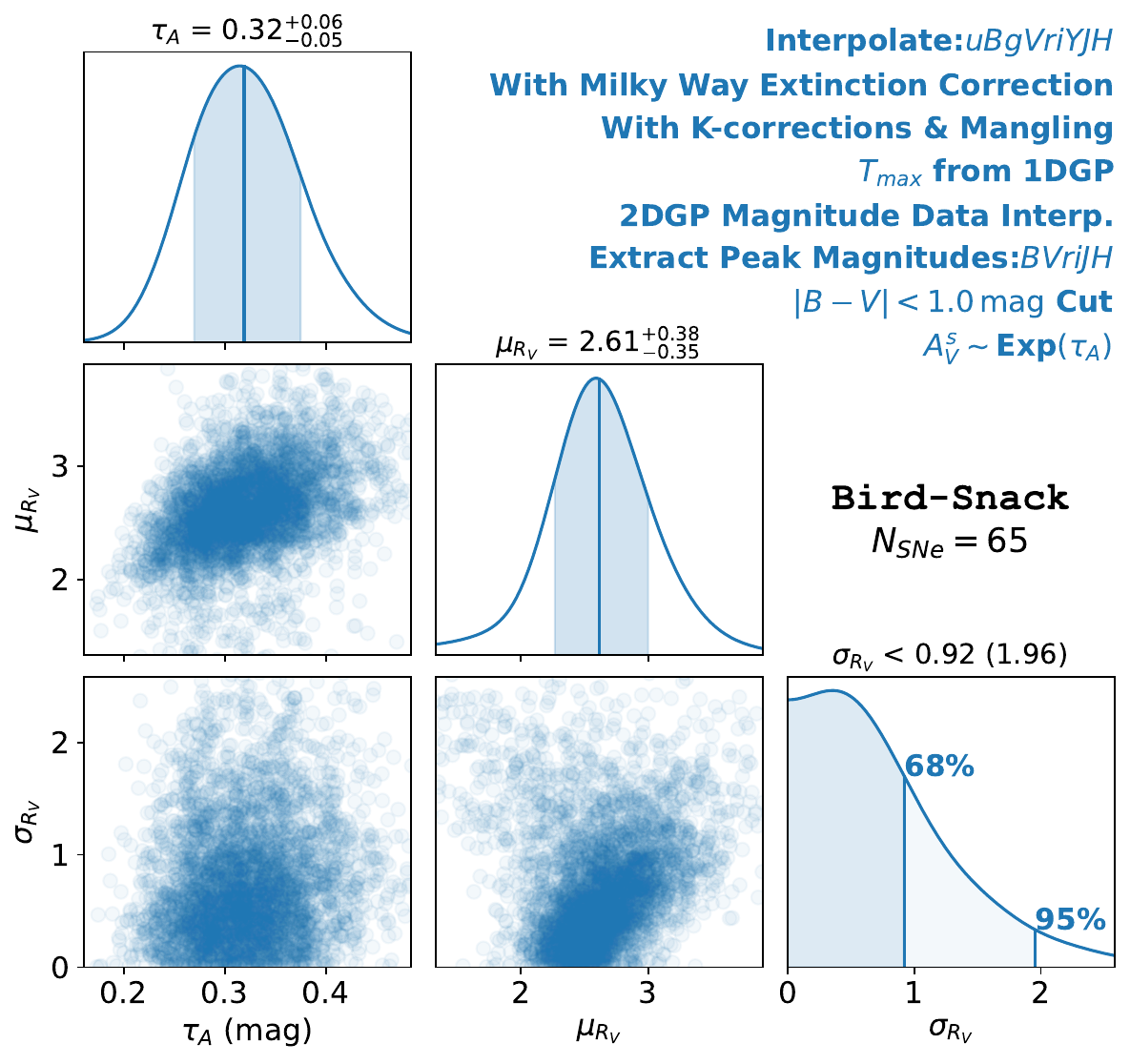}
    \caption{
    Default \textsc{Bird-Snack} inference of dust population hyperparameters, from fitting a sample of 65 SNe~Ia using our intrinsic deviations hierarchical model (\S\ref{S:ModelM2M}), and an exponential $A_V^s$ distribution. The sample comprises 62 SNe with $B-V<0.3$~mag, and 3 SNe with $0.3\lesssim B-V \lesssim 0.5$~mag. The posterior medians are $\tau_A=0.32$~mag, $\mu_{R_V}=2.61$, and $\sigma_{R_V}=0.64$.
    }
    \label{fig:RVInferenceDefault}
\end{figure}

Our default inference in Fig.~\ref{fig:RVInferenceDefault} is thus from fitting a combined sample of 65 SNe~Ia, which comprises 62 low-reddening SNe, and 3 additional moderate-reddening SNe with $0.3\lesssim B-V \lesssim0.5$~mag. We apply the intrinsic deviations model, and the exponential $A_V^s$ distribution. The posterior summaries are $\tau_A=0.32^{+0.06}_{-0.05}$~mag, $\mu_{R_V}=2.61^{+0.38}_{-0.35}$, and $\sigma_{R_V}<0.92(1.96)$, with a posterior median $\sigma_{R_V}\approx 0.64$. These results are consistent with, and marginally improve upon, the inferences from modelling the 3 moderate-reddening objects as censored SNe (Table~\ref{tab:AVspriorCens}).

\begin{figure}
    \centering
    \includegraphics[width=1\linewidth]{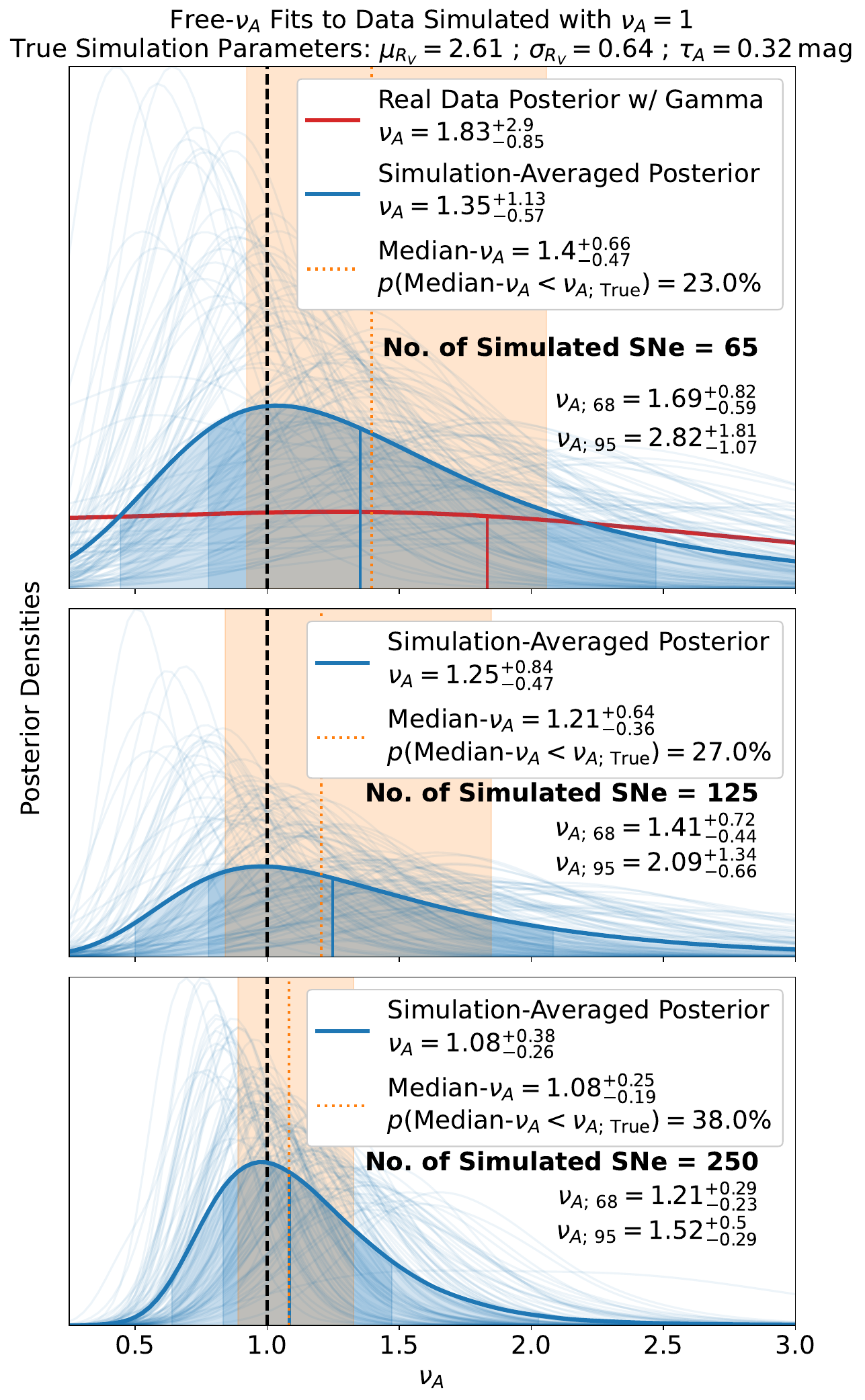}
    \caption{Recovery of $\nu_A$, the gamma $A_V^s$ distribution shape hyperparameter, in simulations. (top panel) $\nu_A$ posteriors from fits to 100 sets of simulated data. Each set comprises 65 SNe synthesised using an exponential $A_V^s$ distribution, and the posterior median hyperparameters from the real-data fit. We also plot the $\nu_A$ posterior from fitting the real-data sample of 65 SNe. Results show $\nu_A$ posteriors are wide, with median point estimates typically greater than the true $\nu_A=1$ value. Therefore, the wide $\nu_A$ posterior, and high $\nu_A\approx 1.83$ point estimate, is not unexpected in the real-data fit if the true $\nu_A=1$. (lower panels) We assess $\nu_A$ recovery for larger samples where the true hyperparameter is $\nu_A=1$. Fits to samples of 125 SNe still yield relatively wide posteriors, whereas 250 SNe typically constrain $\nu_A\lesssim 1.52$ with $\approx 95\%$ posterior probability.
    }
    \label{fig:nuArecovery}
\end{figure}

Table~\ref{tab:AVExpGammafiducial}
shows fitting the gamma dust extinction distribution shape hyperparameter, $\nu_A$, yields results consistent with the exponential-distribution fit. The $\nu_A$ posterior is wide and weakly constraining, with $68\%(95\%)$ posterior upper bounds of $\nu_A<2.85(8.02)$. This posterior is positively skewed towards $\nu_A>1$, but is nonetheless consistent with the fiducial value $\nu_A=1$.

To better understand the free-$\nu_A$ results, we perform additional checks to assess recovery of $\nu_A=1$ for samples of this size. We simulate SNe from our forward model, using the posterior-median hyperparameters from fitting the sample of 65 SNe with the exponential $A_V^s$ distribution. We simulate 100 sets of 65 SNe using the \textit{exponential} $A_V^s$ distribution, and then fit the samples with a \textit{gamma} $A_V^s$ distribution. In addition to the $\nu_A$ posterior medians, and the simulation-averaged posterior (see \S\ref{S:SBC}), we also record the simulations' 68\% and 95\% $\nu_A$ quantiles, $\nu_{A;\,68}, \nu_{A;\,95}$, respectively, and report their resulting medians and 68\% credible intervals. 

Fig.~\ref{fig:nuArecovery} shows wide $\nu_A$ posteriors are expected for this sample. The simulation-averaged posterior is skewed towards high values, with median, 16\% and 84\% quantiles, 
$\nu_A = 1.35_{-0.57}^{+1.13}$. 
Also, the median estimates are distributed at higher values, $\nu_A=1.40^{+0.66}_{-0.47}$,
rather than around $\nu_A=1$. These estimates are consistent with the $\nu_A\approx 1.83$ median point-estimate from the real-data fit. These checks show that a high $\nu_A$ posterior median from fitting a sample of $\approx 65$ SNe does not necessarily indicate that the true $\nu_A>1$.

We conclude this section by assessing recovery of $\nu_A=1$ in larger samples of SNe. Fitting samples of 125 simulated SNe yields slight improvements in $\nu_A$ constraints (Fig.~\ref{fig:nuArecovery}). Samples of 250 SNe yield tight constraints, with a simulation-averaged posterior $\nu_A=1.08^{+0.38}_{-0.26}$, median estimates $\nu_A=1.08^{+0.25}_{-0.19}$, and 95\% posterior upper bounds $\nu_{A;\,95} = 1.52^{+0.50}_{-0.29}$ across simulations. We conclude then that our fiducial sample is too small to tightly constrain $\nu_A$, but larger samples can yield tighter constraints. We proceed to use the exponential $A_V^s$ distribution for the remainder of the analysis. 

In Appendix~\ref{S:appendixmuRV}, we investigate further the trends in the $\mu_{R_V}$ inferences in Tables~\ref{tab:AVspriorCens} and \ref{tab:AVExpGammafiducial}. In particular, we examine why the inferred $\mu_{R_V}$ increases when applying the gamma $A_V^s$ distribution to fit the fiducial or low-reddening samples (compared to applying the exponential distribution), while the opposite trend is found when fitting the full sample. We thus perform simulations to assess $\mu_{R_V}$ recovery under different assumptions. We conclude an increase in $\mu_{R_V}$ when fitting the gamma distribution is typical for the fiducial and low-reddening samples. The decrease in $\mu_{R_V}$ when fitting the gamma distribution to the full sample is likely an artefact resulting from a model misspecification; in particular, the four high-reddening SNe should be assigned their own dust distributions. We fit these high-reddening SNe in isolation whilst fixing their intrinsic distribution to that constrained in the fiducial analysis, and infer $\mu_{R_V}\approx 2$. We then simulate and fit combined samples of SNe drawn from the two distinct dust distributions, and show the high-reddening cluster pulls the common $\mu_{R_V}$ inference down in the full sample fit, and this effect is stronger when fitting with the gamma distribution. 

\begin{table}
\setlength{\tabcolsep}{12pt}
\begin{threeparttable}
    \centering
    \caption{Sensitivity of dust inferences to choice of effective wavelengths.
    }
    \label{tab:lameff}
    \begin{tabular}{l c c c}
    \toprule
        & $\tau_A$ (mag)\,\tnote{a}
        & $\mu_{R_V}$
        & $\sigma_{R_V}$\,\tnote{b}\\
\midrule
$\lambda_{\rm{eff}}$\,\tnote{c} & $0.32^{+0.06}_{-0.05}$ & $2.61^{+0.38}_{-0.35}$ & $< 0.92 (1.96)$ \\ 
$\lambda_{c}$\,\tnote{d} & $0.32^{+0.06}_{-0.05}$ & $2.72^{+0.39}_{-0.34}$ & 
$< 0.92 (1.96)$ \\
\bottomrule
    \end{tabular}
    \begin{tablenotes}
    \item[a] Summaries are the posterior medians and 68\% credible intervals.
    \item[b] The 68\% (95\%) quantiles are recorded for posteriors that peak near the lower prior boundary.
    \item[c] Inference from using the default effective wavelengths (computed using \textsc{BayeSN} simulations; \S\ref{S:lameffsims}; Appendix~\ref{S:appendixlameffcomp}).
    \item[d] Central wavelengths of the passbands (model-independent). 
    \end{tablenotes}
    \end{threeparttable}
\end{table}

\renewcommand{\arraystretch}{1.2}
\setlength{\tabcolsep}{8pt}
\begin{table}
\begin{threeparttable}
    \centering
    \caption{
     Sensitivity of dust inferences to availability of data near peak.
    }
    \label{tab:twindow}
    \begin{tabular}{l c c c c c c}
    \toprule
        $|\Delta t|$\,\tnote{a} & 
        $N_{\rm{SNe}}$\,\tnote{b} &
        $\tau_A$ (mag)\,\tnote{c} &  
        $\mu_{R_V}$ & $\sigma_{R_V}$\,\tnote{d}\\
\midrule
Default\,\tnote{e}& 65 & $0.32^{+0.06}_{-0.05}$ & $2.61^{+0.38}_{-0.35}$ & $< 0.92 (1.96)$ \\
4 days & 59 & $0.33^{+0.06}_{-0.05}$ & $2.62^{+0.35}_{-0.30}$ & $< 0.75 (1.75)$ \\ 
3 days & 52 & $0.36^{+0.07}_{-0.06}$ & $2.69^{+0.39}_{-0.37}$ & $< 0.84 (1.93)$ \\ 
2 days & 41 & $0.35^{+0.08}_{-0.06}$ & $2.51^{+0.41}_{-0.37}$ & $< 0.76 (2.05)$ \\ 

\bottomrule
    \end{tabular}
    \begin{tablenotes}
        \item[a] The phase window around peak in which at least 1 data point is required in each $BVriJH$ passband for the SN to be retained.
        \item[b] The number of SNe in the sub-sample.
        \item[c] Summaries are the posterior medians and 68\% credible intervals.
        \item[d] The 68\% (95\%) quantiles are tabulated for posteriors that peak near the lower prior boundary.
        \item[e] Default is to require 1 data point both before and after peak in all passbands, over the phase range -10 to 40 days.  
    \end{tablenotes}
    \end{threeparttable}
\end{table}

\subsection{Sensitivity to Effective Wavelengths, Data Availability \& Preprocessing Choices}
\label{s:senstolameffpreproc}

Table~\ref{tab:lameff} shows the choice of effective wavelengths has an insignificant effect on dust hyperparameter inferences. The default effective wavelengths are computed by simulating SNe with \textsc{BayeSN}, and including light curve shape variations and residual perturbations. We test these against the passband central wavelengths, which are model-independent. This inference agrees with the default at the level of $\Delta \mu_{R_V}\approx 0.05 \approx 0.1\sigma$. Therefore, our default effective wavelengths, which are the only \textsc{BayeSN}-dependent component of our model, do not affect the dust hyperparameter estimates.

Table~\ref{tab:twindow} shows results are insensitive to fitting sub-samples cut based on the availability of data near peak. We test retaining only SNe with at least 1 data point within 4, 3 or 2 days of peak in each $BVriJH$ passband. While the sample size gets smaller with the narrowing of the phase window, the increase in posterior uncertainties is small, and the posteriors are strongly consistent.

We also find the dust hyperparameter inferences are insensitive to data preprocessing choices. We test different choices of interpolation filters, Milky Way extinction correction method, \textsc{SNooPy} K-corrections, $T_{B;\,\rm{max}}$ estimation method, and interpolation method. The resulting $\mu_{R_V}$ inferences are $|\Delta \mu_{R_V}|\lesssim 0.1-0.2 \approx 0.1-0.3\sigma$ consistent with the default inference. More in Appendix~\ref{S:appendixpreproc}.

\renewcommand{\arraystretch}{1.5}
\setlength{\tabcolsep}{9pt}
\begin{table*}
\begin{threeparttable}
    \centering
    \caption{Fits to sub-samples cut on host galaxy stellar mass. 
    }
    \label{tab:masssplit}
    \begin{tabular}{lc  c c c c}
    \toprule
Sample & Mass Cut & 
$N_{\rm{SNe}}$&
$\tau_A$ (mag)\,\tnote{a} &  
        $\mu_{R_V}$ & $\sigma_{R_V}$\,\tnote{b} \\
\midrule
Default & All & 65 & $0.32^{+0.06}_{-0.05}$ & $2.61^{+0.38}_{-0.35}$ & $< 0.92 (1.96)$ \\ 
\midrule
High-Mass & $\log_{10} M/M_{\odot}=10$ & 44 & $0.33^{+0.07}_{-0.06}$ & $2.77^{+0.58}_{-0.56}$ & $< 1.50 (2.76)$ \\ 
Low-Mass & - & 16 & $0.40^{+0.14}_{-0.10}$ & $2.97^{+0.80}_{-0.65}$ & $< 1.31 (2.79)$ \\ 
\bottomrule
\end{tabular}
    \begin{tablenotes}
    \item[a] Summaries are the posterior medians and 68\% credible intervals.
    \item[b] The 68\% (95\%) quantiles are tabulated for posteriors that peak near the lower prior boundary.
    \end{tablenotes}
    \end{threeparttable}
\end{table*}

\subsection{Host Galaxy Stellar Mass Subsamples}
\label{S:massinferences}
We constrain dust population distributions in low and high host galaxy stellar mass bins, cut at $\log_{10} M/M_{\odot}=10$. The sample of 65 SNe is decomposed as 44 SNe in high stellar mass hosts, 16 in low mass hosts, and 5 with unknown host masses (see Table~\ref{tab:NSNcuts}).

Table~\ref{tab:masssplit} shows the uncertainties on $\mu_{R_V}$ are large, so the $\mu_{R_V}$-mass dependency cannot be tightly constrained. The median estimates, $\mu_{R_V}=2.77,2.97$, in high/low mass hosts, respectively, are consistent within $\Delta \mu_{R_V} \approx 0.2$ -- but the smaller sized sub-samples result in larger $\mu_{R_V}$ uncertainties $\approx 0.6-0.8$. Therefore, larger SN sub-samples with data near peak are required to tightly constrain the $\mu_{R_V}$-mass dependency with our methodologies. 

\subsection{Sensitivity to Intrinsic SN Model and Light Curve Shape}
\label{S:magscolours}

\renewcommand{\arraystretch}{1.2}
\setlength{\tabcolsep}{8pt}
\begin{table}
\begin{threeparttable}
    \centering
    \caption{Sensitivity of inferences to the reference frame in which intrinsic chromatic variations are modelled, the inclusion of light curve shape parameters, $\Delta m^s_{15}(B)$, and the exclusion of intrinsic chromatic population variations.
    }
    \label{tab:intrinsicmodel}
    \begin{tabular}{l c c c}
    \toprule

Intrinsic Model & $\tau_A$ (mag)\,\tnote{a} &  
        $\mu_{R_V}$ & $\sigma_{R_V}$\,\tnote{b} 
        \\
\midrule
\multicolumn{2}{l}{\textbf{Without $\Delta m^s_{15}(B)$ Term}}
\\
\midrule
Deviations & $0.32^{+0.06}_{-0.05}$ & $2.61^{+0.38}_{-0.35}$ & $< 0.92 (1.96)$ \\  
Adjacent Cols. & $0.29^{+0.05}_{-0.05}$ & $2.43^{+0.37}_{-0.33}$ & $< 0.81 (1.90)$ \\ 
$B-X$ Cols. & $0.32^{+0.06}_{-0.05}$ & $2.57^{+0.35}_{-0.29}$ & $< 0.83 (1.82)$ \\ 
$X-H$ Cols. & $0.37^{+0.07}_{-0.05}$ & $3.16^{+0.54}_{-0.55}$ & $1.34^{+0.88}_{-0.66}$ \\ 
\midrule
\multicolumn{2}{l}{\textbf{With $\Delta m^s_{15}(B)$ Term}}
\\
\midrule
Deviations & $0.30^{+0.05}_{-0.04}$ & $2.69^{+0.36}_{-0.34}$ & $< 0.99 (2.07)$ \\ 
Adjacent Cols. & $0.29^{+0.05}_{-0.04}$ & $2.55^{+0.34}_{-0.31}$ & $< 0.84 (1.86)$ \\ 
$B-X$ Cols. & $0.30^{+0.05}_{-0.04}$ & $2.62^{+0.37}_{-0.30}$ & $< 0.93 (1.86)$ \\ 
$X-H$ Cols. & $0.35^{+0.06}_{-0.05}$ & $3.05^{+0.46}_{-0.50}$ & $1.38^{+0.90}_{-0.54}$ \\ 
\midrule
\multicolumn{2}{l}{\textbf{No Intrinsic Variations}} \\
\midrule
Deviations & $0.61^{+0.08}_{-0.08}$ & $3.93^{+0.54}_{-0.78}$ & $2.60^{+0.87}_{-0.57}$ \\ 
Adjacent Cols. & $0.61^{+0.09}_{-0.07}$ & $3.86^{+0.54}_{-0.72}$ & $2.60^{+0.86}_{-0.59}$ \\ 
$B-X$ Cols. & $0.61^{+0.09}_{-0.07}$ & $3.94^{+0.54}_{-0.76}$ & $2.52^{+0.90}_{-0.61}$ \\ 
$X-H$ Cols. & $0.61^{+0.09}_{-0.08}$ & $3.93^{+0.54}_{-0.80}$ & $2.57^{+0.91}_{-0.62}$ \\
\bottomrule
\end{tabular}
    \begin{tablenotes}
    \item[a] Summaries are the posterior medians and 68\% credible intervals.
    \item[b] The 68\% (95\%) quantiles are tabulated for posteriors that peak near the lower prior boundary.
    \end{tablenotes}
    \end{threeparttable}
\end{table}

\begin{figure}
    \centering
    \includegraphics[width=1\linewidth]{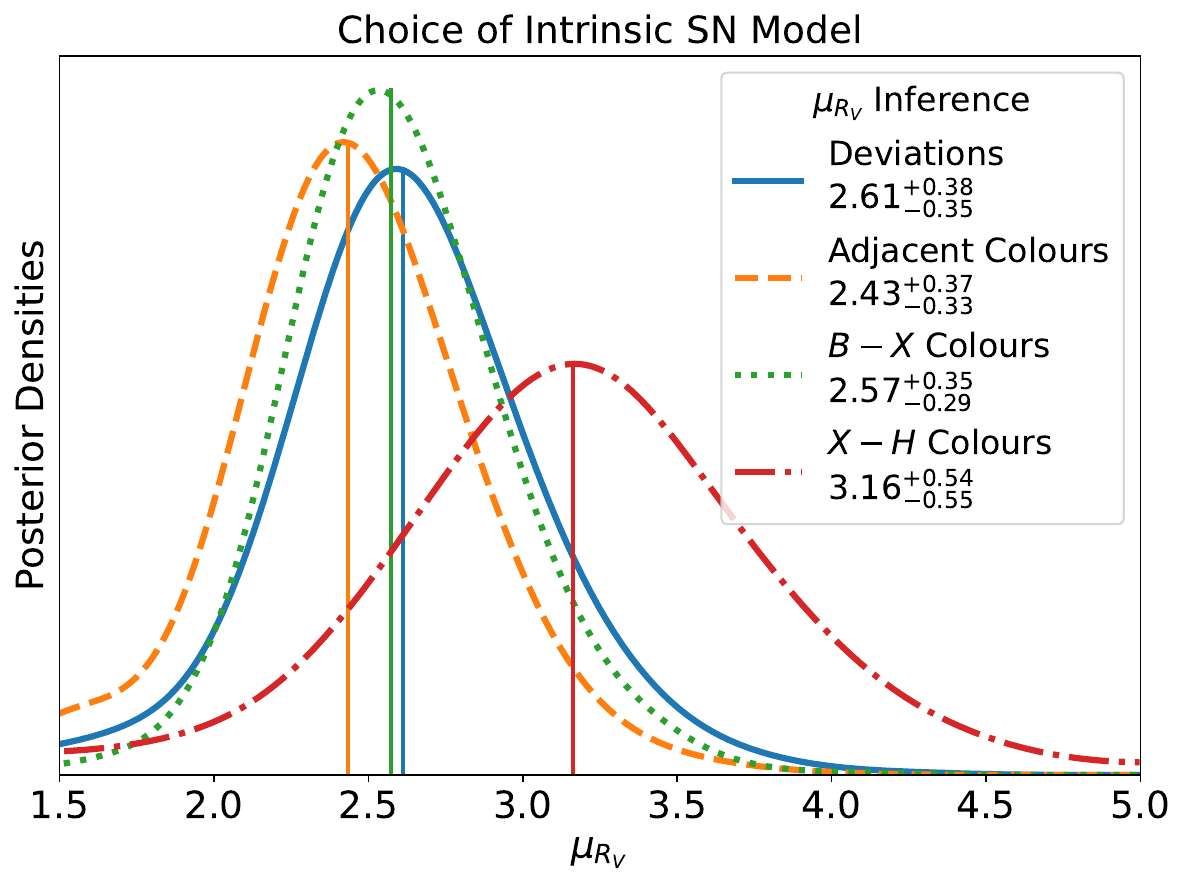}
    \caption{
    Sensitivity of $\mu_{R_V}$ inferences to the reference frame in which intrinsic chromatic variations are modelled. Results from the intrinsic deviations model, which is validated in \S\ref{S:SBC}, are strongly consistent with those from applying the adjacent or $B-X$ intrinsic colour models. The $X-H$ intrinsic colour model yields a higher and wider $\mu_{R_V}$ posterior, but this is still consistent with the others. Simulations in Appendix~\ref{S:appendixintrinsicmodel} show higher $\mu_{R_V}$ posteriors with the $X-H$ model are expected. Therefore, the behaviour of these real-data posteriors is likely a modelling artefact from applying the intrinsic hyperpriors in different reference frames. 
    }
    \label{fig:comparemodeldata}
\end{figure}

We test the sensitivity of dust inferences to the intrinsic SN model. Specifically, we test different reference frames for modelling intrinsic chromatic variations. This \textit{modelling} choice is distinct from the (arbitrary) transformation of the \textit{data}\footnote{Once a reference frame for modelling has been defined, latent parameters can be transformed to fit an arbitrary set of colours data. Inferences are sensitive to the reference frame for modelling, but invariant to the arbitrary transformation of the data.
}.

Fig.~\ref{fig:comparemodeldata} shows modelling intrinsic deviations, adjacent colours or $B-X$ colours, yields consistent results. The deviations reference frame is validated in \S\ref{S:SBC} for samples of this size, so this consistency indicates any one of these reference frames is a reasonable choice. Applying the $X-H$ intrinsic colour model yields a higher $\mu_{R_V}$ median estimate, with a wider uncertainty, $3.16^{+0.54}_{-0.55}$. Nonetheless, this is still consistent with the default estimate, $\mu_{R_V}=2.61^{+0.38}_{-0.35}$. 

We investigate to what extent the sensitivity to intrinsic hyperpriors can be attributed to a dependence on light curve shape in different reference frames. In Table~\ref{tab:intrinsicmodel}, we record dust hyperparameter inferences with and without the modelling of the $\Delta m^s_{15}(B)$ light curve shape parameters (details in \S\ref{S:ModelLCShape}). The light curve shape modelling improves consistency between the 4 model inferences, with a total range in median $\mu_{R_V}$ estimates decreasing from $\Delta \mu_{R_V}\approx 0.73$ to $\Delta \mu_{R_V} \approx 0.50$ (Table~\ref{tab:intrinsicmodel}). However, the $X-H$ model inference is still high, $\mu_{R_V}=3.05^{+0.46}_{-0.50}$, meaning a stronger dependence on light curve shape in this frame can only partially explain the higher/wider $\mu_{R_V}$ inference. Table~\ref{tab:intrinsicmodel} also shows that turning off intrinsic variations yields consistent results across the 4 models; therefore, it must be the hyperpriors on intrinsic hyperparameters that is driving the dependency on reference frame.

In Appendix~\ref{S:appendixintrinsicmodel}, we assess recovery of input dust hyperparameters from applying the intrinsic colour models to data simulated from the deviations model. These simulations show the $X-H$ model is expected to return high $\mu_{R_V}$ estimates for samples of this size. Therefore, the behaviour of the real-data posteriors is consistent with simulations, and is most likely a modelling artefact of applying the intrinsic hyperpriors in different reference frames (more in \S\ref{S:conclusionsfuturework}).

\subsection{Inclusion of $u$-band}
\label{S:ubandanalysis}
Finally, we test including the $u$-band. This broadens the wavelength range, but reduces the sample to 57 SNe, removing 8 low-reddening objects that lack $u$-band data near peak. Analysing their $uBVriJH$ peak magnitude estimates with the deviations model yields: $\mu_{R_V}=2.84^{+0.43}_{-0.40}, \sigma_{R_V}< 1.18 (2.28),  \tau_A=0.35^{+0.07}_{-0.05}$~mag. These results are consistent with the $BVriJH$ inferences\footnote{For these $uBVriJH$ inferences, we use the passband central wavelengths, rather than effective wavelengths computed using \textsc{BayeSN} simulations. We do this because \textsc{BayeSN} is not trained on passbands bluer than the $B$-band, so the simulated SEDs in the $u$-band wavelength range are an extrapolation of the model. The central wavelengths should be sufficient given the insensitivity to this choice in the $BVriJH$ inferences (\S\ref{s:senstolameffpreproc}).}.

\section{Discussion \& Conclusions}
\label{S:Conclusions}

\subsection{Discussion \& Future Work}
\label{S:conclusionsfuturework}
Several outstanding points can be addressed in future work. Our $\mu_{R_V}=2.61^{+0.38}_{-0.35}$ inference is consistent with $\mu_{R_V}\approx 2.4-2.8$ constraints in \cite{Thorp21,Thorp22}. However, our individual $\mu_{R_V}$ inferences in high and low stellar mass host galaxy bins are weakly constraining, which we attribute to the smaller sub-sample sizes. Therefore, we cannot draw any conclusions regarding potential links between dust properties and the mass step.

\begin{figure}
    \centering
\includegraphics[width=1\linewidth]{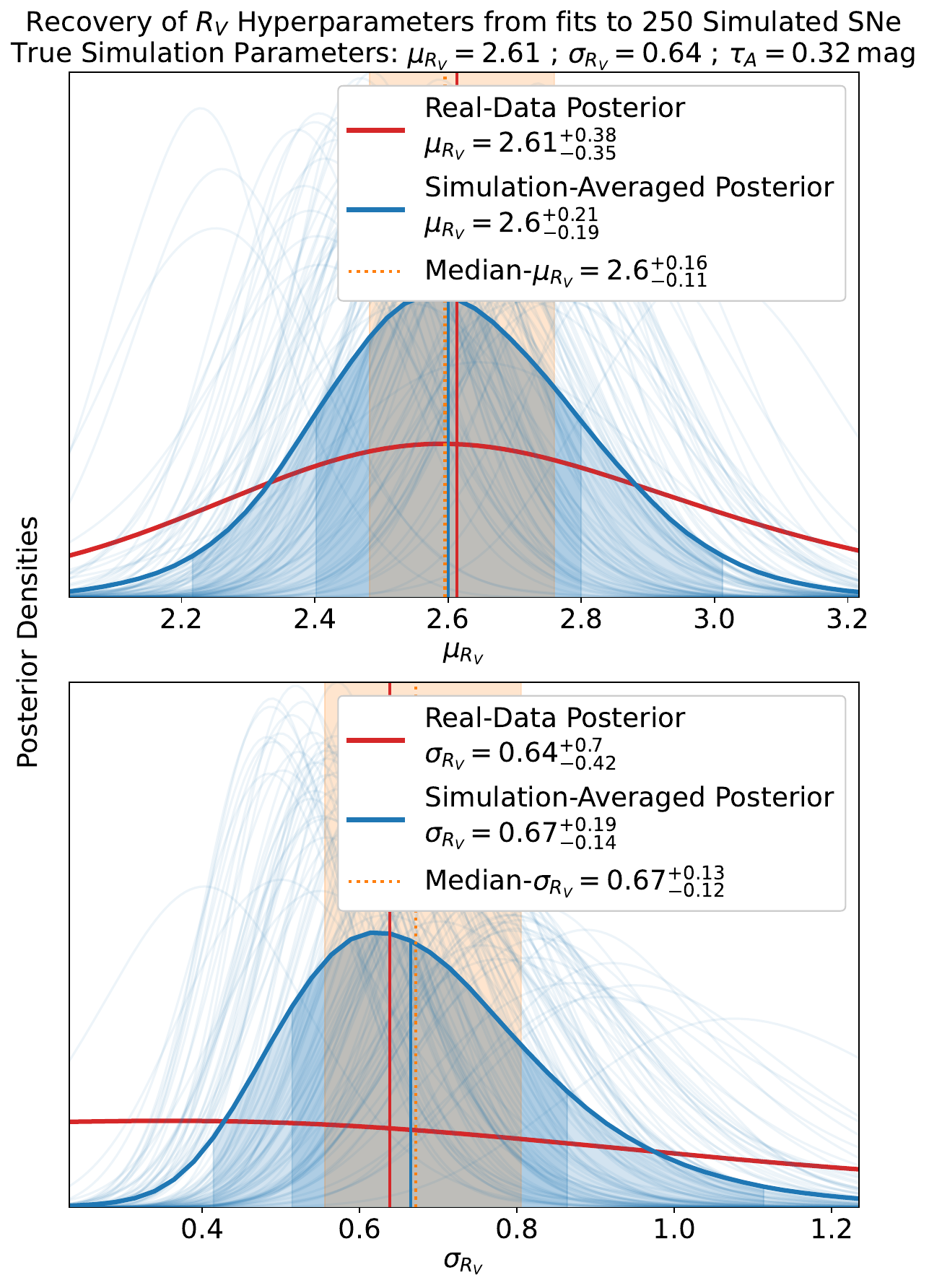}
    \caption{
    Recovery of $\mu_{R_V}=2.61$ and $\sigma_{R_V}=0.64$ from fitting 100 simulated datasets, each comprising $BVriJH$ data of 250 SNe. The constraints on both $\mu_{R_V}$ and $\sigma_{R_V}$ improve compared to the real-data posterior, which itself is obtained by fitting 65 SNe Ia. The $\sigma_{R_V}$ posteriors rule out low values, unlike the real-data posterior which peaks at zero.
    }
    \label{fig:musigRV250rec}
\end{figure}

Ongoing and future optical and near-infrared SN surveys will increase the sample size further, see e.g. CSP-II~\citep{Phillips19}, the FLOWS project~\citep{MullerBravo22} and DEHVILS~\citep{Peterson23}. We explore how constraints will improve with larger samples of 250 SNe, by 
fitting 100 sets of simulated data with \textsc{Bird-Snack}. We simulate $BVriJH$ data using the posterior median hyperparameters from the fiducial fit. Fig.~\ref{fig:musigRV250rec} shows for an input $\mu_{R_V}=2.61$, the simulation-averaged posterior is $\mu_{R_V}=2.60^{+0.21}_{-0.19}$, and the distribution of posterior medians across simulations is $\mu_{R_V}=2.60^{+0.16}_{-0.11}$. The $\sigma_{R_V}$ constraints also improve: the simulation-averaged posterior for an input $\sigma_{R_V}=0.64$ is $\sigma_{R_V}=0.67^{+0.19}_{-0.14}$, and the distribution of posterior medians is $\sigma_{R_V}=0.67^{+0.13}_{-0.12}$. This contrasts with the wide zero-peaking fiducial inference, $\sigma_{R_V}<0.92(1.96)$, from fitting our sample of 65 SNe. Future samples can thus more tightly constrain dust hyperparameters, and better distinguish between zero and non-zero $\sigma_{R_V}$ values. 

It is unclear how the $A_V^s$ population distribution, and/or the treatment of high-reddening objects, affects $\mu_{R_V}$ inferences in SN models that use full light curve information. Dust extinction values should be better constrained with models such as \textsc{BayeSN} and \textsc{SNooPy}, because they leverage spectro-temporal light curve data in the inference; therefore, they can more readily isolate the dust contribution towards the extinguished SED, because it is time-independent. Moreover, these models estimate photometric distances, so they can utilise redshift-based cosmology distances to better inform the absolute normalisations of individual SED terms. We expect then that the gamma $A_V^s$ distribution shape hyperparameter can be more tightly constrained with \textsc{BayeSN} for a given sample size compared to \textsc{Bird-Snack}. The tighter constraints on population hyperparameters also means there may be a stronger sensitivity to the inclusion of moderate-to-high-reddening SNe with \textsc{BayeSN}. Our work shows how posterior predictive checks can be used to assess whether high-reddening objects are reasonably drawn from the population distributions that describe the low-reddening sample. These analysis variants and checks will be important to test in future work.

Despite the above, it is noteworthy that our analysis is largely independent of any SN light curve model. \textsc{Bird-Snack} relies on \textsc{SNooPy} only for K-corrections and Milky Way extinction corrections, and \textsc{BayeSN} for effective wavelengths. We show in \S\ref{s:senstolameffpreproc} that there is no significant sensitivity to these modelling choices (nor any other data preprocessing choices). Further, our analysis shows the $A_V^s$ distribution is consistent with an exponential. Meanwhile, \cite{Thorp21,Thorp22} use an exponential $A_V^s$ population distribution, and yield broadly consistent results. Therefore, we find no reason to reject their conclusions.

Results are weakly sensitive to the reference frame in which intrinsic chromatic variations are modelled. Appendix~\ref{S:appendixintrinsicmodel} shows this dependence is consistent with simulations, implying it is a modelling artefact from placing the intrinsic hyperpriors in different reference frames. For our fiducial sample, the posteriors are consistent, so this is not a dominant systematic uncertainty. With larger samples, we expect reduced sensitivity to the intrinsic hyperpriors; for example, Appendix~\ref{S:appendixintrinsicmodel} shows that $X-H$ model inferences are much closer to the truth, $\Delta \mu_{R_V}\approx 0.1$, when fitting samples of 250 SNe, compared to $\Delta \mu_{R_V}\approx 0.4$ with samples of 65 SNe. However, we cannot rule out that this modelling systematic is insignificant for all regions of parameter space. This should be explored more thoroughly if and when the real-data posteriors from applying different intrinsic models are inconsistent.

\subsection{Conclusions}
We have developed the \href{https://github.com/sam-m-ward/birdsnack/tree/main}{\textsc{Bird-Snack}} model to rapidly infer $R_V$ population distributions in SN Ia host galaxies, and determine which analysis choices significantly impact the population mean hyperparameter, $\mu_{R_V}$. \textsc{Bird-Snack} uses \textsc{SNooPy} to K-correct observer-frame light curves with data near peak, a 2D Gaussian process to estimate peak optical and near-infrared $BVriJH$ apparent magnitudes, and a hierarchical Bayesian model to infer dust population distributions.

\begin{itemize}
    \item
    Fitting a sample of 65 SNe Ia with an exponential $A_V^s$ population distribution, we infer $\mu_{R_V}=2.61^{+0.38}_{-0.35}$, and a Gaussian $R_V$ population dispersion, $\sigma_{R_V}<0.92(1.96)$, with 68\%(95\%) posterior upper bounds, respectively (Fig.~\ref{fig:RVInferenceDefault}). This sample comprises 62 low-reddening SNe ($|B-V|<0.3$~mag), and 3 moderate-reddening SNe with $0.3\lesssim B-V\lesssim0.5$~mag. Inclusion of moderate-to-high-reddening objects is motivated by simulations, which show -- in general -- that analysing low-reddening sub-samples in isolation can affect dust hyperparameter inferences. For example, toy-simulation inferences of the mean extinction hyperparameter, $\tau_A$, are biased low by $\Delta \tau_A\approx-0.2$~mag (Appendix~\ref{S:appcensoreddata}). We use posterior predictive checks to show that the 3 moderate-reddening objects are consistent with the low-reddening sub-sample's population distributions (\S\ref{S:posteriorpredictive}; Appendix~\ref{S:appendixadditionalppc}). Our fiducial inference is insensitive to the availability of data near peak, the set of effective wavelengths, and other data preprocessing choices (\S\ref{s:senstolameffpreproc}).

    \item   
    Fitting with a gamma $A_V^s$ distribution, the shape hyperparameter, $\nu_A$, is weakly constrained, but consistent with $\nu_A=1$ (i.e. an exponential distribution; \S\ref{S:FiducialResults}). Using simulations to assess $\nu_A=1$ recovery with 65 SNe, we find $\nu_A$ posteriors are wide, and the
    posterior medians drift towards $\nu_A>1$. Fig.~\ref{fig:nuArecovery} shows more SNe are required to tightly constrain $\nu_A$.

     \item  We model \textit{intrinsic deviations} from each SN's common achromatic magnitude component, using a multivariate Gaussian population distribution (Fig.~\ref{fig:MagDeviationsCartoon});  this default choice bypasses the need to select an arbitrary set of colours for modelling. We validate this model by fitting \textsc{BayeSN}-simulated data to recover input dust hyperparameters (\S\ref{S:SBC}). Changing the reference frame to model intrinsic colours, e.g. adjacent, $B-X$ or $X-H$ colours, yields consistent results for this sample (Fig.~\ref{fig:comparemodeldata}). 
    
\end{itemize}

\textsc{BayeSN} analyses in \cite{Thorp21, Thorp22} also use an exponential $A_V^s$ distribution, and typically constrain $\mu_{R_V}\approx 2.4-2.8$, consistent with our estimate, so we find no reason to reject their 
conclusions. In future work, larger samples of optical-NIR SN~Ia light curves~\citep[e.g.][]{Phillips19,MullerBravo22,Peterson23} can be analysed with \textsc{Bird-Snack} to more tightly constrain dust hyperparameters, including the gamma dust extinction distribution shape hyperparameter, $\nu_A$. With reduced statistical uncertainties, sensitivity to the intrinsic hyperpriors should be reduced, but this should be assessed in the real-data fits.

Finally, the treatment of high-reddening objects will be important to consider with reduced statistical uncertainties. We have used posterior predictive checks to assess which moderate-to-high-reddening objects are reasonably drawn from the population distributions that describe the low-reddening sub-sample. The 4 objects with $B-V>1.0$~mag form a distinct cluster in parameter space that cannot be described by either the exponential or gamma $A_V^s$ distributions, so they are excluded from the sample. Similar posterior predictive checks in future \textsc{Bird-Snack} and/or \textsc{BayeSN} analyses should be performed to build a complete SN sample that is consistent with the population distributions. Future \textsc{BayeSN} analyses can also test fitting $\nu_A$. Robust constraints on host galaxy dust population distributions will reduce systematic uncertainties in SN~Ia standardisation, and improve cosmological constraints.

\section*{Acknowledgements}
S.M.W. was supported by the UK Science and Technology Facilities Council (STFC). SD acknowledges support from the Marie Curie Individual Fellowship under grant ID 890695 and a Junior Research Fellowship at Lucy Cavendish College. ST was supported by the Cambridge Centre for Doctoral Training in Data-Intensive Science funded by the UK Science and Technology Facilities Council (STFC), and in part by the European Research Council (ERC) under the European Union’s Horizon 2020 research and innovation programme (grant agreement no.\ 101018897 CosmicExplorer). We acknowledge funding from the European Union’s Horizon 2020 research and innovation programme under ERC Grant Agreement No. 101002652 and Marie Skłodowska-Curie Grant Agreement No. 873089.
 
\section*{Data Availability}
All data analysed in this work are publicly available; see \S\ref{S:SNSample}.




\bibliographystyle{mnras}
\bibliography{bib} 

\begin{thebibliography}{}
\makeatletter
\relax
\def\mn@urlcharsother{\let\do\@makeother \do\$\do\&\do\#\do\^\do\_\do\%\do\~}
\def\mn@doi{\begingroup\mn@urlcharsother \@ifnextchar [ {\mn@doi@}
  {\mn@doi@[]}}
\def\mn@doi@[#1]#2{\def\@tempa{#1}\ifx\@tempa\@empty \href
  {http://dx.doi.org/#2} {doi:#2}\else \href {http://dx.doi.org/#2} {#1}\fi
  \endgroup}
\def\mn@eprint#1#2{\mn@eprint@#1:#2::\@nil}
\def\mn@eprint@arXiv#1{\href {http://arxiv.org/abs/#1} {{\tt arXiv:#1}}}
\def\mn@eprint@dblp#1{\href {http://dblp.uni-trier.de/rec/bibtex/#1.xml}
  {dblp:#1}}
\def\mn@eprint@#1:#2:#3:#4\@nil{\def\@tempa {#1}\def\@tempb {#2}\def\@tempc
  {#3}\ifx \@tempc \@empty \let \@tempc \@tempb \let \@tempb \@tempa \fi \ifx
  \@tempb \@empty \def\@tempb {arXiv}\fi \@ifundefined
  {mn@eprint@\@tempb}{\@tempb:\@tempc}{\expandafter \expandafter \csname
  mn@eprint@\@tempb\endcsname \expandafter{\@tempc}}}

\bibitem[\protect\citeauthoryear{{Amanullah} et~al.,}{{Amanullah}
  et~al.}{2014}]{Amanullah14}
{Amanullah} R.,  et~al., 2014, \mn@doi [\apjl] {10.1088/2041-8205/788/2/L21},
  \href {http://adsabs.harvard.edu/abs/2014ApJ...788L..21A} {788, L21}

\bibitem[\protect\citeauthoryear{{Amanullah} et~al.,}{{Amanullah}
  et~al.}{2015}]{Amanullah15}
{Amanullah} R.,  et~al., 2015, \mn@doi [\mnras] {10.1093/mnras/stv1505}, \href
  {http://adsabs.harvard.edu/abs/2015MNRAS.453.3300A} {453, 3300}

\bibitem[\protect\citeauthoryear{{Betancourt}}{{Betancourt}}{2016}]{Betancourt16}
{Betancourt} M.,  2016, arXiv e-prints, \href
  {https://ui.adsabs.harvard.edu/abs/2016arXiv160100225B} {p. arXiv:1601.00225}

\bibitem[\protect\citeauthoryear{{Betancourt}}{{Betancourt}}{2017}]{Betancourt17}
{Betancourt} M.,  2017, \mn@doi [arXiv e-prints] {10.48550/arXiv.1701.02434},
  \href {https://ui.adsabs.harvard.edu/abs/2017arXiv170102434B} {p.
  arXiv:1701.02434}

\bibitem[\protect\citeauthoryear{{Betancourt} \& {Girolami}}{{Betancourt} \&
  {Girolami}}{2013}]{BetancourtGirolami15}
{Betancourt} M.~J.,  {Girolami} M.,  2013, \mn@doi [arXiv e-prints]
  {10.48550/arXiv.1312.0906}, \href
  {https://ui.adsabs.harvard.edu/abs/2013arXiv1312.0906B} {p. arXiv:1312.0906}

\bibitem[\protect\citeauthoryear{{Betancourt}, {Byrne}  \&
  {Girolami}}{{Betancourt} et~al.}{2014}]{Betancourt14}
{Betancourt} M.~J.,  {Byrne} S.,   {Girolami} M.,  2014, \mn@doi [arXiv
  e-prints] {10.48550/arXiv.1411.6669}, \href
  {https://ui.adsabs.harvard.edu/abs/2014arXiv1411.6669B} {p. arXiv:1411.6669}

\bibitem[\protect\citeauthoryear{{Betoule} et~al.,}{{Betoule}
  et~al.}{2014}]{Betoule14}
{Betoule} M.,  et~al., 2014, \mn@doi [\aap] {10.1051/0004-6361/201423413},
  \href {http://adsabs.harvard.edu/abs/2014A%26A...568A..22B} {568, A22}

\bibitem[\protect\citeauthoryear{{Boone}}{{Boone}}{2019}]{Boone19}
{Boone} K.,  2019, \mn@doi [\aj] {10.3847/1538-3881/ab5182}, \href
  {https://ui.adsabs.harvard.edu/abs/2019AJ....158..257B} {158, 257}

\bibitem[\protect\citeauthoryear{{Briday} et~al.,}{{Briday}
  et~al.}{2022}]{Briday22}
{Briday} M.,  et~al., 2022, \mn@doi [\aap] {10.1051/0004-6361/202141160}, \href
  {https://ui.adsabs.harvard.edu/abs/2022A&A...657A..22B} {657, A22}

\bibitem[\protect\citeauthoryear{{Brout} \& {Scolnic}}{{Brout} \&
  {Scolnic}}{2021}]{Brout21}
{Brout} D.,  {Scolnic} D.,  2021, \mn@doi [\apj] {10.3847/1538-4357/abd69b},
  \href {https://ui.adsabs.harvard.edu/abs/2021ApJ...909...26B} {909, 26}

\bibitem[\protect\citeauthoryear{{Brout} et~al.,}{{Brout}
  et~al.}{2022}]{Brout22}
{Brout} D.,  et~al., 2022, \mn@doi [\apj] {10.3847/1538-4357/ac8e04}, \href
  {https://ui.adsabs.harvard.edu/abs/2022ApJ...938..110B} {938, 110}

\bibitem[\protect\citeauthoryear{{Burns} et~al.,}{{Burns}
  et~al.}{2011}]{Burns11}
{Burns} C.~R.,  et~al., 2011, \mn@doi [\aj] {10.1088/0004-6256/141/1/19}, \href
  {http://adsabs.harvard.edu/abs/2011AJ....141...19B} {141, 19}

\bibitem[\protect\citeauthoryear{{Burns} et~al.,}{{Burns}
  et~al.}{2014}]{Burns14}
{Burns} C.~R.,  et~al., 2014, \mn@doi [\apj] {10.1088/0004-637X/789/1/32},
  \href {https://ui.adsabs.harvard.edu/abs/2014ApJ...789...32B} {789, 32}

\bibitem[\protect\citeauthoryear{{Burns} et~al.,}{{Burns}
  et~al.}{2020}]{Burns20}
{Burns} C.~R.,  et~al., 2020, \mn@doi [\apj] {10.3847/1538-4357/ab8e3e}, \href
  {https://ui.adsabs.harvard.edu/abs/2020ApJ...895..118B} {895, 118}

\bibitem[\protect\citeauthoryear{{Carpenter} et~al.,}{{Carpenter}
  et~al.}{2017}]{Carpenter17}
{Carpenter} B.,  et~al., 2017, Journal of Statistical Software, \href
  {https://ui.adsabs.harvard.edu/abs/2017JSS....76....1C} {76, 1}

\bibitem[\protect\citeauthoryear{{Cartier} et~al.,}{{Cartier}
  et~al.}{2014}]{Cartier14}
{Cartier} R.,  et~al., 2014, \mn@doi [\apj] {10.1088/0004-637X/789/1/89}, \href
  {https://ui.adsabs.harvard.edu/abs/2014ApJ...789...89C} {789, 89}

\bibitem[\protect\citeauthoryear{{Childress} et~al.,}{{Childress}
  et~al.}{2013}]{Childress13}
{Childress} M.,  et~al., 2013, \mn@doi [\apj] {10.1088/0004-637X/770/2/108},
  \href {http://adsabs.harvard.edu/abs/2013ApJ...770..108C} {770, 108}

\bibitem[\protect\citeauthoryear{{Childress}, {Wolf}  \& {Zahid}}{{Childress}
  et~al.}{2014}]{Childress14}
{Childress} M.~J.,  {Wolf} C.,   {Zahid} H.~J.,  2014, \mn@doi [\mnras]
  {10.1093/mnras/stu1892}, \href
  {http://adsabs.harvard.edu/abs/2014MNRAS.445.1898C} {445, 1898}

\bibitem[\protect\citeauthoryear{{Cikota}, {Deustua}  \& {Marleau}}{{Cikota}
  et~al.}{2016}]{Cikota16}
{Cikota} A.,  {Deustua} S.,   {Marleau} F.,  2016, \mn@doi [\apj]
  {10.3847/0004-637X/819/2/152}, \href
  {https://ui.adsabs.harvard.edu/abs/2016ApJ...819..152C} {819, 152}

\bibitem[\protect\citeauthoryear{{D'Andrea} et~al.,}{{D'Andrea}
  et~al.}{2011}]{DAndrea11}
{D'Andrea} C.~B.,  et~al., 2011, \mn@doi [\apj] {10.1088/0004-637X/743/2/172},
  \href {https://ui.adsabs.harvard.edu/abs/2011ApJ...743..172D} {743, 172}

\bibitem[\protect\citeauthoryear{{Elias-Rosa} et~al.,}{{Elias-Rosa}
  et~al.}{2006}]{Elias-Rosa06}
{Elias-Rosa} N.,  et~al., 2006, \mn@doi [\mnras]
  {10.1111/j.1365-2966.2006.10430.x}, \href
  {http://adsabs.harvard.edu/abs/2006MNRAS.369.1880E} {369, 1880}

\bibitem[\protect\citeauthoryear{{Elias-Rosa} et~al.,}{{Elias-Rosa}
  et~al.}{2008}]{Elias-Rosa08}
{Elias-Rosa} N.,  et~al., 2008, \mn@doi [\mnras]
  {10.1111/j.1365-2966.2007.12638.x}, \href
  {http://adsabs.harvard.edu/abs/2008MNRAS.384..107E} {384, 107}

\bibitem[\protect\citeauthoryear{{Fitzpatrick}}{{Fitzpatrick}}{1999}]{Fitzpatrick99}
{Fitzpatrick} E.~L.,  1999, \mn@doi [\pasp] {10.1086/316293}, \href
  {https://ui.adsabs.harvard.edu/abs/1999PASP..111...63F} {111, 63}

\bibitem[\protect\citeauthoryear{{Foley} et~al.,}{{Foley}
  et~al.}{2012}]{Foley12:sdss}
{Foley} R.~J.,  et~al., 2012, \mn@doi [\aj] {10.1088/0004-6256/143/5/113},
  \href {http://adsabs.harvard.edu/abs/2012AJ....143..113F} {143, 113}

\bibitem[\protect\citeauthoryear{{Friedman} et~al.,}{{Friedman}
  et~al.}{2015}]{Friedman15}
{Friedman} A.~S.,  et~al., 2015, \mn@doi [\apjs] {10.1088/0067-0049/220/1/9},
  \href {http://adsabs.harvard.edu/abs/2015ApJS..220....9F} {220, 9}

\bibitem[\protect\citeauthoryear{{Gelman} \& {Rubin}}{{Gelman} \&
  {Rubin}}{1992}]{Gelman92}
{Gelman} A.,  {Rubin} D.~B.,  1992, \mn@doi [Statistical Science]
  {10.1214/ss/1177011136}, \href
  {https://ui.adsabs.harvard.edu/abs/1992StaSc...7..457G} {7, 457}

\bibitem[\protect\citeauthoryear{Gelman, Carlin, Stern, Dunson, Vehtari  \&
  Rubin}{Gelman et~al.}{2013}]{Gelman13}
Gelman A.,  Carlin J.~B.,  Stern H.~S.,  Dunson D.,  Vehtari A.,   Rubin D.~B.,
   2013, Bayesian Data Analysis, 3rd Edition.
{Chapman \& Hall/CRC}, New York

\bibitem[\protect\citeauthoryear{{Hicken} et~al.,}{{Hicken}
  et~al.}{2009}]{Hicken09:lc}
{Hicken} M.,  et~al., 2009, \mn@doi [\apj] {10.1088/0004-637X/700/1/331}, \href
  {http://adsabs.harvard.edu/abs/2009ApJ...700..331H} {700, 331}

\bibitem[\protect\citeauthoryear{{Hicken} et~al.,}{{Hicken}
  et~al.}{2012}]{Hicken12}
{Hicken} M.,  et~al., 2012, \mn@doi [\apjs] {10.1088/0067-0049/200/2/12}, \href
  {http://adsabs.harvard.edu/abs/2012ApJS..200...12H} {200, 12}

\bibitem[\protect\citeauthoryear{{Hoang}}{{Hoang}}{2017}]{Hoang17}
{Hoang} T.,  2017, \mn@doi [\apj] {10.3847/1538-4357/836/1/13}, \href
  {https://ui.adsabs.harvard.edu/abs/2017ApJ...836...13H} {836, 13}

\bibitem[\protect\citeauthoryear{Hoffman \& Gelman}{Hoffman \&
  Gelman}{2014}]{Hoffman14}
Hoffman M.~D.,  Gelman A.,  2014, J. Machine Learning Res., 15, 1593

\bibitem[\protect\citeauthoryear{{Hsiao}}{{Hsiao}}{2009}]{Hsiao09}
{Hsiao} E.~Y.,  2009, PhD thesis, Univ. Victoria

\bibitem[\protect\citeauthoryear{{Jha} et~al.,}{{Jha}
  et~al.}{1999}]{Jha99:98bu}
{Jha} S.,  et~al., 1999, \mn@doi [\apjs] {10.1086/313275}, \href
  {http://adsabs.harvard.edu/abs/1999ApJS..125...73J} {125, 73}

\bibitem[\protect\citeauthoryear{{Jha}, {Riess}  \& {Kirshner}}{{Jha}
  et~al.}{2007}]{Jha07}
{Jha} S.,  {Riess} A.~G.,   {Kirshner} R.~P.,  2007, \mn@doi [\apj]
  {10.1086/512054}, \href {http://adsabs.harvard.edu/abs/2007ApJ...659..122J}
  {659, 122}

\bibitem[\protect\citeauthoryear{{Johansson} et~al.,}{{Johansson}
  et~al.}{2021}]{Johansson21}
{Johansson} J.,  et~al., 2021, \mn@doi [\apj] {10.3847/1538-4357/ac2f9e}, \href
  {https://ui.adsabs.harvard.edu/abs/2021ApJ...923..237J} {923, 237}

\bibitem[\protect\citeauthoryear{{Jones} et~al.,}{{Jones}
  et~al.}{2022}]{Jones22}
{Jones} D.~O.,  et~al., 2022, \mn@doi [\apj] {10.3847/1538-4357/ac755b}, \href
  {https://ui.adsabs.harvard.edu/abs/2022ApJ...933..172J} {933, 172}

\bibitem[\protect\citeauthoryear{{Kelly}, {Hicken}, {Burke}, {Mandel}  \&
  {Kirshner}}{{Kelly} et~al.}{2010}]{Kelly10}
{Kelly} P.~L.,  {Hicken} M.,  {Burke} D.~L.,  {Mandel} K.~S.,   {Kirshner}
  R.~P.,  2010, \mn@doi [\apj] {10.1088/0004-637X/715/2/743}, \href
  {http://adsabs.harvard.edu/abs/2010ApJ...715..743K} {715, 743}

\bibitem[\protect\citeauthoryear{{Krisciunas}, {Hastings}, {Loomis},
  {McMillan}, {Rest}, {Riess}  \& {Stubbs}}{{Krisciunas}
  et~al.}{2000}]{Krisciunas00}
{Krisciunas} K.,  {Hastings} N.~C.,  {Loomis} K.,  {McMillan} R.,  {Rest} A.,
  {Riess} A.~G.,   {Stubbs} C.,  2000, \mn@doi [\apj] {10.1086/309263}, \href
  {http://adsabs.harvard.edu/abs/2000ApJ...539..658K} {539, 658}

\bibitem[\protect\citeauthoryear{{Krisciunas} et~al.,}{{Krisciunas}
  et~al.}{2001}]{Krisciunas01}
{Krisciunas} K.,  et~al., 2001, \mn@doi [\aj] {10.1086/322120}, \href
  {https://ui.adsabs.harvard.edu/abs/2001AJ....122.1616K} {122, 1616}

\bibitem[\protect\citeauthoryear{{Krisciunas} et~al.,}{{Krisciunas}
  et~al.}{2003}]{Krisciunas03}
{Krisciunas} K.,  et~al., 2003, \mn@doi [\aj] {10.1086/345571}, \href
  {http://adsabs.harvard.edu/abs/2003AJ....125..166K} {125, 166}

\bibitem[\protect\citeauthoryear{{Krisciunas} et~al.,}{{Krisciunas}
  et~al.}{2004a}]{Krisciunas04:4sn}
{Krisciunas} K.,  et~al., 2004a, \mn@doi [\aj] {10.1086/381911}, \href
  {https://ui.adsabs.harvard.edu/abs/2004AJ....127.1664K} {127, 1664}

\bibitem[\protect\citeauthoryear{{Krisciunas} et~al.,}{{Krisciunas}
  et~al.}{2004b}]{Krisciunas04:7sn}
{Krisciunas} K.,  et~al., 2004b, \mn@doi [\aj] {10.1086/425629}, \href
  {http://adsabs.harvard.edu/abs/2004AJ....128.3034K} {128, 3034}

\bibitem[\protect\citeauthoryear{{Krisciunas} et~al.,}{{Krisciunas}
  et~al.}{2007}]{Krisciunas07}
{Krisciunas} K.,  et~al., 2007, \mn@doi [\aj] {10.1086/509126}, \href
  {http://adsabs.harvard.edu/abs/2007AJ....133...58K} {133, 58}

\bibitem[\protect\citeauthoryear{{Krisciunas} et~al.,}{{Krisciunas}
  et~al.}{2017}]{Krisciunas17}
{Krisciunas} K.,  et~al., 2017, \mn@doi [\aj] {10.3847/1538-3881/aa8df0}, \href
  {http://adsabs.harvard.edu/abs/2017AJ....154..211K} {154, 211}

\bibitem[\protect\citeauthoryear{{Lampeitl} et~al.,}{{Lampeitl}
  et~al.}{2010}]{Lampeitl10:host}
{Lampeitl} H.,  et~al., 2010, \mn@doi [\apj] {10.1088/0004-637X/722/1/566},
  \href {http://adsabs.harvard.edu/abs/2010ApJ...722..566L} {722, 566}

\bibitem[\protect\citeauthoryear{Lewandowski, Kurowicka  \& Joe}{Lewandowski
  et~al.}{2009}]{LKJ09}
Lewandowski D.,  Kurowicka D.,   Joe H.,  2009, \mn@doi [Journal of
  Multivariate Analysis] {https://doi.org/10.1016/j.jmva.2009.04.008}, 100,
  1989

\bibitem[\protect\citeauthoryear{{Mandel}, {Wood-Vasey}, {Friedman}  \&
  {Kirshner}}{{Mandel} et~al.}{2009}]{Mandel09}
{Mandel} K.~S.,  {Wood-Vasey} W.~M.,  {Friedman} A.~S.,   {Kirshner} R.~P.,
  2009, \mn@doi [\apj] {10.1088/0004-637X/704/1/629}, \href
  {http://adsabs.harvard.edu/abs/2009ApJ...704..629M} {704, 629}

\bibitem[\protect\citeauthoryear{{Mandel}, {Narayan}  \& {Kirshner}}{{Mandel}
  et~al.}{2011}]{Mandel11}
{Mandel} K.~S.,  {Narayan} G.,   {Kirshner} R.~P.,  2011, \mn@doi [\apj]
  {10.1088/0004-637X/731/2/120}, \href
  {http://adsabs.harvard.edu/abs/2011ApJ...731..120M} {731, 120}

\bibitem[\protect\citeauthoryear{{Mandel}, {Foley}  \& {Kirshner}}{{Mandel}
  et~al.}{2014}]{Mandel14}
{Mandel} K.~S.,  {Foley} R.~J.,   {Kirshner} R.~P.,  2014, \mn@doi [\apj]
  {10.1088/0004-637X/797/2/75}, \href
  {http://adsabs.harvard.edu/abs/2014ApJ...797...75M} {797, 75}

\bibitem[\protect\citeauthoryear{{Mandel}, {Scolnic}, {Shariff}, {Foley}  \&
  {Kirshner}}{{Mandel} et~al.}{2017}]{Mandel17}
{Mandel} K.~S.,  {Scolnic} D.~M.,  {Shariff} H.,  {Foley} R.~J.,   {Kirshner}
  R.~P.,  2017, \mn@doi [\apj] {10.3847/1538-4357/aa6038}, \href
  {http://adsabs.harvard.edu/abs/2017ApJ...842...93M} {842, 93}

\bibitem[\protect\citeauthoryear{{Mandel}, {Thorp}, {Narayan}, {Friedman}  \&
  {Avelino}}{{Mandel} et~al.}{2022}]{Mandel22}
{Mandel} K.~S.,  {Thorp} S.,  {Narayan} G.,  {Friedman} A.~S.,   {Avelino} A.,
  2022, \mn@doi [\mnras] {10.1093/mnras/stab3496}, \href
  {https://ui.adsabs.harvard.edu/abs/2022MNRAS.510.3939M} {510, 3939}

\bibitem[\protect\citeauthoryear{{Marion} et~al.,}{{Marion}
  et~al.}{2015}]{Marion15}
{Marion} G.~H.,  et~al., 2015, \mn@doi [\apj] {10.1088/0004-637X/798/1/39},
  \href {http://adsabs.harvard.edu/abs/2015ApJ...798...39M} {798, 39}

\bibitem[\protect\citeauthoryear{{Matheson} et~al.,}{{Matheson}
  et~al.}{2012}]{Matheson12}
{Matheson} T.,  et~al., 2012, \mn@doi [\apj] {10.1088/0004-637X/754/1/19},
  \href {http://adsabs.harvard.edu/abs/2012ApJ...754...19M} {754, 19}

\bibitem[\protect\citeauthoryear{{Meldorf} et~al.,}{{Meldorf}
  et~al.}{2023}]{Meldorf23}
{Meldorf} C.,  et~al., 2023, \mn@doi [\mnras] {10.1093/mnras/stac3056}, \href
  {https://ui.adsabs.harvard.edu/abs/2023MNRAS.518.1985M} {518, 1985}

\bibitem[\protect\citeauthoryear{{M{\"u}ller-Bravo} et~al.,}{{M{\"u}ller-Bravo}
  et~al.}{2022}]{MullerBravo22}
{M{\"u}ller-Bravo} T.~E.,  et~al., 2022, \mn@doi [\aap]
  {10.1051/0004-6361/202243845}, \href
  {https://ui.adsabs.harvard.edu/abs/2022A&A...665A.123M} {665, A123}

\bibitem[\protect\citeauthoryear{{Neill} et~al.,}{{Neill}
  et~al.}{2009}]{Neill09}
{Neill} J.~D.,  et~al., 2009, \mn@doi [\apj] {10.1088/0004-637X/707/2/1449},
  \href {https://ui.adsabs.harvard.edu/abs/2009ApJ...707.1449N} {707, 1449}

\bibitem[\protect\citeauthoryear{{Nicolas} et~al.,}{{Nicolas}
  et~al.}{2021}]{Nicolas21}
{Nicolas} N.,  et~al., 2021, \mn@doi [\aap] {10.1051/0004-6361/202038447},
  \href {https://ui.adsabs.harvard.edu/abs/2021A&A...649A..74N} {649, A74}

\bibitem[\protect\citeauthoryear{{Nobili} \& {Goobar}}{{Nobili} \&
  {Goobar}}{2008}]{Nobili08}
{Nobili} S.,  {Goobar} A.,  2008, \mn@doi [\aap] {10.1051/0004-6361:20079292},
  \href {https://ui.adsabs.harvard.edu/abs/2008A&A...487...19N} {487, 19}

\bibitem[\protect\citeauthoryear{{Pan} et~al.,}{{Pan} et~al.}{2014}]{Pan14}
{Pan} Y.-C.,  et~al., 2014, \mn@doi [\mnras] {10.1093/mnras/stt2287}, \href
  {http://adsabs.harvard.edu/abs/2014MNRAS.438.1391P} {438, 1391}

\bibitem[\protect\citeauthoryear{{Perlmutter} et~al.,}{{Perlmutter}
  et~al.}{1999}]{Perlmutter99}
{Perlmutter} S.,  et~al., 1999, \mn@doi [\apj] {10.1086/307221}, \href
  {http://adsabs.harvard.edu/abs/1999ApJ...517..565P} {517, 565}

\bibitem[\protect\citeauthoryear{{Peterson} et~al.,}{{Peterson}
  et~al.}{2023}]{Peterson23}
{Peterson} E.~R.,  et~al., 2023, \mn@doi [\mnras] {10.1093/mnras/stad1077},
  \href {https://ui.adsabs.harvard.edu/abs/2023MNRAS.522.2478P} {522, 2478}

\bibitem[\protect\citeauthoryear{{Phillips}}{{Phillips}}{1993}]{Phillips93}
{Phillips} M.~M.,  1993, \mn@doi [\apjl] {10.1086/186970}, \href
  {http://adsabs.harvard.edu/abs/1993ApJ...413L.105P} {413, L105}

\bibitem[\protect\citeauthoryear{{Phillips} et~al.,}{{Phillips}
  et~al.}{2019}]{Phillips19}
{Phillips} M.~M.,  et~al., 2019, \mn@doi [\pasp] {10.1088/1538-3873/aae8bd},
  \href {https://ui.adsabs.harvard.edu/abs/2019PASP..131a4001P} {131, 014001}

\bibitem[\protect\citeauthoryear{{Pignata} et~al.,}{{Pignata}
  et~al.}{2008}]{Pignata08:02dj}
{Pignata} G.,  et~al., 2008, \mn@doi [\mnras]
  {10.1111/j.1365-2966.2008.13434.x}, \href
  {http://adsabs.harvard.edu/abs/2008MNRAS.388..971P} {388, 971}

\bibitem[\protect\citeauthoryear{{Ponder}, {Wood-Vasey}, {Weyant}, {Barton},
  {Galbany}, {Liu}, {Garnavich}  \& {Matheson}}{{Ponder}
  et~al.}{2021}]{Ponder21}
{Ponder} K.~A.,  {Wood-Vasey} W.~M.,  {Weyant} A.,  {Barton} N.~T.,  {Galbany}
  L.,  {Liu} S.,  {Garnavich} P.,   {Matheson} T.,  2021, \mn@doi [\apj]
  {10.3847/1538-4357/ac2d99}, \href
  {https://ui.adsabs.harvard.edu/abs/2021ApJ...923..197P} {923, 197}

\bibitem[\protect\citeauthoryear{{Popovic}, {Brout}, {Kessler}  \&
  {Scolnic}}{{Popovic} et~al.}{2023}]{Popovic23}
{Popovic} B.,  {Brout} D.,  {Kessler} R.,   {Scolnic} D.,  2023, \mn@doi [\apj]
  {10.3847/1538-4357/aca273}, \href
  {https://ui.adsabs.harvard.edu/abs/2023ApJ...945...84P} {945, 84}

\bibitem[\protect\citeauthoryear{{Riess} et~al.,}{{Riess}
  et~al.}{1998}]{Riess98:lambda}
{Riess} A.~G.,  et~al., 1998, \aj, \href
  {http://adsabs.harvard.edu/cgi-bin/nph-bib_query?bibcode=1998AJ....116.1009R&db_key=AST}
  {116, 1009}

\bibitem[\protect\citeauthoryear{{Riess} et~al.,}{{Riess}
  et~al.}{2022}]{Riess22}
{Riess} A.~G.,  et~al., 2022, \mn@doi [\apjl] {10.3847/2041-8213/ac5c5b}, \href
  {https://ui.adsabs.harvard.edu/abs/2022ApJ...934L...7R} {934, L7}

\bibitem[\protect\citeauthoryear{{Rigault} et~al.,}{{Rigault}
  et~al.}{2013}]{Rigault13}
{Rigault} M.,  et~al., 2013, \mn@doi [\aap] {10.1051/0004-6361/201322104},
  \href {http://adsabs.harvard.edu/abs/2013A%26A...560A..66R} {560, A66}

\bibitem[\protect\citeauthoryear{{Rose}, {Garnavich}  \& {Berg}}{{Rose}
  et~al.}{2019}]{Rose19}
{Rose} B.~M.,  {Garnavich} P.~M.,   {Berg} M.~A.,  2019, \mn@doi [\apj]
  {10.3847/1538-4357/ab0704}, \href
  {https://ui.adsabs.harvard.edu/abs/2019ApJ...874...32R} {874, 32}

\bibitem[\protect\citeauthoryear{{Salim}, {Boquien}  \& {Lee}}{{Salim}
  et~al.}{2018}]{Salim18}
{Salim} S.,  {Boquien} M.,   {Lee} J.~C.,  2018, \mn@doi [\apj]
  {10.3847/1538-4357/aabf3c}, \href
  {https://ui.adsabs.harvard.edu/abs/2018ApJ...859...11S} {859, 11}

\bibitem[\protect\citeauthoryear{{Schlafly} et~al.,}{{Schlafly}
  et~al.}{2016}]{Schlafly16}
{Schlafly} E.~F.,  et~al., 2016, \mn@doi [\apj] {10.3847/0004-637X/821/2/78},
  \href {https://ui.adsabs.harvard.edu/abs/2016ApJ...821...78S} {821, 78}

\bibitem[\protect\citeauthoryear{{Scolnic} et~al.,}{{Scolnic}
  et~al.}{2018}]{Scolnic18:ps1}
{Scolnic} D.~M.,  et~al., 2018, \mn@doi [\apj] {10.3847/1538-4357/aab9bb},
  \href {http://adsabs.harvard.edu/abs/2018ApJ...859..101S} {859, 101}

\bibitem[\protect\citeauthoryear{{Scolnic} et~al.,}{{Scolnic}
  et~al.}{2019}]{Scolnic19}
{Scolnic} D.,  et~al., 2019, Astro2020: Decadal Survey on Astronomy and
  Astrophysics, \href {https://ui.adsabs.harvard.edu/abs/2019astro2020T.270S}
  {2020, 270}

\bibitem[\protect\citeauthoryear{{Stan Development Team}}{{Stan Development
  Team}}{2020}]{Stan20}
{Stan Development Team} 2020, Stan Modelling Language Users Guide and Reference
  Manual v.2.25.
\url {https://mc-stan.org}

\bibitem[\protect\citeauthoryear{{Stanishev} et~al.,}{{Stanishev}
  et~al.}{2007}]{Stanishev07:03du}
{Stanishev} V.,  et~al., 2007, \mn@doi [\aap] {10.1051/0004-6361:20066020},
  \href {http://adsabs.harvard.edu/abs/2007A%26A...469..645S} {469, 645}

\bibitem[\protect\citeauthoryear{{Sullivan} et~al.,}{{Sullivan}
  et~al.}{2010}]{Sullivan10}
{Sullivan} M.,  et~al., 2010, \mn@doi [\mnras]
  {10.1111/j.1365-2966.2010.16731.x}, \href
  {http://adsabs.harvard.edu/abs/2010MNRAS.406..782S} {406, 782}

\bibitem[\protect\citeauthoryear{{Talts}, {Betancourt}, {Simpson}, {Vehtari}
  \& {Gelman}}{{Talts} et~al.}{2018}]{Talts18}
{Talts} S.,  {Betancourt} M.,  {Simpson} D.,  {Vehtari} A.,   {Gelman} A.,
  2018, \mn@doi [arXiv e-prints] {10.48550/arXiv.1804.06788}, \href
  {https://ui.adsabs.harvard.edu/abs/2018arXiv180406788T} {p. arXiv:1804.06788}

\bibitem[\protect\citeauthoryear{{Thorp} \& {Mandel}}{{Thorp} \&
  {Mandel}}{2022}]{Thorp22}
{Thorp} S.,  {Mandel} K.~S.,  2022, \mn@doi [\mnras] {10.1093/mnras/stac2714},
  \href {https://ui.adsabs.harvard.edu/abs/2022MNRAS.517.2360T} {517, 2360}

\bibitem[\protect\citeauthoryear{{Thorp}, {Mandel}, {Jones}, {Ward}  \&
  {Narayan}}{{Thorp} et~al.}{2021}]{Thorp21}
{Thorp} S.,  {Mandel} K.~S.,  {Jones} D.~O.,  {Ward} S.~M.,   {Narayan} G.,
  2021, \mn@doi [\mnras] {10.1093/mnras/stab2849}, \href
  {https://ui.adsabs.harvard.edu/abs/2021MNRAS.508.4310T} {508, 4310}

\bibitem[\protect\citeauthoryear{{Tripp}}{{Tripp}}{1998}]{Tripp98}
{Tripp} R.,  1998, \aap, \href
  {http://adsabs.harvard.edu/abs/1998A%26A...331..815T} {331, 815}

\bibitem[\protect\citeauthoryear{{Uddin} et~al.,}{{Uddin}
  et~al.}{2020}]{Uddin20}
{Uddin} S.~A.,  et~al., 2020, \mn@doi [\apj] {10.3847/1538-4357/abafb7}, \href
  {https://ui.adsabs.harvard.edu/abs/2020ApJ...901..143U} {901, 143}

\bibitem[\protect\citeauthoryear{{Valentini} et~al.,}{{Valentini}
  et~al.}{2003}]{Valentini03}
{Valentini} G.,  et~al., 2003, \mn@doi [\apj] {10.1086/377448}, \href
  {http://adsabs.harvard.edu/abs/2003ApJ...595..779V} {595, 779}

\bibitem[\protect\citeauthoryear{{Vehtari}, {Gelman}, {Simpson}, {Carpenter}
  \& {B{\"u}rkner}}{{Vehtari} et~al.}{2019}]{Vehtari21}
{Vehtari} A.,  {Gelman} A.,  {Simpson} D.,  {Carpenter} B.,   {B{\"u}rkner}
  P.-C.,  2019, \mn@doi [arXiv e-prints] {10.48550/arXiv.1903.08008}, \href
  {https://ui.adsabs.harvard.edu/abs/2019arXiv190308008V} {p. arXiv:1903.08008}

\bibitem[\protect\citeauthoryear{{Wang} et~al.,}{{Wang}
  et~al.}{2008}]{Wang08:06x}
{Wang} X.,  et~al., 2008, \mn@doi [\apj] {10.1086/526413}, \href
  {http://adsabs.harvard.edu/abs/2008ApJ...675..626W} {675, 626}

\bibitem[\protect\citeauthoryear{{Ward} et~al.,}{{Ward} et~al.}{2022}]{Ward22}
{Ward} S.~M.,  et~al., 2022, arXiv e-prints, \href
  {https://ui.adsabs.harvard.edu/abs/2022arXiv220910558W} {p. arXiv:2209.10558}

\bibitem[\protect\citeauthoryear{{Wojtak}, {Hjorth}  \& {Hjortlund}}{{Wojtak}
  et~al.}{2023}]{Wojtak23}
{Wojtak} R.,  {Hjorth} J.,   {Hjortlund} J.~O.,  2023, \mn@doi [\mnras]
  {10.1093/mnras/stad2590}, \href
  {https://ui.adsabs.harvard.edu/abs/2023MNRAS.525.5187W} {525, 5187}

\bibitem[\protect\citeauthoryear{{Wood-Vasey} et~al.,}{{Wood-Vasey}
  et~al.}{2008}]{Wood-Vasey08}
{Wood-Vasey} W.~M.,  et~al., 2008, \mn@doi [\apj] {10.1086/592374}, \href
  {http://adsabs.harvard.edu/abs/2008ApJ...689..377W} {689, 377}

\bibitem[\protect\citeauthoryear{{Zhang} et~al.,}{{Zhang}
  et~al.}{2016}]{Zhang16:11fe}
{Zhang} K.,  et~al., 2016, \mn@doi [\apj] {10.3847/0004-637X/820/1/67}, \href
  {http://adsabs.harvard.edu/abs/2016ApJ...820...67Z} {820, 67}

\makeatother
\end{thebibliography}




\appendix

\newcommand\at[2]{\left.#1\right|_{#2}}

\section{Rest-Frame Peak Apparent Magnitude Estimates}
Table~\ref{tab:peakinterpolations} records estimates of rest-frame peak apparent magnitudes using the default pre-processing choices (interpolate $uBgVriYJH$, use \textsc{SNooPy} to apply Milky Way extinction and mangled K-corrections, use a 1DGP to estimate $T_{B;\,\rm{max}}$, and use a 2DGP to interpolate rest-frame data to peak time). The full table is available online, and at \href{https://github.com/sam-m-ward/birdsnack/tree/main}{\texttt{https://github.com/birdsnack}}.

\renewcommand{\arraystretch}{1.2}
\setlength{\tabcolsep}{7pt}
\begin{table*}
\begin{threeparttable}
    \centering
    \caption{Estimates of SN~Ia rest-frame peak apparent magnitudes with Milky Way extinction corrections applied.
    }
    \label{tab:peakinterpolations}
    \begin{tabular}{l c c c c c c c c c c c c c}
    \toprule
    SN & Source\,\tnote{a} & $B$ & $\sigma_B$ & $V$ & $\sigma_V$ & $r$ & $\sigma_r$ & $i$ & $\sigma_i$ & $J$ & $\sigma_J$ & $H$ & $\sigma_H$  \\
    \midrule
2004eo & CSP & 15.120 & 0.005 & 15.132 & 0.008 & 15.129 & 0.016 & 15.683 & 0.006 & 15.588 & 0.018 & 15.850 & 0.021  \\
2004ey & CSP & 14.772 & 0.005 & 14.824 & 0.005 & 14.932 & 0.006 & 15.589 & 0.008 & 15.558 & 0.024 & 15.906 & 0.042  \\
2005el & CSP & 14.891 & 0.057 & 14.808 & 0.106 & 14.966 & 0.12 & 15.501 & 0.069 & 15.655 & 0.031 & 15.856 & 0.04  \\
2005iq & CSP & 16.796 & 0.011 & 16.826 & 0.01 & 16.920 & 0.011 & 17.529 & 0.013 & 17.460 & 0.027 & 17.642 & 0.058  \\
2005kc & CSP & 15.577 & 0.01 & 15.394 & 0.009 & 15.352 & 0.009 & 15.805 & 0.009 & 15.555 & 0.031 & 15.743 & 0.034 \\
\multicolumn{13}{c}{...} \\
\bottomrule
\end{tabular}
\begin{tablenotes}
        \item[a] Source of SN~Ia light curves. CSP: \cite{Krisciunas17}, CfA: \cite{Wood-Vasey08,Hicken09:lc, Hicken12, Friedman15}, RATIR: \cite{Johansson21}, Misc: J99: \cite{Jha99:98bu}, K00: \cite{Krisciunas00}, K03: \cite{Krisciunas03}, V03: \cite{Valentini03} K04a: \cite{Krisciunas04:4sn}, K04b: \cite{Krisciunas04:7sn}, ER06: \cite{Elias-Rosa06}, M12+Z16: \cite{Matheson12,Zhang16:11fe}, C14: \cite{Cartier14}, M15: \cite{Marion15}, B20: \cite{Burns20}.
    \end{tablenotes}
    \end{threeparttable}
\end{table*}

\section{Computation of Effective Wavelengths}
\label{S:appendixlameffcomp}

\renewcommand{\arraystretch}{1}
\setlength{\tabcolsep}{3pt}
\begin{table}
\begin{threeparttable}
    \centering
    \caption{The changes in effective wavelength, with respect to the passband central wavelength, $\lambda_c$, computed using simulations of SED-integrated extinguished and intrinsic apparent magnitudes, and Eq.~\ref{eq:xieff} (more in \S\ref{S:EffectiveLambda}).
    }
    \label{tab:leffs2}
    \begin{tabular}{c c c c c}
    \toprule
    Passband & $\lambda_{c}$ (\AA)&
        \multicolumn{3}{c}{Changes in Effective Wavelength (\AA)\,\tnote{a}}  \\
\cmidrule{3-5}
 &  &$\theta^s\sim \mathcal{N}(0,1)$\,\tnote{b} & $\theta^s=0$\,\tnote{c} & $\theta^s, \sigma_{\rm{int}} = 0$\,\tnote{d}\\
\midrule
$B$ &  $4402.1$ & $-35.6\pm5.6$ & $-38.2\pm4.2$ & $-38.8\pm4.2$ \\ 
$V$ &  $5389.3$ & $-27.9\pm4.1$ & $-28.1\pm3.7$ & $-28.2\pm4.0$ \\ 
$r$ &  $6239.9$ & $\,\,\,-90.8\pm24.3$ & $\,\,\,-89.6\pm23.8$ & $\,\,\,-90.8\pm24.2$ \\ 
$i$ &  $7631.1$ & $\,\,\,-84.6\pm14.0$ & $\,\,\,-83.6\pm13.2$ & $\,\,\,-83.3\pm14.5$ \\ 
$J$ &  $12516.3$ & $\,\,\,-78.1\pm13.8$ & $\,\,\,-78.1\pm12.9$ & $\,\,\,-80.2\pm14.0$ \\ 
$H$ &  $16277.2$ & $-138.1\pm10.7$ & $-142.9\pm8.4\,\,\,$ & $-143.6\pm9.0\,\,\,$ \\
\midrule
 &  &$\tau_A = 0.2$\,\tnote{e} & $\tau_A = 0.5$ & $\tau_A = 0.8$\\
 \cmidrule{3-5}
 $B$ &  $-$ & $-42.0\pm4.7$ & $-36.9\pm3.6$ & $-32.0\pm3.2$ \\ 
 $V$ &  $-$ & $-30.0\pm2.6$ & $-27.5\pm3.5$ & $-25.7\pm3.9$ \\ 
 $r$ &  $-$ & $\,\,\,-97.6\pm25.1$ & $\,\,\,-93.8\pm25.4$ & $\,\,\,-87.9\pm22.9$ \\ 
 $i$ &  $-$ & $\,\,\,-87.7\pm14.7$ & $\,\,\,-85.4\pm15.5$ & $\,\,\,-82.5\pm14.4$ \\ 
 $J$ &  $-$ & $\,\,\,-79.9\pm15.1$ & $\,\,\,-78.4\pm15.4$ & $\,\,\,-77.2\pm13.8$ \\ 
 $H$ &  $-$ & $-139.2\pm10.2$ & $-137.7\pm11.3$ & $-135.9\pm10.6$ \\
\bottomrule
    \end{tabular}
    \begin{tablenotes}
    \item[a] Uncertainties denote the sample standard deviation of effective wavelengths over 1000 simulations.
    \item[b] Population variations in light curve shape, $\theta^s$, are included in simulations.
    \item[c] Light curve shape is fixed to zero for all simulated SEDs.
    \item[d] Residual perturbations are also fixed to zero.
    \item[e] Keeping light curve shape variations and residual perturbations turned on, we test fixing the dust extinction hyperparameter.
    \end{tablenotes}
    \end{threeparttable}
\end{table}

We detail here our procedure for computing effective wavelengths from SN light curves simulated with \textsc{BayeSN}~\citep{Mandel22, Thorp21, Ward22}. To simulate an intrinsic SN SED, we include the \cite{Hsiao09} spectral template, and the \textsc{BayeSN} zeroth order population mean SED component, $W_0(t,\lambda)$. We also incorporate light curve shape variations using the first functional principal component, $W_1(t,\lambda)$. The light curve shape parameters, $\theta^s$, are drawn from a standard normal, and we impose $\theta^s \in [-1.5,2.0]$, to match the \citetalias{Mandel22} training sample. 

At peak time, we then add on residual perturbations and host galaxy dust extinction. For the residuals, we synthesise correlation matrices using the $LKJ(1)$ distribution as in Eq.~\ref{eq:LKJ}, and intrinsic dispersions from a uniform distribution:
\begin{equation}  
\sigma_{\rm{int},\,i} \sim U(0,0.25).  
\end{equation}
For the dust hyperparameters, we draw:
\begingroup
\allowdisplaybreaks
\begin{align}
    \tau_A &\sim U(0.1,1), 
    \\
    \mu_{R_V} &\sim U(1,5),
    \\
    \sigma_{R_V} &\sim U(0.1,2).
\end{align}
\endgroup
We then draw $(A_V^s, R_V^s)$ as in Eqs.~(\ref{eq:drawAV},~\ref{eq:drawRV}), the residuals using a zero-mean multivariate Gaussian, and integrate the resulting intrinsic and extinguished SEDs over the $BVriJH$ passbands. 

For each simulation, we synthesise 100 SNe Ia, and compute the median set of effective wavelengths. We then repeat for 1000 simulations, and record the median and sample standard deviations of effective wavelengths in Table~\ref{tab:leffs2}. For all passbands, we find the effective wavelengths are lower than the passband central wavelengths. This aligns with expectations, because SN flux generally decreases with increasing wavelength, so the flux-weighted average wavelength should be lower than the nominal passband central wavelength. These offsets, of order $\sim 10-100$\AA, are significantly non-zero ($\approx 4-17\sigma$). The form of the intrinsic SEDs has a minimal effect on the effective wavelengths, with shifts $\lesssim0.1\sigma$ when excluding light curve shape population variations ($\theta^s = 0$), and when also removing residual perturbations ($\theta^s, \sigma_{\rm{int}} = 0$). To explore how this insensitivity can be attributed to averaging over the $\tau_A$ simulation distribution, we perform three additional sets of simulations, fixing $\tau_A = (0.2,0.5,0.8)$~mag, while including light curve shape variations and residuals. The shifts in effective wavelengths for different choices of $\tau_A$ are still insignificant, of order $\approx0.1-1 \sigma$. We show in \S\ref{s:senstolameffpreproc} that dust inferences are insensitive to the choice of either the default effective wavelengths or the passband central wavelengths. 

\section{Censored Data Simulations}
\label{S:appcensoreddata}

\begin{figure*}
\begin{subfigure}{0.5\textwidth}
\centering
\includegraphics[width=1\linewidth]{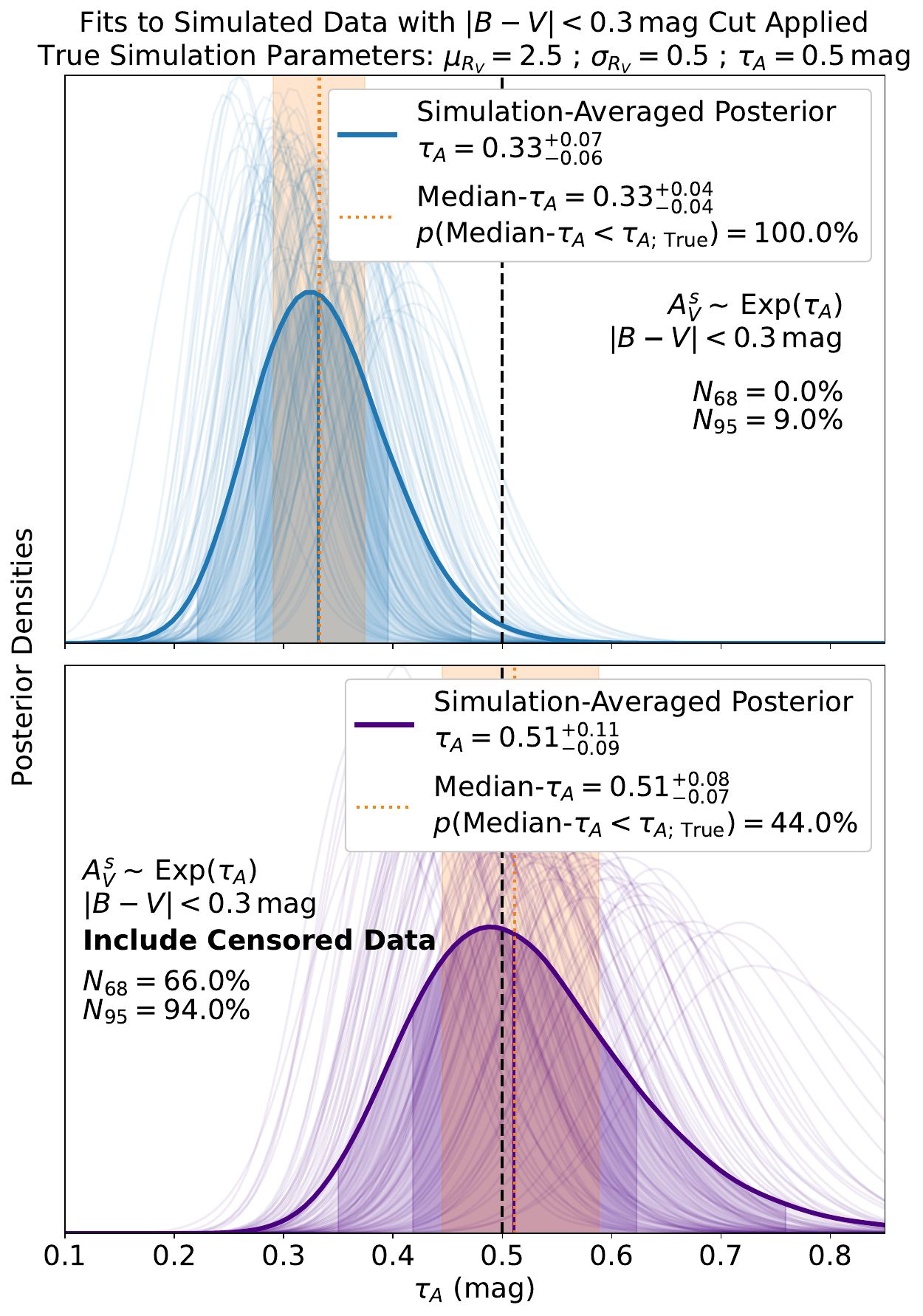}
\caption{Recovery of $\tau_A$.}
\end{subfigure}%
\begin{subfigure}{0.5\textwidth}
\centering
\includegraphics[width=1\linewidth]{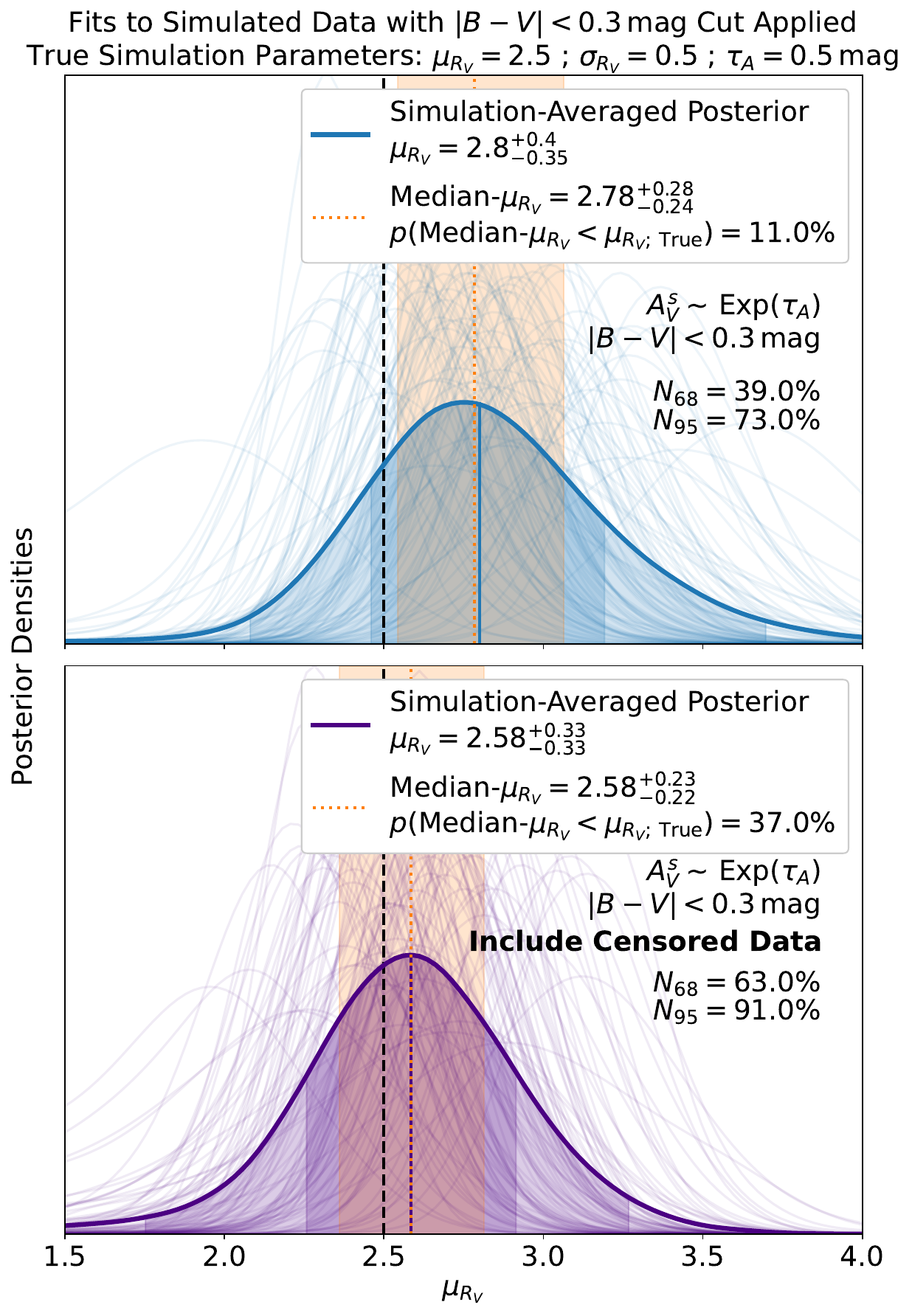}
\caption{Recovery of $\mu_{R_V}$.}
\end{subfigure}
    \caption{
    Recovery of dust hyperparameters,  $\tau_A=0.5$~mag and $\mu_{R_V}=2.5$, in simulated low-reddening samples, when censored data are either ignored or included in the model. Each panel shows posteriors from fits to 100 simulations. Each simulation comprises 100 SNe Ia, with dust extinction values drawn from an exponential $A_V^s$ distribution.
    The $|B-V|<0.3$~mag cut is applied to fit a low-reddening sample (which typically comprises $\approx 75-85$ SNe). 
    Sub-figure(a) shows $\tau_A$ inferences are biased low when fitting the low-reddening sub-sample in isolation, to account for the lack of high-reddening objects in the sample; however, modelling censored data leads to robust $\tau_A$ inferences. Sub-figure (b) shows the $\mu_{R_V}$ inferences are also affected by the censoring process. When censored SNe are ignored, $\mu_{R_V}$ inferences are $\Delta \mu_{R_V}\approx 0.3$ higher than the input value, but including the censored data leads to robust $\mu_{R_V}$ recovery.
    }
    \label{fig:censoredsbc}
\end{figure*}
 
We show using simulations that analysing low-reddening sub-samples in isolation can bias dust hyperparameter inferences. We simulate 100 samples of 100 SNe, fit the low-reddening sub-samples
(by applying the $|B-V|<0.3$~mag cut), and assess hyperparameter recovery when censored data (\S\ref{S:CensoredDataInferences}) are either included or ignored. Each SN sample is simulated using dust hyperparameters: $\mu_{R_V}=2.5$, $\sigma_{R_V}=0.5$ and $\tau_A=0.5$~mag. We use the exponential $A_V^s$ distribution for simulating and fitting, and simulate using the posterior median $\bm{\mu}_{\rm{int}}, \bm{\Sigma}_{\rm{int}}$ from fitting the real low-reddening sample with 3 censored SNe under the exponential $A_V^s$ distribution. 

Fig.~\ref{fig:censoredsbc} shows $\tau_A$ is biased low when censored data are ignored, $\Delta \tau_A \approx -0.2$~mag, to account for the lack of high-reddening objects in the sample. Consequently, the $\mu_{R_V}$ inferences are also affected, and are $\Delta \mu_{R_V}\approx 0.3$ higher than the input value. However, by modelling censored SNe, population inferences are robust. For this specific set of input hyperparameters and sample size, $\tau_A$ is significantly offset from the truth when excluding censored SNe, while the $\mu_{R_V}$ inferences are consistent with the truth given the posterior uncertainties. However, with reduced statistical uncertainties, and/or different input hyperparameters, the $\mu_{R_V}$ offset may be significant. 

These simulations serve to demonstrate the underlying principle, that analysing low-reddening sub-samples in isolation can bias dust hyperparameter inferences. Therefore, to robustly infer dust hyperparameters in low-reddening cosmological samples, information about SNe cut as a result of the censoring process should be incorporated. The minimum information required is the censored data (the number of cut SNe, and their $B-V$ measurement errors). To model censored data is to assume the censored SNe are drawn from the same population distributions as the low-reddening sub-sample. Therefore, the best constraints are obtained by including all available data of the moderate-to-high-reddening SNe that are consistent with the low-reddening population distributions.

\vspace{-0.5cm}
\section{Number of Censored SNe}
\label{S:appendixadditionalppc}

\begin{figure*}
\centering
    \begin{subfigure}[b]{0.49\textwidth}
        \centering
        \includegraphics[width=\textwidth]{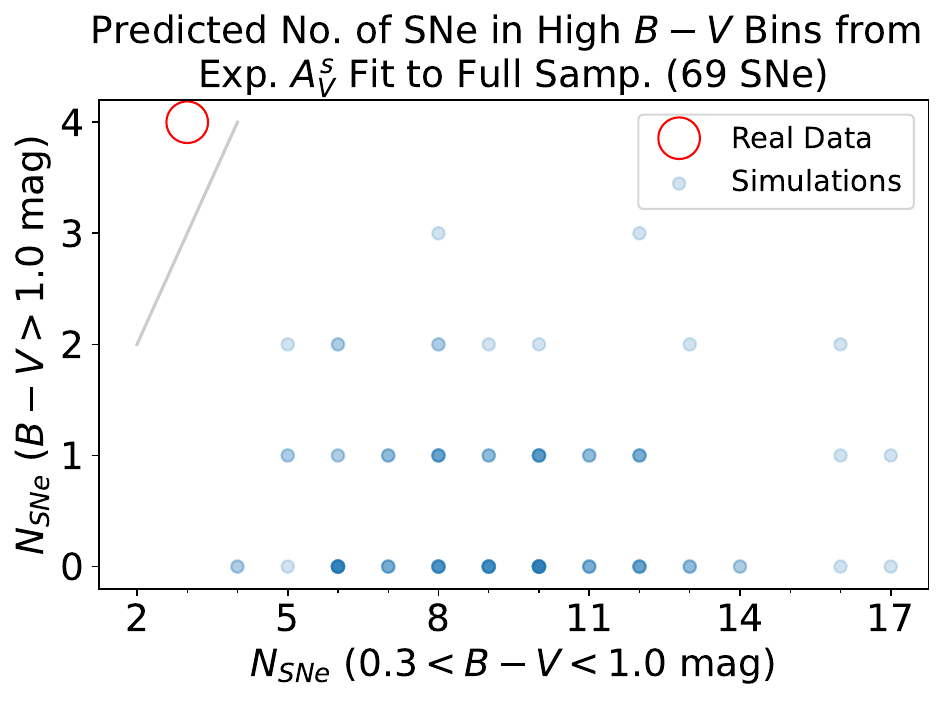}
        \includegraphics[width=\textwidth]{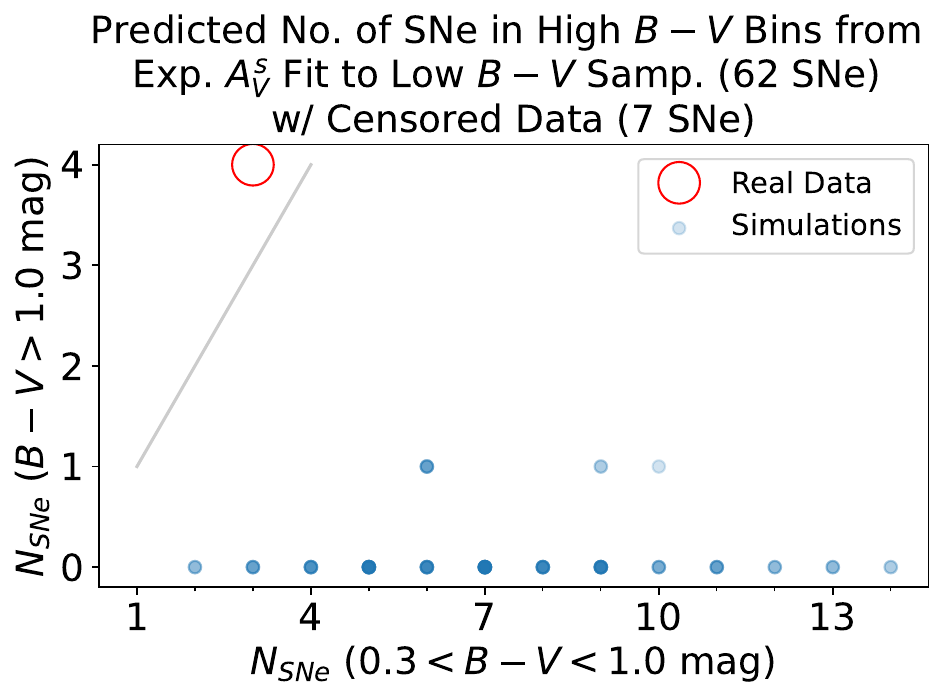}
        \includegraphics[width=\textwidth]{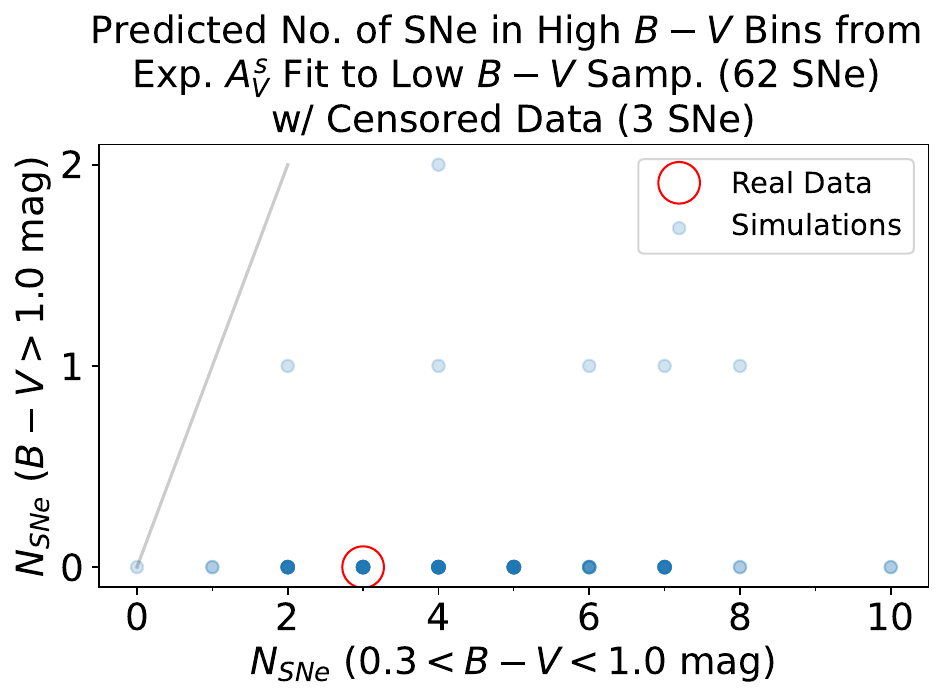}
        \caption{Predicted $N_{\rm{SNe}}$ with $A_V^s\sim$Exp$(\tau_A)$.}
    \end{subfigure}
    \hfill
    \begin{subfigure}[b]{0.49\textwidth}
        \centering
        \includegraphics[width=\textwidth]{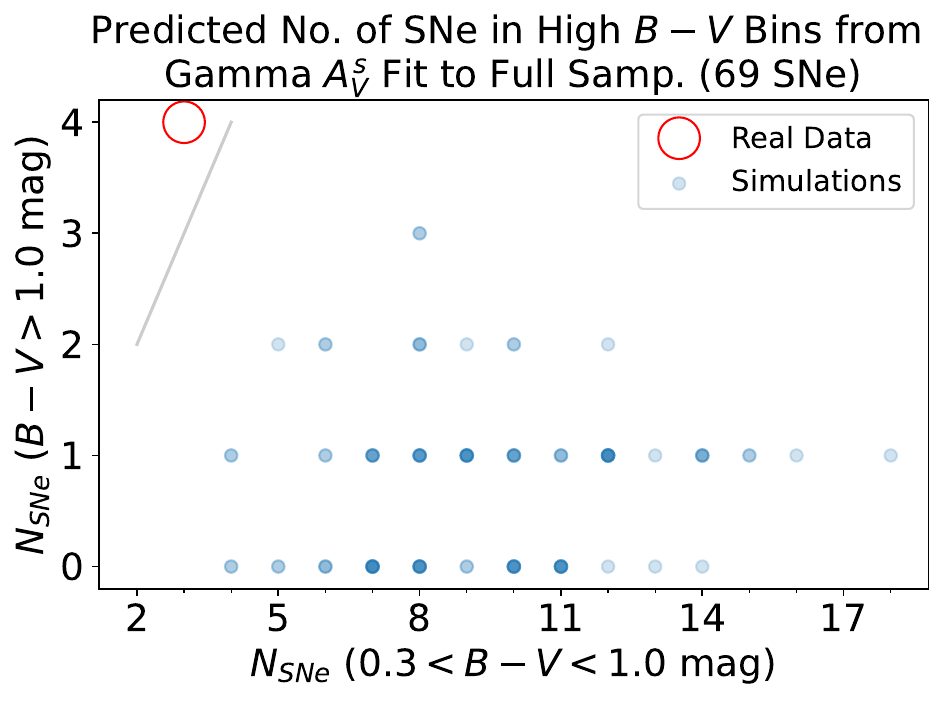}
        \includegraphics[width=\textwidth]{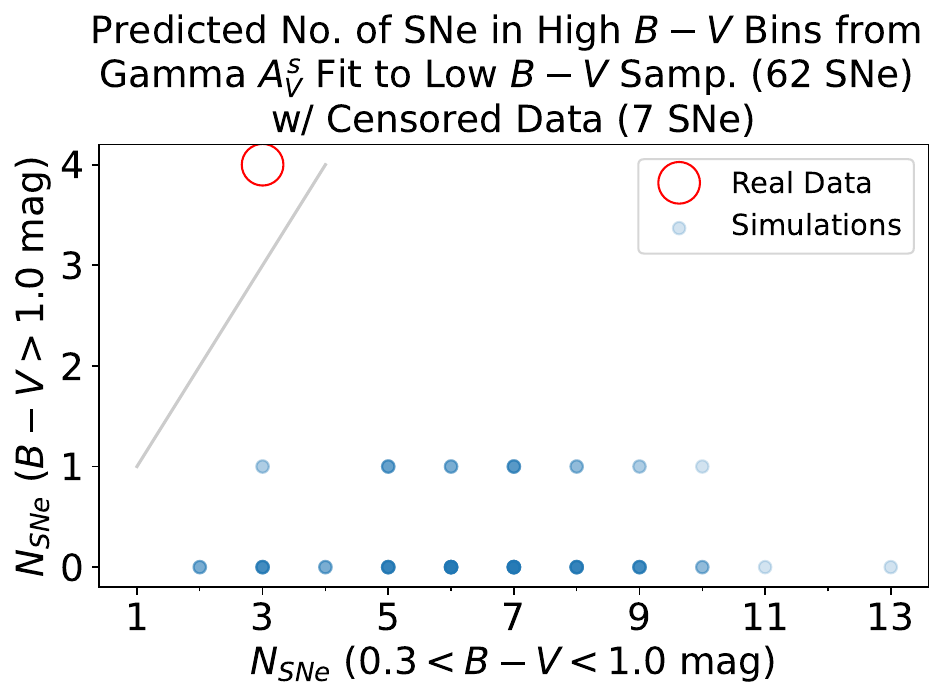}
        \includegraphics[width=\textwidth]{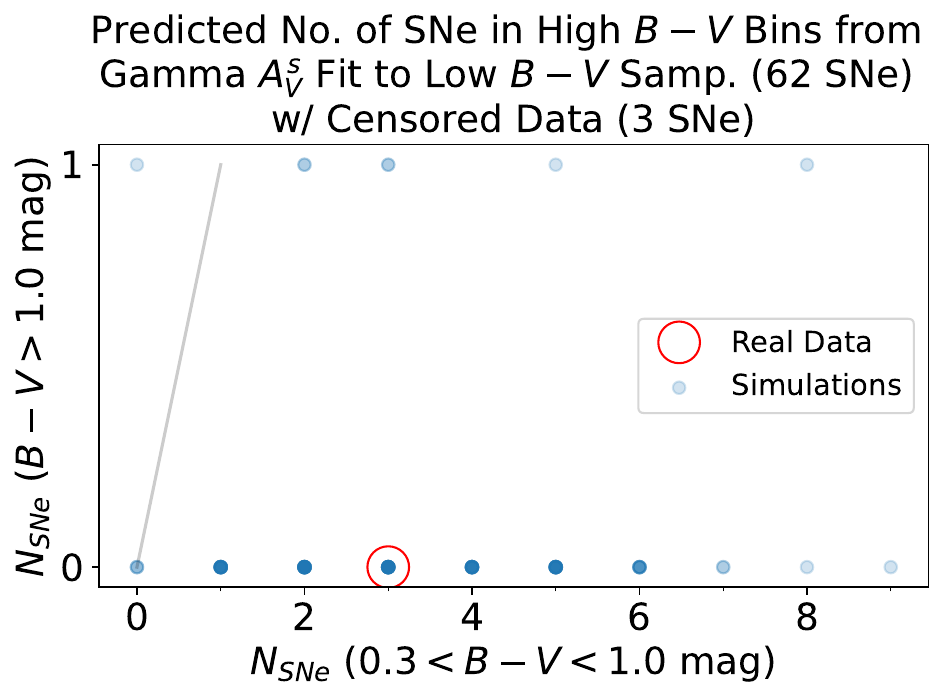}
        \caption{Predicted $N_{\rm{SNe}}$ with $A_V^s\sim$ Gamma$(\nu_A,\tau_A)$.}
    \end{subfigure}
    \caption{Using the posterior median hyperparameters from fits to real data, we simulate 100 new samples of SNe, and predict the number of SNe with true rest-frame apparent $B-V$ colour in the ranges $0.3<B-V<1.0$~mag, and $B-V>1.0$~mag. The boldness of the scatter points indicates the number of simulations with high $B-V$ SNe at that grid location.    
    We then compare to the data, which is denoted by the \textcolor{red}{red circle}. Left and right columns of panels show predicted number of SNe from fitting with the exponential or gamma distributions, respectively; from top to bottom, rows show predictions from fitting the full sample, or the low-reddening sample with 7 censored SNe, or 3 censored SNe, respectively; the grey faint lines show the axis of symmetry between the $B-V$ bins. These results show exponential/gamma models with 3 censored SNe are viable descriptions of our dataset. These $A_V^s$ distributions are unable to replicate the asymmetric cluster of 4 highly reddened SNe with $B-V>1.0$~mag, seen in the real data; therefore, we prefer to exclude these 4 objects from the dataset.
    }
    \label{fig:Nppc}
\end{figure*}

We use posterior predictive checks ~\citep[e.g][]{Mandel11} to assess which high-reddening SNe should be modelled in the fiducial analysis as censored SNe. We use the posterior samples from fits to real data to predict how many SNe with high $B-V$ apparent colours are expected. We extract the posterior medians of the dust and intrinsic deviation hyperparameters, and simulate new $(\bm{\delta N}^s, A_V^s, R_V^s)$ samples. We then plot the number of SNe with latent $B-V$ apparent colours in the range $0.3<B-V<1.0$~mag, against the number with $B-V>1.0$~mag. This is compared against the data, which has 3 and 4 SNe in each bin, respectively. We do this for 100 simulations.

Fig.~\ref{fig:Nppc} shows both the exponential and gamma $A_V^s$ distributions are unable to predict the $B-V$ distribution observed in the full sample. The posterior always predicts there are more SNe in the lower $0.3<B-V<1.0$~mag bin than in the higher $B-V>1.0$~mag bin, unlike the real data. Moreover, the number of SNe with $0.3<B-V<1.0$~mag is typically larger than 3, numbering $\approx 5-10$. However, when these $A_V^s$ distributions are applied to model only the sample of 65 SNe, with the 4 $B-V>1.0$~mag objects excluded, the models successfully predict 3 SNe should occupy the $0.3<B-V<1.0$~mag bin. These posterior predictive checks (Fig.~\ref{fig:Nppc}) show the 3 censored SNe with $0.3<B-V<1.0$~mag should be included in the model, while the 4 SNe with $B-V>1.0$~mag should be excluded altogether.

\section{Sensitivity of $\mu_{R_V}$ to high-reddening objects, $A_V^s$ prior, and $B-V$ cut}
\label{S:appendixmuRV}
We explore how applying the gamma $A_V^s$ distribution affects $\mu_{R_V}$ inferences when fitting the: 
\begin{itemize}
    \item Full sample ($N_{\rm{SNe}}=69$; includes the 4 high-reddening SNe),
    \item Fiducial low-to-moderate reddening sample $(N_{\rm{SNe}}=65)$,
    \item Low-reddening sample $(N_{\rm{SNe}}=62)$.
\end{itemize}
In particular, we note that applying the gamma distribution to fit either the fiducial or low-reddening samples leads to a higher posterior median $\mu_{R_V}$, compared to applying the exponential distribution. Meanwhile, fitting the full sample with the gamma distribution leads to a lower $\mu_{R_V}$. We aim to determine with simulations which behaviours are typical, and the cause of $\mu_{R_V}$ moving in opposite directions for different samples when fitting the gamma distribution.

We start by investigating the $R_V$ distribution in the high-reddening cluster of 4 SNe ($B-V>1.0$~mag). Appendix~\ref{S:appendixadditionalppc} shows these SNe are unlikely to belong to the same exponential or gamma $A_V^s$ distributions as the remaining 65 SNe. Therefore, we model these 4 SNe using Gaussian $A_V^s$ and $R_V^s$ population distributions. We further assume that the intrinsic distribution is the same as the low-to-moderate reddening sample, so fix the intrinsic hyperparameters to the posterior medians from the fiducial fit. Our hyperprior on $\sigma_{A}$, the Gaussian $A_V^s$ dispersion, is a Half-Normal distribution with a dispersion of 1~mag, and the hyperpriors on all other hyperparameters are unchanged.
\begingroup
\allowdisplaybreaks
\begin{align}
    \sigma_{A} \sim \textrm{Half-}\mathcal{N}(0,1^2)
    \\
    A_V^s \sim \mathcal{N}(\tau_A, \sigma^2_A)
\end{align}
\endgroup
This fit yields: $\mu_{R_V}=2.00^{+0.07}_{-0.08}$, $\sigma_{R_V}<0.10(0.29)$, $\tau_A=2.30^{+0.11}_{-0.12}$~mag and $\sigma_{A}=0.17^{+0.16}_{-0.08}$~mag. With the caveat that we have assumed a common intrinsic distribution across reddening bins, our low $\mu_{R_V}$ inference is broadly consistent with other literature studies of high-reddening SNe~Ia, which typically find $R_V\approx 1.5-2$~\citep[e.g.][]{Elias-Rosa08, Wang08:06x, Mandel11, Amanullah14, Hoang17}. 

The assumption of a common intrinsic distribution may be invalid, seeing as the dust population distributions differ between the low and high reddening bins. Nonetheless, the resulting low $R_V^s$ values may pull the $\mu_{R_V}$ inference down in the full sample fits. To investigate this, and the effect of fitting the gamma distribution, we simulate 100 sets of 69 SNe, by combining the population distributions for the fiducial and high-reddening samples. For 65 SNe, we simulate using the posterior median hyperparameters from the fiducial fit, and for the remaining 4 SNe, we use the same intrinsic hyperparameters combined with the posterior median extrinsic hyperparameters from the Gaussian $A_V^s$ distribution fit. We then fit the combined samples using either an exponential or gamma $A_V^s$ distribution, and record the simulation-averaged posteriors, and the number of simulations where the posterior median $\mu_{R_V}$ is higher with the gamma distribution than with the exponential.

Results in Table~\ref{tab:simmuRVposteriors} show that the $\mu_{R_V}$ inferences are higher with the gamma distribution in only 7\% of simulations, which aligns with the decrease in $\mu_{R_V}$ in the real-data inferences. The trend is thus likely an artefact from modelling two distinct samples with a single set of distributions, and apparently low $R_V^s$ values at high $A_V^s$.

We now investigate the fiducial and low-reddening sample inferences. As above, we perform simulations using the posterior median hyperparameters from the fiducial fit. Table~\ref{tab:simmuRVposteriors} shows the posterior median $\mu_{R_V}$ is higher when fitting the gamma distribution in 75\% of fiducial-sample simulations. This increases to 90\% for the low-reddening subsamples. Therefore, an increase in $\mu_{R_V}$ when fitting the gamma distribution is typical behaviour.

\renewcommand{\arraystretch}{1.5}
\setlength{\tabcolsep}{3pt}
\begin{table*}
\begin{threeparttable}
    \centering
    \caption{
    Simulation-averaged and real-data $\mu_{R_V}$ posteriors. 100 sets of $N_{\rm{SNe}}$ objects are simulated and fitted. 
    }
    \label{tab:simmuRVposteriors}
    \begin{tabular}{l c c c c c c}
    \toprule
        
        & \multicolumn{3}{c}{Sim. Posterior\,\tnote{a}} & \multicolumn{3}{c}{Data Posterior\,\tnote{b}}\\
        \cmidrule(lr){2-4}\cmidrule(lr){5-7}
\midrule
Sample & Full\,\tnote{c} & Fiducial\,\tnote{d} & Low-reddening\,\tnote{e} & Full & Fiducial & Low-reddening \\ 
        Colour Cuts
        &  No Cut & $|B-V|<1.0$~mag & $|B-V|<0.3$~mag 
        &  No Cut & $|B-V|<1.0$~mag & $|B-V|<0.3$~mag \\
$N_{\rm{SNe}}$ & 69 & 65 & $\approx$62  & 69 & 65 & 62\\
\midrule
$A_V^s$ Distribution\,\tnote{f} \\
\midrule
$A^s_V \sim \rm{Exp}(\tau_A)$ &  $2.56^{+0.42}_{-0.37}$ &  $2.80^{+0.53}_{-0.47}$  & $3.08^{+0.63}_{-0.57}$ & $2.42^{+0.37}_{-0.31}$ & $2.61^{+0.38}_{-0.35}$ & $2.61^{+0.41}_{-0.35}$\\ 
$A_V^s \sim $ Gamma$(\nu_A,\tau_A)$ & $2.42^{+0.42}_{-0.42}$ & $2.89^{+0.55}_{-0.47}$ & $3.26^{+0.63}_{-0.52}$ & $2.15^{+0.29}_{-0.18}$ & $3.00^{+0.78}_{-0.53}$ & $3.43^{+0.85}_{-0.83}$\\
\midrule
$p_{\rm{median}}$\,\tnote{g} & 7/100 & 75/100 & 90/100 & - & - & - \\
\bottomrule
    \end{tabular}
    \begin{tablenotes}
        \item[a] The simulation-averaged posterior. 100 sets of $N_{\rm{SNe}}$ supernovae are simulated and fitted, and the posterior samples are collated and summarised using the posterior median and the 16\% and 84\% credible intervals. 
        \item[b] The posterior estimates from fitting the real-data (sub-)samples (c.f. Tables~\ref{tab:AVspriorCens}, \ref{tab:AVExpGammafiducial}).
        \item[c] 
        The 4 high-reddening objects ($B-V>1.0$~mag) are simulated using the same \textit{intrinsic} posterior median hyperparameters from the fiducial fit, but using \textit{extrinsic} posterior median hyperparameters from fitting the 4 SNe with a Gaussian $A_V^s$ distribution whilst fixing the intrinsic hyperparameters at the fiducial-fit values.
        \item[d] The low-to-moderate reddening SNe are simulated using the posterior median hyperparameters from fitting the fiducial sample of 65 SNe with the exponential $A_V^s$ prior.
        \item[e] SN samples are simulated as for the low-to-moderate reddening sample, but then cut on $B-V<0.3$~mag, which retains $\approx 62$ out of 65 simulated SNe.
        \item[f] The $A_V^s$ prior used to fit simulated samples, and the real-data samples.
        \item[g] The number of simulations where the $\mu_{R_V}$ posterior median from fitting the gamma distribution is higher than from fitting the exponential.
    \end{tablenotes}
    \end{threeparttable}
\end{table*}

\vspace{-0.5cm}
\section{Data Preprocessing Results}
\label{S:appendixpreproc}

\renewcommand{\arraystretch}{1.2}
\setlength{\tabcolsep}{8pt}
\begin{table}
\begin{threeparttable}
    \centering
    \caption{
     Sensitivity of dust inferences to data preprocessing choices.}
    \label{tab:preprocessingeffect}
    \begin{tabular}{l c c c c c c}
    \toprule
        Hyperparameters &
        $\tau_A$ (mag)\,\tnote{a} &  
        $\mu_{R_V}$ & $\sigma_{R_V}$\,\tnote{b}\\
\midrule
\bf{Default}\,\tnote{c} &  $0.32^{+0.06}_{-0.05}$ & $2.61^{+0.38}_{-0.35}$ & $< 0.92 (1.96)$ \\ 
\midrule
\midrule
Interp. Filters & 
\multicolumn{3}{c}{\bf{Default} : $uBgVriYJH$} \\
\midrule
$BVriJH$ & $0.34^{+0.06}_{-0.05}$ & $2.76^{+0.41}_{-0.35}$ & $< 0.99 (2.07)$ \\
\midrule
\midrule
EBVMW Method & 
\multicolumn{3}{c}{\bf{Default} : \textsc{SNooPy} $E(B-V)_{\rm{MW}}$} \\
\midrule
Pre-subtract & $0.32^{+0.06}_{-0.05}$ & $2.59^{+0.34}_{-0.29}$ & $< 0.69 (1.71)$ \\ 
\midrule
\midrule
K-corrections & \multicolumn{3}{c}{\bf{Default} : With Mangling } \\
\midrule
No Mangling & $0.34^{+0.06}_{-0.05}$ & $2.60^{+0.45}_{-0.39}$ & $< 1.17 (2.33)$ \\ 
\midrule
\midrule
$T_{B;\,\rm{max}}$ Method & \multicolumn{3}{c}{\bf{Default} : 1DGP } \\
\midrule  
2DGP & $0.32^{+0.06}_{-0.05}$ & $2.69^{+0.40}_{-0.33}$ & $< 0.99 (2.06)$ \\ 
\textsc{SNooPy} & $0.33^{+0.06}_{-0.05}$ & $2.80^{+0.48}_{-0.52}$ & $1.28^{+0.77}_{-0.52}$ \\ 
\midrule
\midrule
Interp. Method & \multicolumn{3}{c}{\bf{Default} : 2DGP Mag. Interp.} \\
\midrule 
2DGP Flux Interp. & $0.32^{+0.06}_{-0.05}$ & $2.69^{+0.35}_{-0.29}$ & $< 0.87 (1.73)$ \\ 
1DGP Mag Interp. & $0.32^{+0.06}_{-0.05}$ & $2.60^{+0.39}_{-0.31}$ & $< 0.90 (1.96)$ \\ 
1DGP Flux Interp. & $0.32^{+0.05}_{-0.04}$ & $2.70^{+0.33}_{-0.31}$ & $< 0.86 (1.58)$ \\ 
\bottomrule
    \end{tabular}
    \begin{tablenotes}
        \item[a] Summaries are the posterior medians and 68\% credible intervals.
        \item[b] The 68\% (95\%) quantiles are tabulated for posteriors that peak near the lower prior boundary.
        \item[c] Default data preprocessing choices are in \textbf{bold}. Left column denotes alternative choices, described in full in \S\ref{s:senstolameffpreproc}.
    \end{tablenotes}
    \end{threeparttable}
\end{table}

We test the sensitivity of dust hyperparameter inferences to the data preprocessing choices. 
These include the choice of: interpolation filters ($uBgVriYJH$ or $BVriJH$), 
method for computing Milky Way extinction corrections (using \textsc{SNooPy}, or pre-subtracting Milky Way extinction using the \texttt{extinction}\footnote{\url{https://extinction.readthedocs.io/en/latest/}} package before feeding in to \textsc{SNooPy}), choice of K-corrections (with or without mangling), method for estimating $T_{B;\,\rm{max}}$ (1DGP, 2DGP or \textsc{SNooPy}-fit estimate),
and method for interpolating rest-frame data (2DGP or 1DGP fit to magnitude or flux data). 

Table~\ref{tab:preprocessingeffect} shows the data-processing choices have an insignificant effect on dust hyperparameter inferences. Most entries in Table~\ref{tab:preprocessingeffect} have $|\Delta \mu_{R_V}|\lesssim 0.1$ compared to the default inference, and all entries have $|\Delta \mu_{R_V}| < 0.2$. These $\lesssim 0.1-0.3\sigma$ differences are insignificant compared to uncertainties, $\approx 0.5$. Similarly, the $\tau_A$ and $\sigma_{R_V}$ values are strongly consistent. The $\sigma_{R_V}$ posterior peaks away from zero when using the \textsc{SNooPy} $T_{B;\,\rm{max}}$ estimates, implying they are less accurate. The 1DGP $T_{B;\,\rm{max}}$ estimates are typically earlier than the \textsc{SNooPy} estimates, by $\Delta T_{B;\,\rm{max}} = -0.41\pm0.86$~days (median and sample standard deviation). Following visual inspection of the light curves, we judge the 1DGP yields a more robust, data-driven $T_{B;\,\rm{max}}$ estimate compared to \textsc{SNooPy}.

\section{Intrinsic Model Results}
\label{S:appendixintrinsicmodel}

\begin{figure}
    \centering
    \includegraphics[width=1\linewidth]{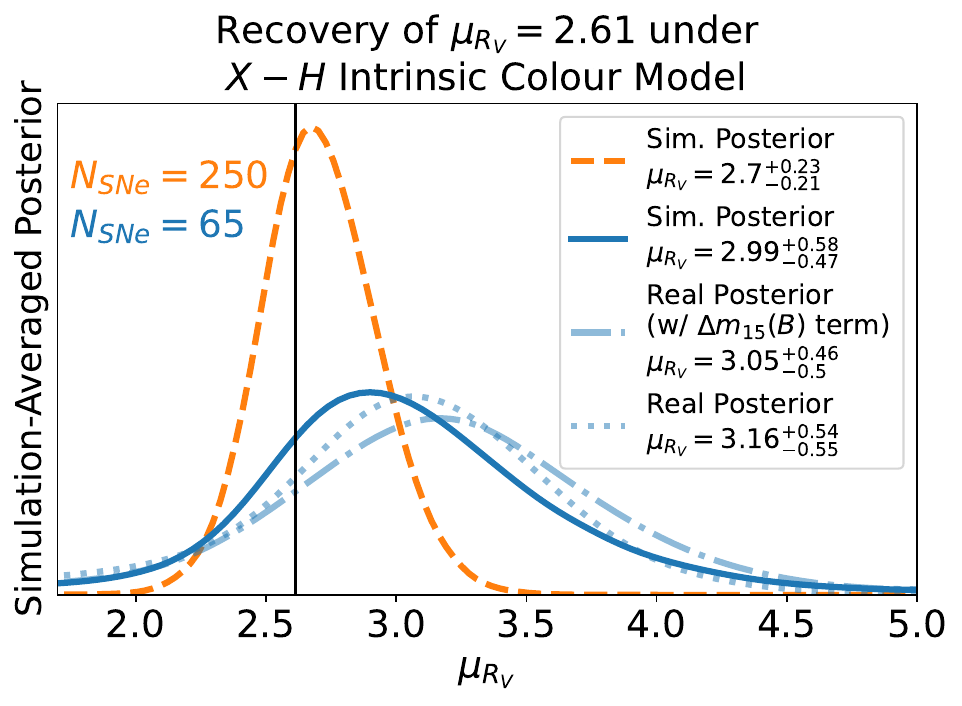}
    \caption{
    Recovery of input dust hyperparameter $\mu_{R_V}=2.61$ when applying the $X-H$ intrinsic colour model to fit SNe simulated from the intrinsic deviations model. For 65 SNe, the $X-H$ model yields high and wide $\mu_{R_V}$ posteriors, strongly consistent with the real-data posterior. For 250 SNe, the simulation-averaged posterior narrows to $\mu_{R_V}=2.70^{+0.23}_{-0.21}$,
    much closer to the input value $\mu_{R_V}=2.61$.
    }
    \label{fig:XHModelPPC}
\end{figure}

In \S\ref{S:magscolours}, we show dust hyperparameter inferences have a dependency on the intrinsic SN model. More specifically, the choice of reference frame in which the multivariate Gaussian hyperpriors are placed (to model the intrinsic chromatic variations) affects the posterior inferences. We proceed to assess recovery of input dust hyperparameters, by applying the intrinsic colour models to SNe simulated using the deviations model. As in \S\ref{S:AVspriorCens}, we simulate using the posterior median hyperparameters from fits to the real data. 

The simulation-averaged $\mu_{R_V}$ posteriors under the adjacent, $B-X$, and $X-H$ colour models are: $\mu_{R_V} = 2.62 ^{+0.52}_{-0.50}, 2.70^{+0.49}_{-0.46}, 2.99^{+0.58}_{-0.47}$, respectively. This ordering of medians from lowest to highest is the same as for the real data inferences. Moreover, the $X-H$ simulations are consistent with the real-data inference, which is shown visually in Fig.~\ref{fig:XHModelPPC}. These results indicate the high real-data $X-H$-model $\mu_{R_V}$ inference is likely a consequence of the intrinsic hyperpriors. 

To assess the dominance of this systematic uncertainty when statistical uncertainties are reduced, we fit larger simulated samples of 250 SNe. Fig.~\ref{fig:XHModelPPC} shows the resulting $X-H$-model simulation-averaged posterior, $\mu_{R_V}=2.70^{+0.23}_{-0.21}$, is still consistent with $\mu_{R_V}=2.61$. This shifting of posteriors towards the true values with larger samples aligns with our expectations, because the improved statistics should reduce the sensitivity to the hyperprior. Nonetheless, we cannot rule out that this systematic is significant for some regions of hyperparameter space. This should be investigated thoroughly when the real-data posteriors for different choices of reference frame are inconsistent.

\bsp	
\label{lastpage}
\end{document}